\documentclass[12pt,aps,prb,amsmath,amssymb,sort&compress,comma,super,nofootinbib,floatfix]{revtex4-1}

\usepackage{graphics}
\usepackage{epsfig}
\usepackage{subfigure}  

\begin{document}

\newcommand{\etal}{{\it et al.}\/}
\newcommand{\gtwid}{\mathrel{\raise.3ex\hbox{$>$\kern-.75em\lower1ex\hbox{$\sim$}}}}
\newcommand{\ltwid}{\mathrel{\raise.3ex\hbox{$<$\kern-.75em\lower1ex\hbox{$\sim$}}}}

\title{A Common Thread: the pairing interaction for the unconventional superconductors}

\author{D.~J.~Scalapino}
\email{djs@physics.ucsb.edu}
\affiliation{Department of Physics, University of California, Santa Barbara, CA 93106-9530 USA}


\begin{abstract}
The structures, the phase diagrams, and the appearance of a neutron resonance
signaling an unconventional superconducting state provide phenomenological
evidence relating the cuprates, the Fe-pnictides/chalcogenides as well as some
heavy fermion and actinide materials. Single- and multi-band Hubbard models
have been found to describe a number of the observed properties of these
materials so that it is reasonable to examine the origin of the pairing
interaction in these models. In this review, based on the experimental phenomenology
and studies of the pairing interaction for Hubbard-like models, it is proposed
that spin-fluctuation mediated pairing is the common thread linking a
broad class of superconducting materials.
\end{abstract}


\maketitle
\tableofcontents

\section{Introduction}\label{sec:1}

\citet{ref:Fisk} have noted that a striking aspect of superconducting materials is
the ``remarkable amount of phase space they inhabit: superconductivity is
everywhere but sparse. So the central question in superconductivity and the
search for new superconducting materials is whether there is anything common
to the known superconductors." This review addresses this question by examining
common features of the cuprate and iron superconductors as well as some
heavy fermion and actinide superconductors to see what they tell us about the
pairing mechanism in these materials.\footnote{A brief account of this was given
in the Proceedings of the M2S -- IX Conference, Physica C {\bf 470}, 51-54 (2010).}

We begin in Sec.~\ref{sec:2} by looking at the crystal structures, the phase
diagrams, the coexistence and interplay of antiferromagnetism and
superconductivity and a neutron scattering spin resonance which is observed in
the superconducting phase. One finds that these materials come in families which
have quasi-2D layers containing square arrays of $d$- or $f$-electron cations. Their
temperature-doping and magnetic field phase diagrams show antiferromagnetism
in close proximity, or in some cases coexisting, with superconductivity. A
variety of experiments show that the antiferromagnetism and superconductivity
are strongly coupled. A spin resonance peak, which is observed in inelastic
neutron scattering experiments in the superconducting phase, provides evidence
of unconventional pairing. The similarity of the structures, the phase diagrams,
the interplay of antiferromagnetism and superconductivity, and the unconventional
nature of the superconductivity seen in these materials suggest they share a
common underlying physics.

Sec.~\ref{sec:3} contains a discussion of models that have been used to
describe these materials. These are minimal models in which the cuprates
are described by a single-band 2-dimensional Hubbard model while the heavy
fermion and Fe materials involve orbital degenerate multi-band models.
Various numerical calculations as well as approximate analytic calculations
find that these models exhibit a number of phenomena which are experimentally
observed in these materials. In particular, the close proximity of an
antiferromagnetic or spin-density-wave phase to an unconventional $d$-wave or
sign changing $s$-wave superconducting phase is found to be a
common feature. A second important common feature is the dual character of the
$3d$ or $4f$ electrons in these models. These electrons are involved in both
the magnetism and the superconductivity. The models can exhibit behavior ranging from
local moments and insulating antiferromagnetic order to itinerant magnetism,
stripes and superconductivity. Furthermore the models show the close relationship
between $d$-wave and $s^\pm$-wave pairing.

Motivated by this, the momentum, frequency and orbital dependence of the
interaction which is responsible for pairing in these models is examined
in Section~\ref{sec:4}. The ``same electrons" that are associated with the
magnetism and superconductivity are found to give rise to a spin-fluctuation
mediated pairing interaction. The short range near-neighbor antiferromagnetic
fluctuations give rise to a sign changing gap (${\rm Sgn}\Delta(k+Q)=-{\rm Sgn}\Delta(k)$)
for large momentum transfers. Appendix~A contains a comparison of the traditional
electron-phonon-Coulomb pairing interaction with this interaction. Based on the
experimental phenomenology and the analysis of the models, it is proposed that
this spin-fluctuation pairing interaction is the common thread that links this
class of unconventional superconducting materials. Although the organic Bechgaard
salts \citep{ref:Bechgaard} will not be discussed, they clearly are also part of
this class of materials \citep{ref:Taill28,ref:Taillefer,ref:Doiron}.
Section~\ref{sec:5} contains a brief summary and an outlook regarding the guidance
this brings to the search for higher $T_c$ materials.

\section{Common Features of a Class of Unconventional Superconductors}\label{sec:2}

In this section we begin by looking at similarities in the structures and the
phase diagrams of some heavy fermion, cuprate and iron-based superconductors.
Following this, experimental evidence of the interplay of antiferromagnetism
and superconducting and the dominant role of spin-fluctuation scattering in
these materials will be discussed. The section concludes with an experimental
definition of what we will call ``unconventional superconductors" in this review.

\subsection{Structures}

\begin{figure}[!htbp]
\includegraphics[height=8.5cm]{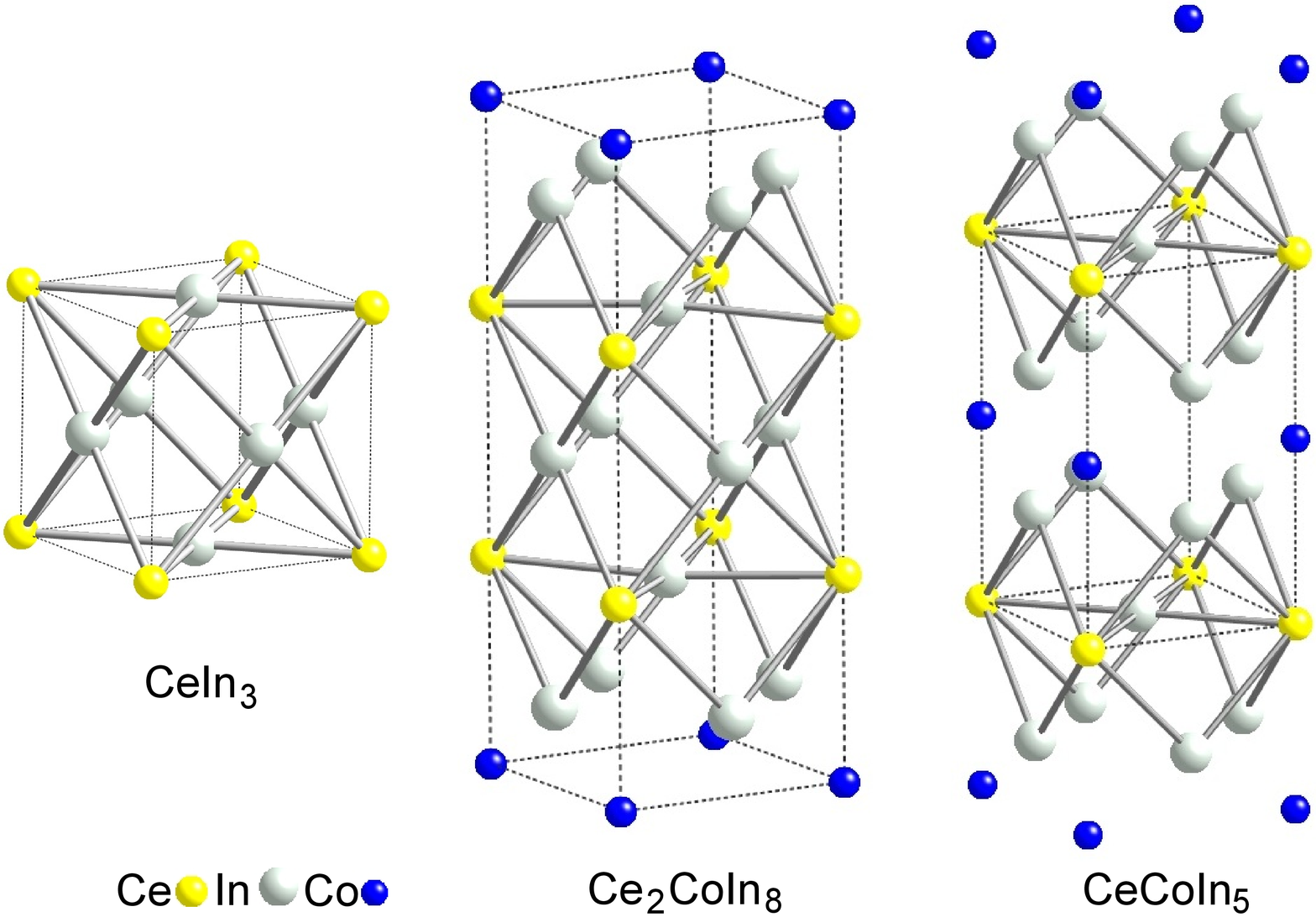}
\caption{Some members of the Ce family of heavy fermion superconductors. The key
structural element is the quasi-two-dimensional layer of Ce$^{3+}$ ions which sit at the
center of a tetragon formed by 12 near neighbor In$^-$ anions. ($T_c\sim0.2$ K
CeIn$_3$ \protect\citep{ref:Shishido}, 1.0 K Ce$_2$CoIn$_8$ \protect\citep{ref:Chen},
2.3 K CeCoIn$_5$ \protect\citep{ref:CeCoIn5})\label{fig:1}}
\end{figure}

\begin{figure}[!htbp]
\includegraphics[height=8.5cm]{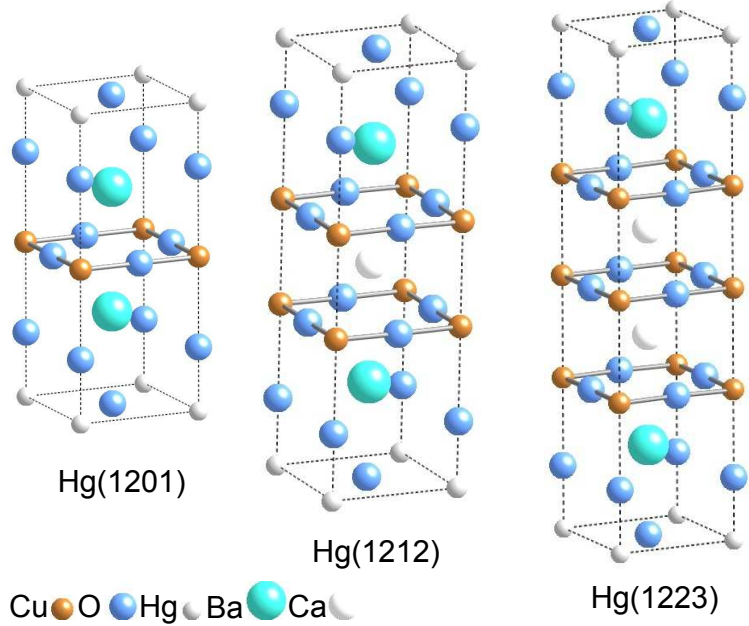}
\caption{The key element of the Hg-cuprate superconductors is the CuO$_2$ layer.
The 1201 structure on the left has apical O's above and below the Cu sites
while the inner CuO$_2$ layer of the 1223 structure on the right has no apex
oxygens (optimally doped $T_c\sim94$ K Hg(1201), 127 K Hg(1212), 135 K Hg(1223)
\protect\citep{ref:Wagner}).\label{fig:2}}
\end{figure}

\begin{figure}[!htbp]
\includegraphics[height=8.5cm]{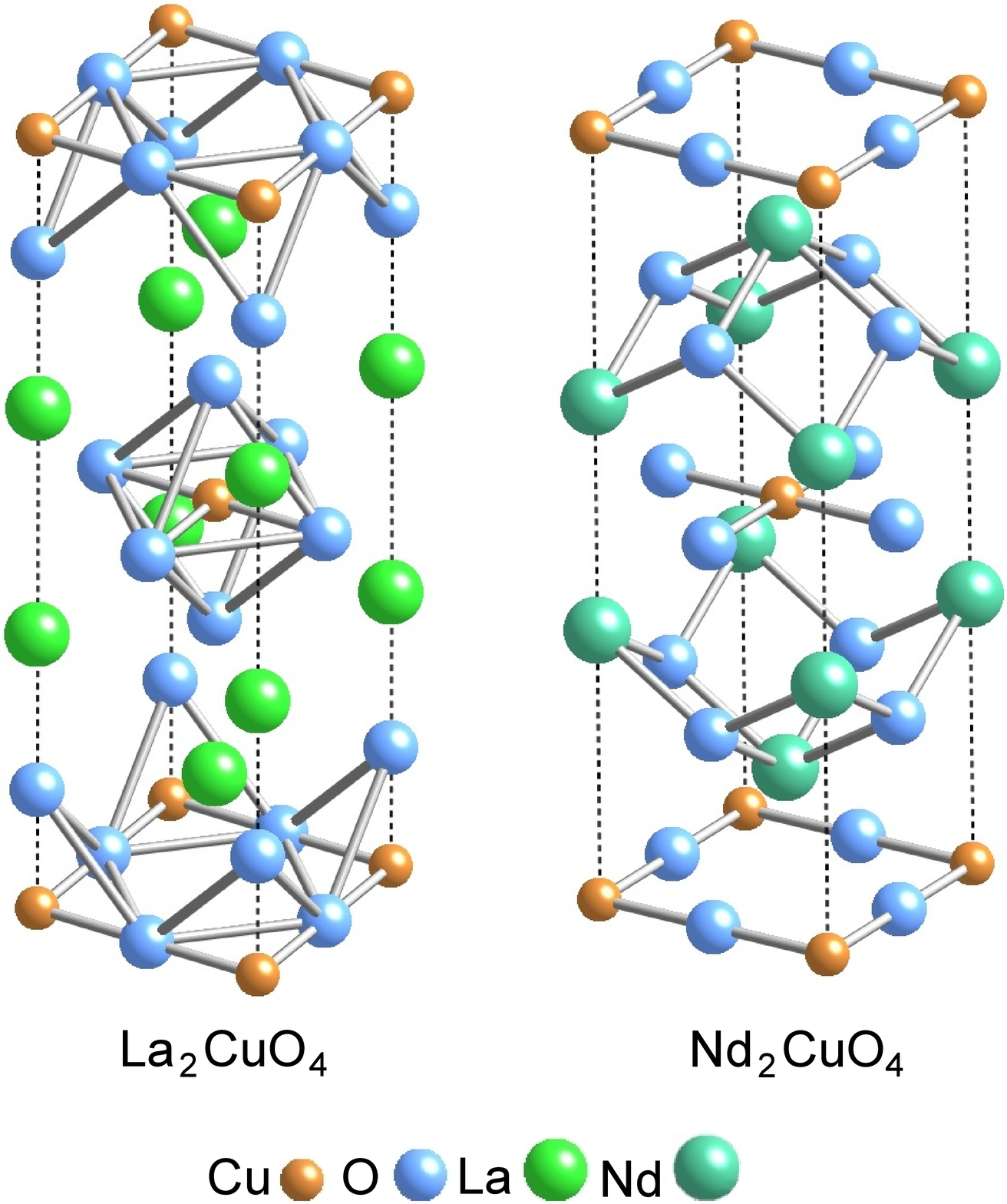}
\caption{The 214 cuprate structures La$_2$CuO$_4$ and Nd$_2$CuO$_4$. The former
can be hole doped and the latter structure which is missing the apex oxygens
can be electron doped. ($T_c\sim38$ K La$_{1.85}$Sr$_{0.15}$CuO$_4$
\protect\citep{ref:Takagi}, 25 K Nd$_{1.85}$Ce$_{0.15}$CuO$_4$ \protect\citep{ref:Yamada})
\label{fig:3a}}
\end{figure}

\begin{figure}[!htbp]
\includegraphics[height=8.5cm]{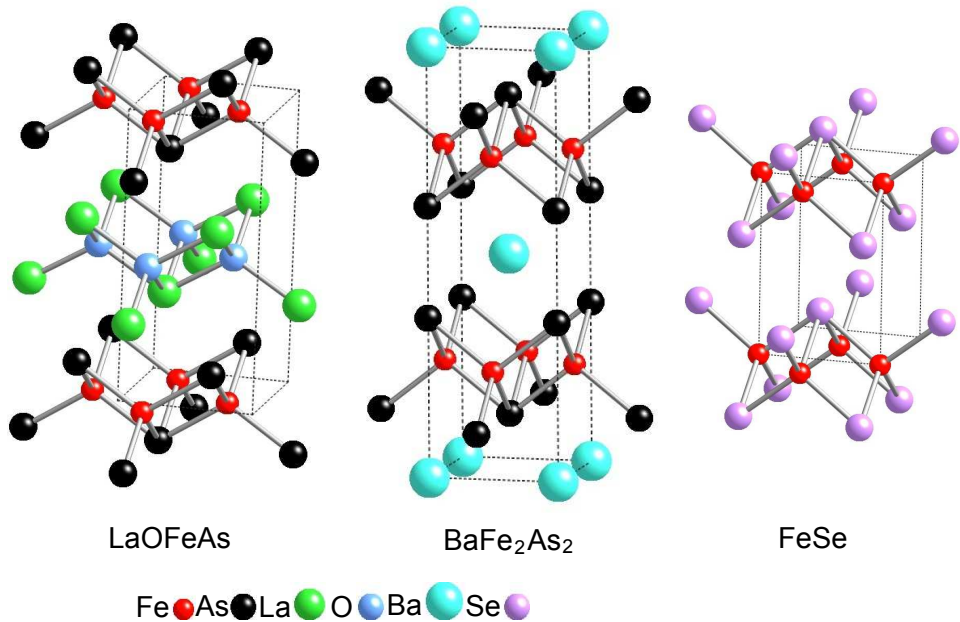}
\caption{Examples of the Fe-based superconductors. Here the key element is the
Fe-pnictide or chalcogen layer. ($T_c\sim26$ K La(O$_{0.92}$F$_{0.08}$)FeAs
\protect\citep{ref:Cruz}, 22 K Ba(Fe$_{0.92}$Co$_{0.08})_2$As$_2$ \protect\citep{ref:Delaire},
38 K (Ba$_{0.6}$K$_{0.4}$)Fe$_2$As$_2$ \protect\citep{ref:Rotter}, 13.6 K to 37 K
(4.5GPa)FeSe \protect\citep{ref:Okabe})\label{fig:3}}
\end{figure}

As illustrated\footnote{These illustrations were made by N.~Ghimire using a
CrystalMaker 8.5 software package. D.R.~Harshman, A.T.~Fiory and J.D.~Dow,
{\it J. Phys: Condens. Matter} {\bf 23}, 295701 (2011) contains a useful
tabulation of $T_c$ values.} in Figs.~\ref{fig:1}-\ref{fig:3}, these materials
come in
families and the common structural element is a quasi 2-dimensional layer with
metallic $d$ or $f$ cations arranged on a nominally square planar set of lattice
sites. Surrounding these sites are an array of ligand anions which provide a
local crystal field and a hybridization network. Three members of the heavy
fermion CeIn$_3$ family are shown in Fig.~\ref{fig:1}. On the left is the unit
cell of the so-called infinite layered  ($T_c\sim0.2$K) material in which
CeIn$_3$ layers are stacked one on top of another \citep{ref:Shishido}. The middle
structure consists of a similar stack of CeIn$_3$ layers in which a CoIn$_2$
layer is inserted after every two CeIn$_3$ layers. This is called a 218
structure corresponding to (CeIn$_3$)$_2$(CoIn$_2$)$_1$=Ce$_2$Co$_1$In$_8$ and
has a superconducting transition temperature \citep{ref:Chen} $T_c\sim1K$. On the
right is the 115 structure which consists of alternating CeIn$_3$ and CoIn$_2$
layers giving (CeIn$_3$)(CoIn$_2$)=CeCoIn$_5$ ($T_c\sim2.3$K) \citep{ref:CeCoIn5}.
In addition, there are materials \citep{ref:CeRhIn5,ref:CeIrIn5} in which Co is
replaced by Rh or Ir, or Cd is substituted for In. The heavy-fermion actinide
PuMGa$_5$ materials have a similar structure to the 115 CeCoIn$_5$ with Pu
replacing Ce and Ga replacing In. In this case one has PuCoGa$_5$ with a
superconducting transition temperature \citep{ref:PuCoGa5} $T_c=18.5$K, PuRhGa$_5$ with
$T_c=8.7$K \citep{ref:PuRuGa5} as well as mixtures such as Pu(Co$_{1-x}$Rh$_x$)Ga$_5$.
Recently it has been reported \citep{ref:Zhu} that PuCoIn$_5$ becomes
superconducting with $T_c=2.5$K.

For the cuprates there are the well-known Hg, Tl and Bi families with different
numbers of CuO$_2$ layers. The one, two and three layer members of the Hg family
are shown in Fig.~\ref{fig:2}.
In this case the naming
scheme involves four numbers. For example, for the three CuO$_2$ layer Hg 1223
compound \citep{ref:Hg1223,ref:Wagner} with $T_c\sim135$K shown on the right, the first
index denotes the number of HgO planes, the second the number of spacing BaO
layers, the third is the number of separating Ca atom layers and the final the
number of CuO$_2$ layers. Thus one has the
(HgO)$_1$(BaO)$_2$(Ca)$_2$(CuO$_2$)$_3$=HgBa$_2$Ca$_2$Cu$_3$O$_9$ ``1223" three
layer material on the right and the
(HgO)$_1$(BaO)$_2$(CuO$_2$)$_1$=HgBa$_2$CuO$_5$ ``1201"
structure \citep{ref:Hg1201} with $T_c\sim94$K \citep{ref:Wagner} on the left.
Some of the O sites in the Hg layer are only partially occupied giving the
usual chemical formulae HgBa$_2$CuO$_{4+\delta}$ and
HgBa$_2$Ca$_2$Cu$_3$O$_{8+\delta}$. A Cu in the CuO$_2$ layer of the single
layer 1201 material has two apical O, while a Cu in the middle layer of the 1223
material has none. There are also the so-called 214 families such as La$_2$CuO$_4$
which can be hole doped La$_{2-x}$M$_x$CuO$_4$ with M=Sr or Ba and Nd$_{2-x}$CuO$_4$
which can be electron doped Nd$_{2-x}$Ce$_x$CuO$_4$. These latter electron doped
cuprates have structures in which the apical O is absent. (Fig.~\ref{fig:3a}) 
There are also the
so-called infinite layer electron doped cuprates \citep{ref:inflayer} in which the
CuO$_2$ planes are separated by Sr$_{1-x}$Ln$_x$ layers with Ln a lanthanide
such as La, Sm or Nd.

Figure~\ref{fig:3} shows some examples of the recently
discovered \citep{ref:LaOFeP,ref:LaFeAsO} Fe-superconducting families which are
built up from Fe/pnictide or chalcogen layers. In these layers the Fe ions sit
on a planar two-dimensional square lattice and the pnictide or chalcogen sit at
the centers of the squares, alternatively above or below the plane formed
by the Fe ions. Again these layers can be stacked in a variety of ways leading
to the LaOFeAs, Ba(FeAs)$_2$ and FeSe structures illustrated in Fig.~\ref{fig:3}.
These are called the (1111), (122) and (11) Fe-based materials, respectively.
The alternating arrangement of the pnictides or chalcogens leads to a
doubling of the unit cell compared with the square Fe lattice. In LaOFeAs, the
Fe is tetrahedrally coordinated with four As forming square pyramids. The
LaO layer has the same type of structure but with the O forming the square
planar array. There are many equiatomic quaternary pnictide oxides of this
type \citep{ref:Ozawa}. The phosphorus version of this material \citep{ref:LaOFeP}
LaOFeP has a superconducting transition of 6K. When the As version is electron
doped by replacing some of the O with F giving LaO$_{1-x}$F$_x$FeAs, it can
become superconducting with a $T_c=26$K \citep{ref:LaFeAsO} and replacing La with
Sm has given $T_c=55$K \citep{ref:1111Sm}. In the BaFe$_2$As$_2$ (122) compound,
the Fe$_2$As$_2$ layers are separated by Ba$^{2+}$ ions. In this case the system
can be hole doped \citep{ref:BaKFe2As2} Ba$_{1-x}$K$_x$Fe$_2$As$_2$ with an optimal
$T_c\sim38$K or electron doped \citep{ref:CoBa122} Ba(Fe$_{1-x}$Co$_x$)$_2$As$_2$ with $T_c\sim22$K.
The third Fe(Se,Te) family shown on the right hand side of Fig.~\ref{fig:3} is
essentially the infinite layer member of the family and has a $T_c\sim13.6$K$-37$K
depending upon the Se/Te composition and the pressure \citep{ref:Okabe,ref:FeSeTe}.

\begin{figure}[!htbp]
\includegraphics[height=8.5cm]{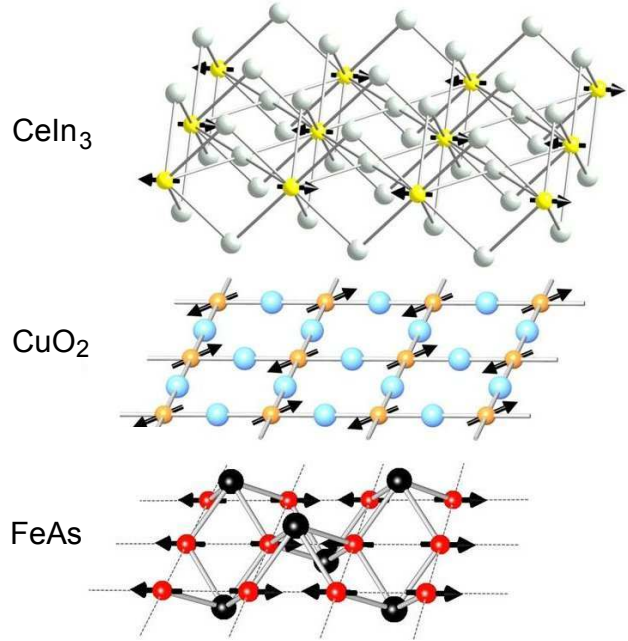}
\caption{The active layers of the Ce, Cu and Fe families. The antiferromagnetic
spin order of the undoped groundstates are shown.\label{fig:4}}
\end{figure}

The active layers of these Ce, Cu and Fe families are illustrated in Fig.~\ref{fig:4}.
For the actinide Pu family, the active layer is similar to the Ce layer with Pu
replacing Ce and Ga replacing In or as recently found for the PuCoIn$_5$ 115
compound, one can simply replace Ce with Pu. In each case, these layers contain
a square sheet of metallic $d$ or $f$ cations surrounded by ligand anions. However,
the spacing of the metallic ions in these compounds are significantly different with the Ce$^{3+}$
ions separated by approximately $4.6^\circ$A, the Cu$^{2+}$ ions by $3.8^\circ$A
and the Fe$^{2+}$ ions by $2.7^\circ$A. The Fe$^{2+}$ ions are close enough that
there is a direct Fe--Fe hopping which along with the d-p hybridization through
the pnictogen or chalcogen anions leads to a metallic groundstate with the
possibility of itinerate striped SDW antiferromagnetism and/or superconductivity.
Observations of quantum oscillations originating from the Shubnikov-de~Haas effect
\citep{ref:QoscFe,ref:Coldea,ref:Qosc} provide clear evidence of well defined Fermi
surfaces in the parent Fe-based compounds as well as the doped materials.

In contrast to this itinerant electron behavior, the undoped cuprate materials
are Mott charge-transfer antiferromagnetic insulators. In the undoped CuO$_2$
layer, one has Cu$^{2+}$ in a $(3\rm d)^9$ configuration. The crystal field is
such that the $d_{x^2-y^2}$ orbital has the highest energy and is half-filled.
The onsite Cu Coulomb interaction energy is large leading to the formation of
local moments. The O orbital mediates an exchange interaction \citep{ref:J} between the Cu spins and
the groundstate has long range antiferromagnetic order. In the three dimensional
crystal, the interlayer exchange coupling leads to a finite N\'eel temperature.
The undoped system is a charge-transfer insulator with a gap set by the
difference in energy between the 2p state of the O and the $d_{x^2-y^2}$ state
of the Cu. In order to have metallic behavior and the possibility of
superconductivity, the CuO$_2$ planes need to be doped. The occupancy of the
oxygen site in the Hg layer typically controls the hole doping of the CuO$_2$
in the Hg cuprates while cation substitution or O doping excess or depletion can
provide hole or electron doping for the 214 cuprates.

In the heavy fermion materials one has the largest ion separation but in this case the
conduction band of the ligands gives rise to a metallic state. The 14-fold
degenerate $f$ electronic states of the $(4f)^1$ configuration of Ce$^{3+}$ are
split by a large spin-orbit coupling into a low lying $j=5/2$ sextet and a
higher energy $j=7/2$ octet. The one electron states of the $j=5/2$ sextet are
further split by the crystalline electric field of the In ligand anions into
three sets of Kramer's doublets \citep{ref:Hotta}. Then, depending upon the strength of the
hybridization, these states are localized or delocalized. For example,
CeRhIn$_5$ has an antiferromagnetic groundstate in which the $4f$-electron of
Ce is localized with a magnetic moment only slightly reduced from its full
atomic value \citep{ref:CeIrIn5}. The system is metallic due to the conduction band associated with
the ligands. Under sufficient pressure, 1.7 GPa, the $4f$-electron takes
on some itinerant character and the system becomes superconducting \citep{ref:CeRh}.
In CeCoIn$_5$ and CeIrIn$_5$, at low temperatures the $4f$ electron are
delocalized through their coupling with the ligand conduction band and these
systems become superconducting at atmospheric pressure \citep{ref:CeCoIn5,ref:CeIrIn5}.
Replacing a small amount of In with a few percent of Cd leads to a metallic
antiferromagnetic state \citep{ref:CeCd,ref:CeCdphdiag}. The two-dimensional
character of the Ce ion layers lead to nearly cylindrical Fermi surfaces which
are seen in de~Haas and van~Alphen measurements. The cyclotron masses are large
consistent with the fact that the $4f$ electrons make a contribution to the
Fermi surface states \citep{ref:FSofCe}.

\subsection{Phase diagrams}
These materials exhibit a range of different phases. There are tetragonal and
orthorhombic lattice phases, nematic electronic phases, charge density wave and
striped magnetic phases, charge-transfer antiferromagnetic Mott insulating as
well as metallic spin density wave phases, and of course superconductivity. Via
temperature, doping, chemical or hydrostatic pressure, or the application of a
magnetic field one can change the phase of these materials. However, the feature
that is striking in the phase diagrams for all of these materials is the proximity
of the antiferromagnetic or spin density wave and superconducting phases. These
phases may in some cases coexist or alternatively there may be a first order
transition from the AF state to the superconducting state. Then as noted by
\citet{ref:Emery1999}, Coulomb frustrated phase separation can lead to a
mesoscopic phase in which a lightly doped locally AF and a more heavily hole doped region
are in close contact. It has been suggested that this type of inhomogeneity may
in fact lead to an optimal superconducting transition temperature \citep{ref:KivFrad}.

\begin{figure}[!htbp]
\begin{center}
\subfigure{
\includegraphics[width=7.5cm]{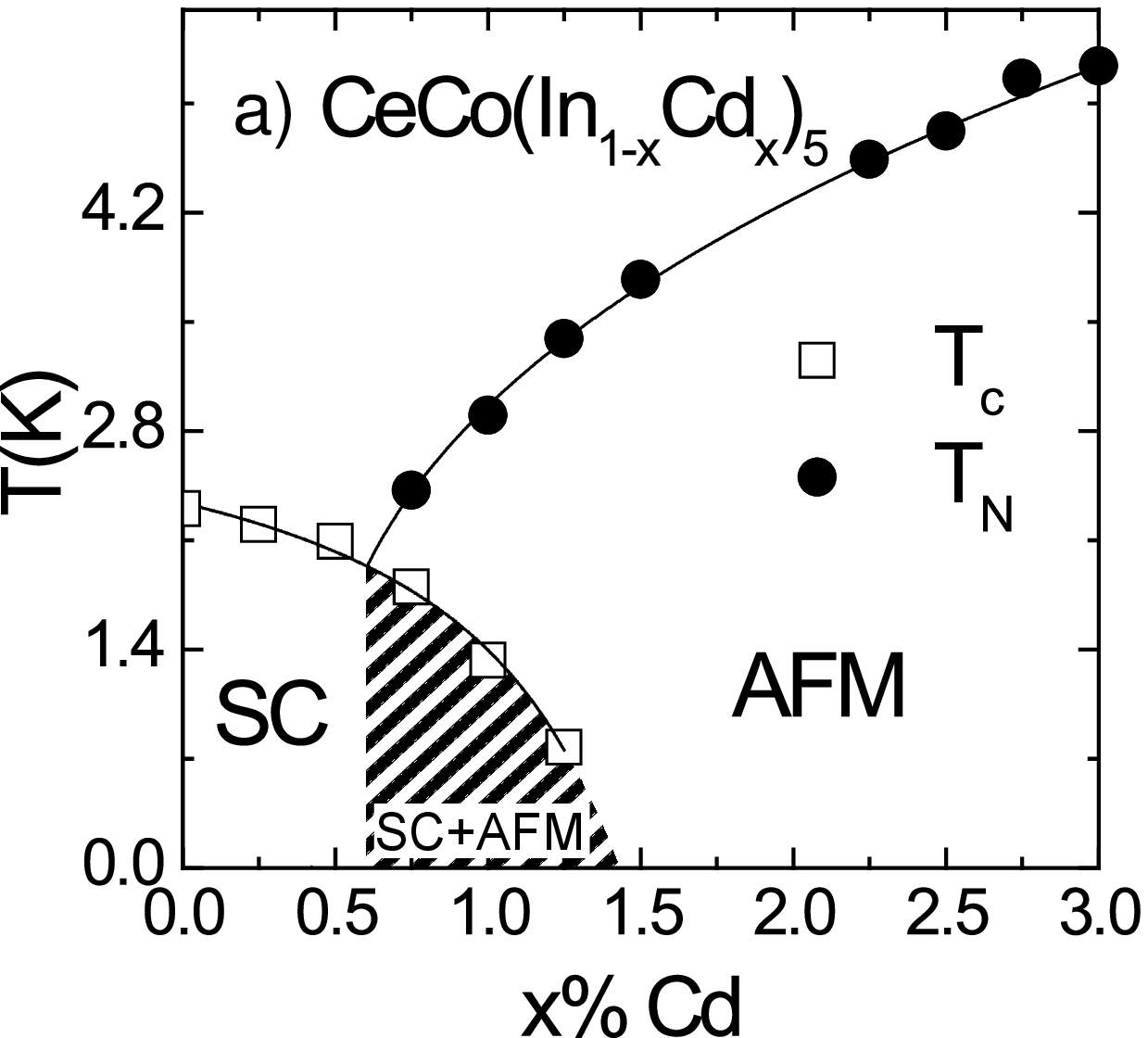}}
\subfigure{
\includegraphics[width=7.5cm]{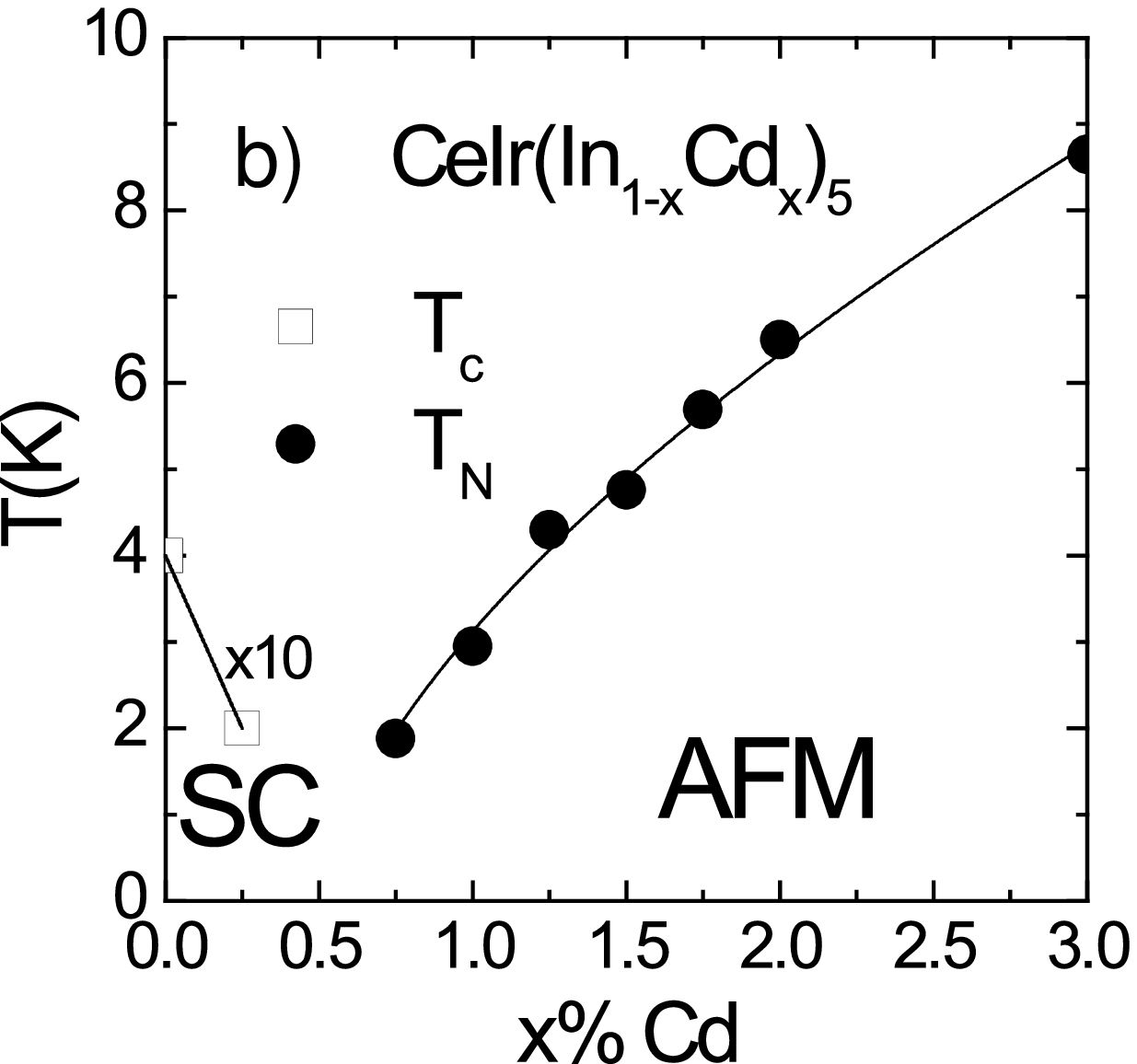}}
\caption{Phase diagrams for two heavy fermion Ce-115 systems:
(a) CeCo(In$_{1-x}$Cd$_x)_5$ (after Nicklas et al.\protect\cite{ref:CeCdphdiag})
and (b) CeIr(In$_{1-x}$Cd$_x)_5$ (after Pham et al.\protect\cite{ref:CeCd}).
Note that $T_c$ is multiplied by a factor of 10 for CeIr(In$_{1-x}$Cd$_x$)$_5$.
In both cases one sees the close proximity of superconductivity and
antiferromagnetism. For the Co compound there is a region of coexistence.\label{fig:5}}
\end{center}
\end{figure}

\begin{figure}[!htbp]
\includegraphics[height=8.5cm]{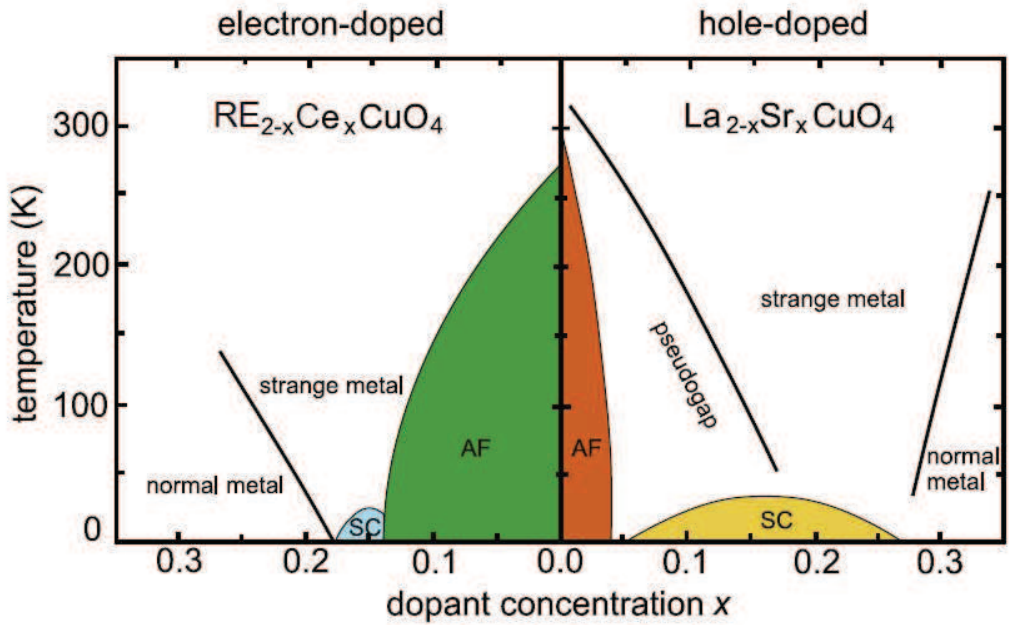}
\caption{Schematic phase diagrams for hole doped La$_{2-x}$Sr$_x$CuO$_4$ and
electron-doped RE$_{2-x}$Ce$_x$CuO$_4$ (RE = La, Pr, Nd) cuprates (after
R.L.~Greene and Kui Jin). In the electron-doped case, the AF region extends to
the superconducting region, while in the hole-doped case a pseudogap region intervenes.\label{fig:6}}
\end{figure}

\begin{figure}[!htbp]
\includegraphics[height=8.5cm]{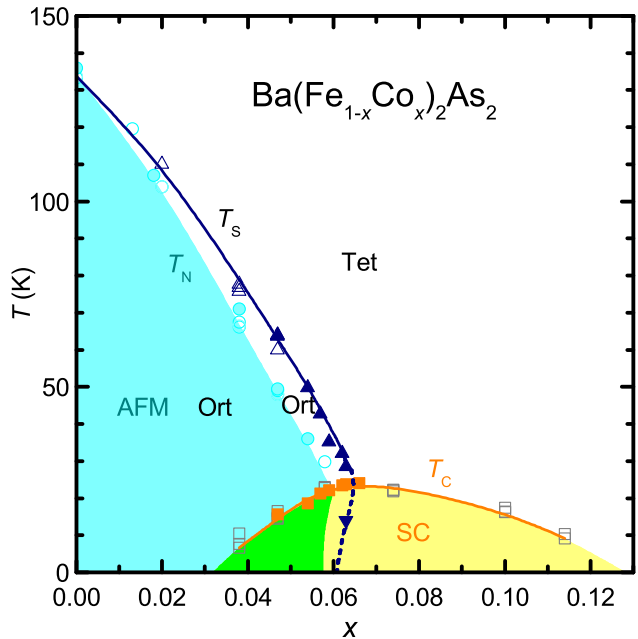}
\caption{The phase diagram for Ba(Fe$_{1-x}$Co$_x$)As$_2$
(after Fernandes et al.\protect\cite{ref:Fernandes}). There
appears a coexistence region similar to CeCo(In$_{1-x}$Cd$_x)_5$
shown in Fig.~\protect\ref{fig:5}.\label{fig:7}}
\end{figure}

Examples of phase diagrams for the heavy fermion, cuprate and Fe-based materials
are shown in Figs.~\ref{fig:5}-\ref{fig:7}. The phase diagram for the 115 heavy
fermion system \citep{ref:CeCd} CeCo$(\rm In_{1-x}Cd_x)_5$ is shown in Fig.~\ref{fig:5}a.
For $x=0$, CeCoIn$_5$ becomes superconducting at temperatures below approximately
2.3K. Then as the Cd concentration increases, one enters a region where
the system first becomes antiferromagnetic and then below the superconducting $T_c$
there is a coexistence regime. Finally, for Cd concentration $x\gtwid0.15$,
superconductivity is absent and the N\'eel temperature $T_N$ continues to
increase. A similar phase diagram for the case in which Co is replaced by Ir is
shown in Fig.~\ref{fig:5}b. In this case, while the N\'eel temperatures are
comparable to those of the Co material, the superconducting $T_c$ is significantly smaller.

Figure~\ref{fig:6} shows the phase diagrams of La$_{2-x}$Sr$_x$CuO$_4$ and
Nd$_{2-x}$Ce$_x$CuO$_4$ \citep{ref:Green}.
Undoped La$_2$CuO$_4$ and Nd$_2$CuO$_4$ are charge-transfer insulators which
undergo antiferromagnetic N\'eel transitions as the temperature drops below 300K.
Replacing a small amount of La with Sr leads to a hole doping of the CuO$_2$
layer, while replacing Nd with Ce leads to an electron doped CuO$_2$ layer.
As the hole doping $x$ increases, the N\'eel temperature is suppressed and at
low temperatures the system passes through a spin glass phase 
in which local charge and spin ordered regions may be pinned. In the hole
doped case, the doping for optimal superconductivity is well separated from
the onset of antiferromagnetism.
The antiferromagnetic order extends much further out for the electron
doped system and appears adjacent to the superconducting phase.

The phase diagram for one of the Fe-based superconductors \citep{ref:Fernandes}
Ba(Fe$_{1-x}$Co$_x$)As$_2$ is shown in Fig.~\ref{fig:7}.
The parent compound BaFe$_2$As$_2$ is metallic and undergoes a structural
tetragonal to orthorhombic transition and at the same temperature an
antiferromagnetic SDW transition. In the SDW phase the moments are oriented
antiferromagnetically along the longer $a_0$ axis of the orthorhombic 2Fe/cell
and ferromagnetically along the $b_0$ axis giving a stripe-like structure. As
Co is added, the system is electron doped and the structural and SDW transitions
are suppressed. The structural transition is found to occur at temperatures
slightly above the SDW transition. For dopings $x\gtwid0.07$, the
structural and SDW transitions are completely suppressed and the system goes
into a superconducting state below $T_c$. However, for a range of smaller
dopings $0.03\ltwid x\ltwid 0.06$ the system enters a region in which there is
microscopic coexistence of superconductivity, SDW and orthorhombic order. As
will be discussed, evidence for this is seen in the temperature dependence of
the SDW Bragg peak intensity and the orthorhombic distortion. It is also
possible to hole dope this compound \citep{ref:BaKFe2As2} by substituting K for Ba,
Ba$_{1-x}$K$_x$Fe$_2$As$_2$. Here again, as $x$ increases the structural and
SDW transition are suppressed and superconductivity onsets \citep{ref:Paglione}.

\subsection{Coexistence and interplay of antiferromagnetism and superconductivity}

\begin{figure}[!htbp]
\begin{center}
\subfigure{
\includegraphics[width=8.5cm]{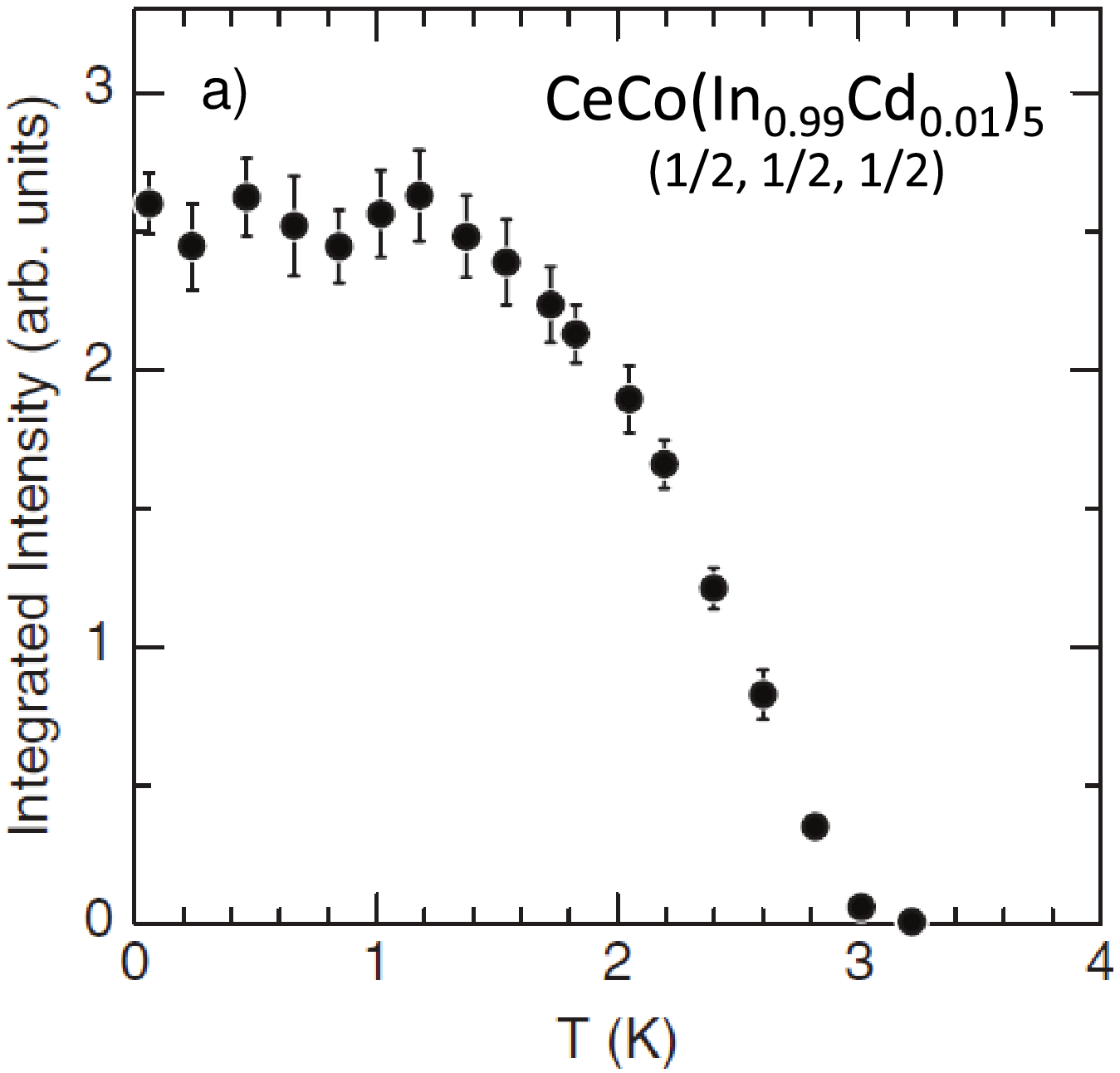}}
\subfigure{
\includegraphics[width=6.7cm]{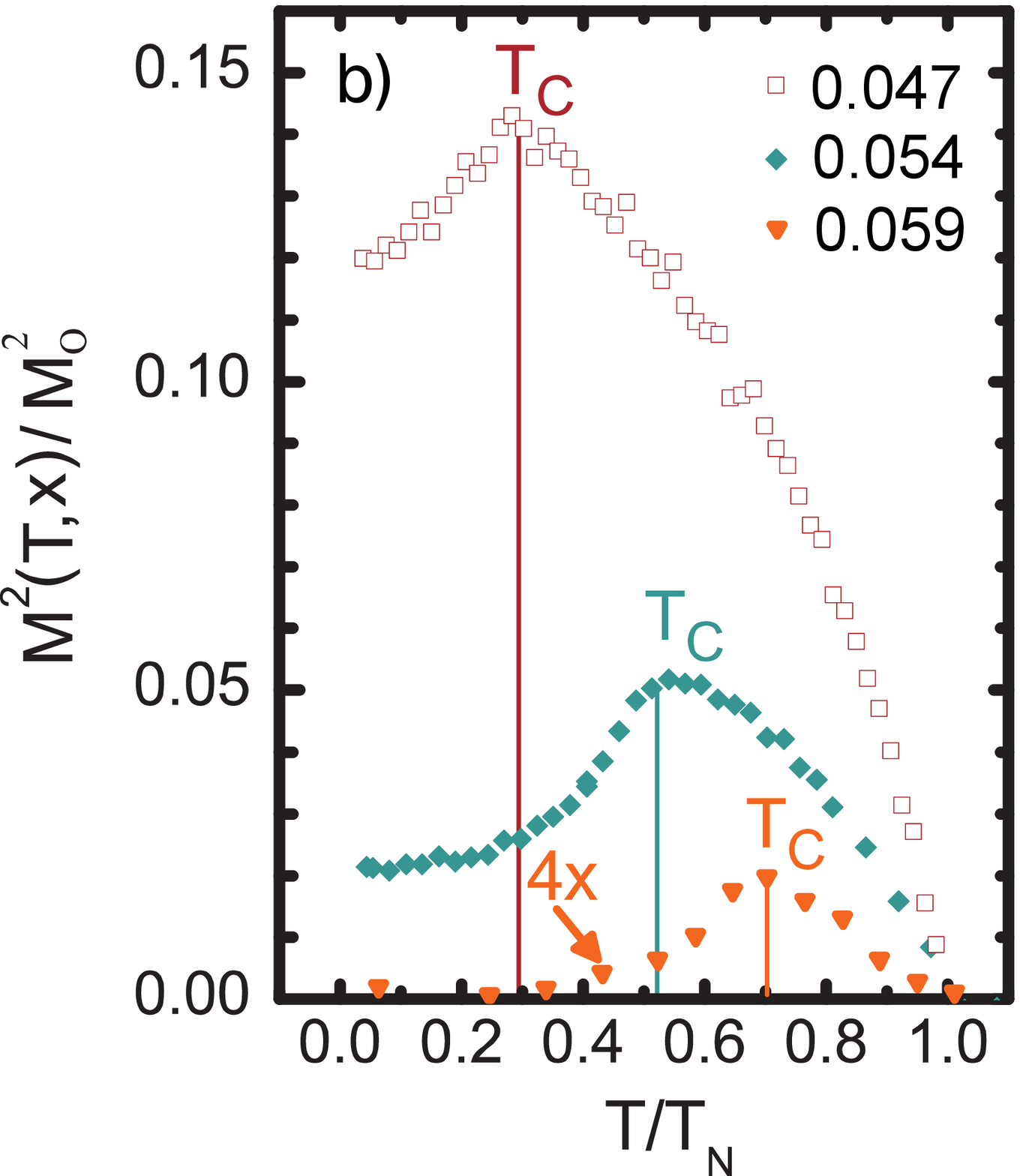}}
\caption{The interplay of antiferromagnetism and superconductivity is seen in
the temperature dependence of the Bragg scattering. (a) The integrated Bragg
scattering intensity for the 115 heavy fermion superconductor
CeCo(In$_{0.99}$Cd$_{0.01}$)$_5$ at $Q_{\rm AF}$ versus temperature
(after Nicklas et al.\protect\cite{ref:CeCdphdiag}).
(b) The integrated Bragg scattering intensity for Ba(Fe$_{1-x}$Co$_x$)As$_2$ at
$Q_{\rm SDW}$ versus the temperature for various values of $x$
(after Fernandes et al.\protect\cite{ref:Fernandes}). In both cases, the
strength of the Bragg scattering from the magnetic order is clearly altered by
the onset of the superconductivity.\label{fig:7a}}
\end{center}
\end{figure}

NMR as well as neutron scattering measurements have provided evidence that the
observed coexistence regions in some systems represent microscopic coexistence in which the same
electrons are involved with both the superconductivity and the antiferromagnetism.
For example, elastic neutron scattering measurements \citep{ref:CeCd} on
$\rm CeCo(In_{0.9}Cd_{0.1})_5$ find the integrated magnetic intensity at
the antiferromagnetic wave vector
$Q_{\rm AF}$ versus temperature shown in Fig.~\ref{fig:7a}a. This intensity is a
measure of the square of the ordered magnetic moment and onsets at the N\'eel
temperature $T_N$. As seen in Fig.~\ref{fig:7a}a, $M^2(T)$ initially increases as $T$
decreases below $T_N$, but then as $T$ drops below the superconducting
transition temperature $T_c$, it saturates. Similar data for $\rm Ba(Fe_{1-x}Co_x)As_2$
at three different dopings are shown in Fig.~\ref{fig:7a}b.
In this case, below $T_c$ the ordered moment is reduced as the superconducting
order increases. Both these examples reflect the competition of
superconductivity and antiferromagnetism \citep{ref:Fernandes,ref:Vorontsov}.
This competition is also believed to be responsible for the anomalous
suppression of the orthorhombic distortion in Ba(Fe$_{1-x}$Co$_x$)As$_2$ as the
temperature decreases below $T_c$ \citep{ref:Nandi}. Evidence for
atomic scale coexistence of superconductivity and antiferromagnetism for
Ba(Fe$_{1-x}$Co$_x$)$_2$As$_2$ with $x=0.06$ was reported by
\citet{ref:LaPlace}. Here volume susceptibility measurements showed a
superconducting fraction greater than 95\%. Then measurements of the homogeneous
broadening of the $^{75}$As NMR spectrum showed that frozen moments remained
on all of the Fe atoms for $T$ less than $T_c$ while at the same time, the spin-lattice
relaxation rate $T^{-1}_1$ of $^{75}$As showed that the Fe electrons also exhibited
superconductivity. Since the As nuclei are coupled to only the four near neighbor
Fe sites, this experiment provided evidence of homogeneous coexistence on a
unit cell scale.

In addition to the ordered antiferromagnetic (N\'eel) phase, there are a variety
of incommensurate spin density wave striped phases that compete and interact
with the superconducting phase. Evidence of this is seen in neutron scattering
experiments on La$_{2-x}$Sr$_x$CuO$_4$ which reveal a strong enhancement of
spin-stripe order at low energies produced by modest magnetic fields
\citep{ref:Khaykovich,ref:Lake}. This behavior has been modeled by Landau-Ginzburg
theories in which the incommensurate antiferromagnetic order is coupled to the
$d$-wave superconducting order \citep{ref:Demler,ref:Kivelson}. This mutual
coupling of SDW and $d$-wave scattering processes has also been found
in renormalization group calculations \citep{ref:HalbothPRL,ref:Honerkamp,ref:Zhai,ref:Platt}.

A particularly striking example of the coexistence and interplay of
antiferromagnetism and $d$-wave superconductivity is seen in La$_{2-x}$Ba$_x$CuO$_4$
near a doping $x\sim1/8$ \citep{ref:QLi,ref:Tranquada08}. Here a combination
of tunneling and photoemission measurements along with transport studies provide
evidence that two-dimensional $d$-wave superconducting correlations coexist with
$\pi$-phase shifted antiferromagnetic stripes at temperatures below 40K. The
observation that macroscopic 2D superconductivity persists at temperatures well
above the 3D transition temperature suggests that the pairing correlations form
a pair density wave with a wavevector which is the same as that of the spin-density
wave \citep{ref:Berg07,ref:Himeda}. That is, the amplitude of the $d$-wave
superconducting order parameter is enhanced in the hole-rich regions of the
striped system and the phase of the adjacent superconducting stripes are opposite
in sign (antiphase). In this case, the structurally driven orthogonal orientation
of the stripes in neighboring planes leads to a frustration of the Josephson
coupling between planes allowing for the possibility of a Berezinskii-Kosterlitz-Thouless
transition in the 3D crystal.

The interplay between the antiferromagnetic spin fluctuations and the
superconducting pairs is also seen in the change in the exchange energy
$\Delta E_{\rm ex}$ between the superconducting and normal states
\citep{ref:ScalWhite1998}. For a material with a near neighbor exchange
coupling $J$, the change in exchange energy $\Delta E_{\rm ex}(T)$ is given by
\begin{equation}
\Delta E_{\rm ex}(T)=2J\left(\left\langle{\bf S}_{i+x}\cdot{\bf S}_i\right\rangle_N
-\left\langle{\bf S}_{i+x}\cdot{\bf S}_i\right\rangle_S\right)
\label{eq:1}
\end{equation}
with
\begin{equation}
\left\langle{\bf S}_{i+x}\cdot{\bf S}_i\right\rangle_{S(N)}=\frac{1}{g^2\mu^2_\beta}
\int^\infty_0\frac{d\omega}{\pi}(n(\omega)+1)\langle\cos(q_xa)\chi^{\prime\prime}_{S(N)}
(q,\omega)\rangle_{BZ}
\label{eq:2a}
\end{equation}
Here $n(\omega)$ is the usual Bose factor, the momentum $q$ is summed over the
Brillouin zone and $\chi^{\prime\prime}_{S(N)}(q,\omega)$ is the imaginary part
of the wavevector and frequency dependent spin susceptibility in the
superconducting ($S$) and normal ($N$) phases respectively, measured at temperature $T$. Additional
next-near-neighbor exchange terms appropriate to a given material can be added
to Eq.~(\ref{eq:1}). In initial studies of YBa$_2$Cu$_3$O$_{6.95}(T_c=92.5$K), a
low temperature value of $\Delta E_{\rm ex}$ was estimated from measurements of
$\chi^{\prime\prime}_s (q,\omega)$ at $T=15$K and $\chi^{\prime\prime}_N(q,\omega)$
taken at 100K. This estimate gave a change in the exchange energy which was
approximately 15 times larger than the superconducting condensation
energy \citep{ref:Woo}. Recent measurements of the heavy fermion superconductor
CeCu$_2$Si$_2$ found a change of the exchange energy which was of order 20 times
larger than its low temperature superconducting condensation
energy \citep{ref:Stockert}. In this case, the lower $T_c\sim0.6$K of this heavy fermion
systems allowed direct access at this same temperature to the putative normal
state using a $2.5T$ magnetic field. While the superconducting condensation
energy $U_c$ arises from a cancellation between this change in the
exchange energy $\Delta E_{\rm ex}$ and other electronic energies, the important
point is that $\Delta E_{\rm ex}$ is large compared with $U_c$ so that
antiferromagnetic fluctuations clearly have the strength to drive the
superconducting pairing. In addition, we note that $\Delta E_{\rm ex}/U_c$ is
similar in size for YBa$_2$Cu$_3$O$_{6.95}$ and  CeCu$_2$Si$_2$.

The similarities of the suppression of the Bragg scattering intensity $M^2$ in the
coexisting antiferromagnetic and superconducting state, the magnetic field
induced SDW in the superconducting state and the change of the exchange energy
between the superconducting and normal paramagnetic states not only serve to
establish a relationship between these different materials but in addition
provide evidence that the antiferromagnetism and superconductivity in these
materials are strongly coupled. Further evidence of this is also clearly seen
in NMR studies of the spin-lattice relaxation time $T_1$
of FeSe \citep{ref:Imai} and inelastic neutron scattering measurements of overdoped
LSCO \citep{ref:Wakimoto}. Measurements of $(T_1T)^{-1}$ probe the Brillouin zone
average of Im$\chi(q,\omega_0)/\omega_0$ weighted with the square of the
hyperfine form factor. Here $\omega_0$ is a low frequency set by the nuclear Zeeman
energy. As the pressure is increased on FeSe, $(T_1T)^{-1}$ and $T_c$ are both
enhanced. Similarly, the strength of the low-energy incommensurate antiferromagnetic
spin fluctuations in overdoped LSCO is observed to decrease \citep{ref:Wakimoto} as the doping
increases and $T_c$ is reduced.

While neutron scattering measurements provide evidence of the $q$-$\omega$
spin-fluctuation spectral weight for the underdoped materials \citep{ref:Woo},
one is of course also interested in the optimally as well as the overdoped
materials. Recently \citep{ref:LeTacon}, resonant inelastic x-ray scattering (RIXS)
experiments have provided such information over a wide energy-momentum region
for YBa$_2$Cu$_4$O$_8$, YBa$_2$Cu$_3$O$_{6+x}$, and Nd$_{1.2}$Ba$_{1.8}$Cu$_3$O$_{6+x}$.
These experiments clearly show, for a range of dopings covering underdoped,
optimal as well as over-doped materials, the existence of damped, dispersive
magnetic excitations, which have significant spectral weight in an appropriate
spectral range to produce pairing.

\begin{figure}[!htbp]
\includegraphics[width=16cm]{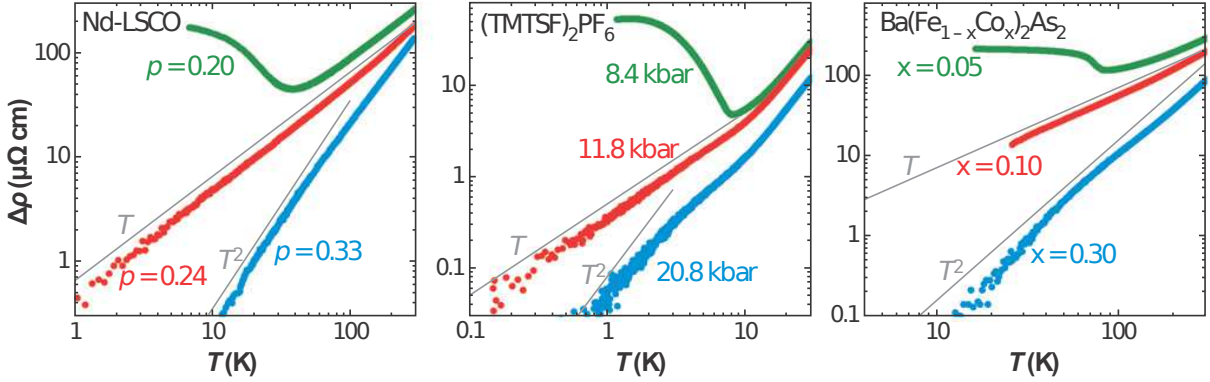}
\caption{The temperature dependent part $\Delta\rho(T)$ of the in-plane
resistivity versus $T$ on a log-log scale for the cuprate Nd-LSCO, the organic
Bechgaard salt (TMTSF)$_2$PF$_6$ and the Fe-pnictide Ba(Fe$_{1-x}$Co$_x)_2$As$_2$.
As a relevant tuning parameter, doping or pressure, is changed, the temperature
dependence of $\Delta\rho(T)$ for all three systems pass from a $T^2$ dependence
to an approximately linear $T$ dependence and then to an upturn associated with
a Fermi surface reconstruction (after Taillefer\protect\cite{ref:Taillefer}).
\label{fig:8}}
\end{figure}

There is also resistivity data which provides evidence of the strong coupling
of the spin-fluctuations and the quasiparticles in the regions of the phase diagram
where superconductivity appears. \citeauthor*{ref:Taillefer} has emphasized
the similar behavior of the temperature dependent part of the in-plane normal
state resistivity of the cuprate Nd-LSCO, the organic Bechgaard salt
(TMTSF)$_2$PF$_6$ and the Fe-pnictide Ba(Fe$_{1-x}$Co$_x$)As$_2$ shown in Fig.~\ref{fig:8}.
Here the linear $T$ dependence of the resistivity of Nd-LSCO is associated
with a hole doping 0.24 at which the stripe-ordered antiferromagnetic phase
ends \citep{ref:Daou2009}. Likewise, a Co concentration $\sim0.10$ for Co-Ba122 and a pressure
$\gtwid10$ kbar for (TMTSF)$_2$PF$_6$ mark the ends of the SDW phases for these
materials. As the doping (or pressure for the Bechgaard salt) is increased,
the anomalous $T$ dependence is replaced by a Fermi-liquid $T^2$ dependence
and the superconducting $T_c$ goes to zero. At low doping or under pressure, the
upturn in $\Delta\rho$ shows evidence of a Fermi-surface reconstruction due
to the occurrence of an ordered phase. Based on transport and NMR measurements
on the (TMTSF)$_2$X materials as a function of pressure, \citet{ref:Doiron,ref:DR-arxiv}
argue that the linear $T$ dependence of the resistivity is associated with
scattering from antiferromagnetic spin fluctuations at the border of
antiferromagnetic order and that this scattering is directly linked to $T_c$.
\citet{ref:Hartnoll} have argued that a quantum critical response arises
from spin-fluctuation scattering and umklapp processes as the spin density wave
phase of a 2D metal is approached.

A similar connection between spin-fluctuation scattering of the carriers and the
basal plane resistivity of La$_{2-x}$Ce$_x$CuO$_4$ films has been reported by
\citet{ref:Jin}. These authors carried out low temperature resistivity
experiments as a function of doping and magnetic field. They found a correlation
between the strength of the low temperature linear-in-$T$ resistivity and the
superconducting $T_c$ as a function of doping. They noted that this electron
doped cuprate provided a particularly interesting case since there is no
pseudogap phase in the underdoped region of its phase diagram, leaving the spin
fluctuations as the dominant link to the temperature dependence of the resistivity.

A magnetic field-tuned quantum critical response is also seen in the heavy fermion
CeCoIn$_5$ system \citep{ref:PaglionePRL} as well as other heavy fermion
materials. Of particular interest, as \citet{ref:Si} have
discussed for CaCu$_2$Si$_2$ and CePd$_d$Si$_2$, are the antiferromagnetic to
paramagnetic quantum critical transitions. Here, the critical degrees of freedom are the SDW fluctuations.
The role of the quantum critical point and the interplay between antiferromagnetism and the
resulting temperature, carrier concentration and magnetic field phase diagram
have been discussed by Sachdev and Metlitski (2010). 
To summarize,
the possible coexistence of antiferromagnetism and $d$-wave superconductivity,
the change in the exchange energy upon entering the superconducting phase and
the importance of spin-fluctuation scattering are characteristic of the class
of materials being discussed.

\subsection{A neutron spin resonance}
\begin{figure}[!htbp]
\includegraphics[height=8cm]{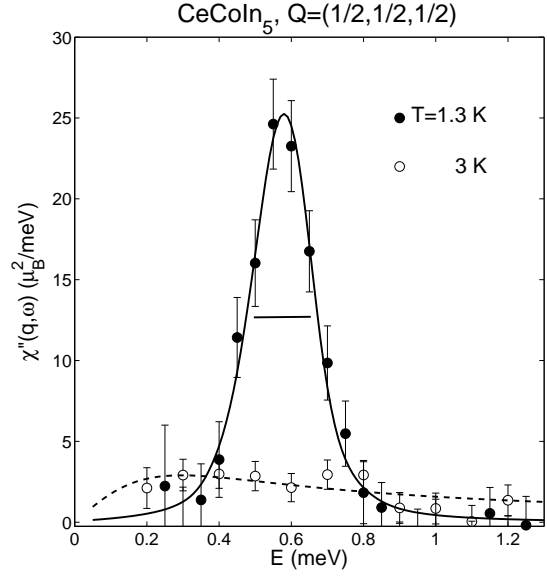}
\caption{The neutron scattering spin resonance in the normal (dashed) and
superconducting (solid) phases observed for the 115 Ce heavy fermion material
CeCoIn$_5$(T$_c=2.3$ K) (after Stock et al.\protect\cite{ref:Stock}).\label{fig:8a}}
\end{figure}

\begin{figure}[!htbp]
\includegraphics[height=8cm]{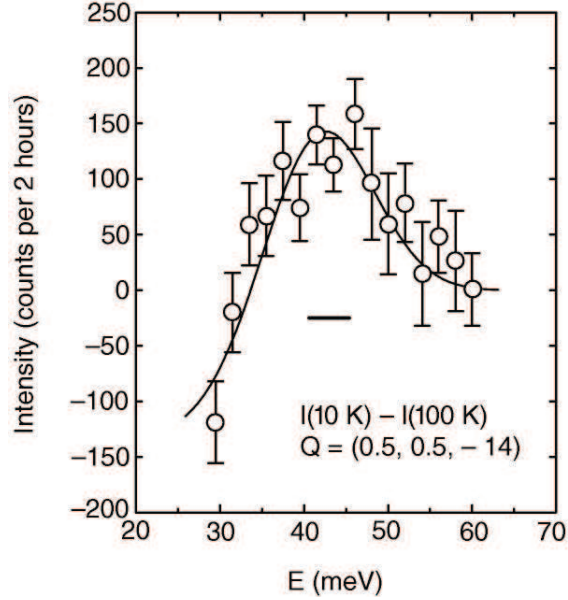}
\caption{Difference spectrum of the neutron scattering intensities from
Bi$_2$Sr$_2$CaCu$_2$O$_{8+\delta}$(T$_c=91$ K) at $T=10$K and 100K at wavevector
$Q=(\pi/a,\pi/a)$ showing the spin resonance at $\sim43$meV. The horizontal bar
represents the instrumental energy resolution and the solid curve is a guide to
the eye (after Fong et al.\protect\cite{ref:Fong1999}).\label{fig:9}}
\end{figure}

\begin{figure}[!htbp]
\begin{center}
\subfigure{
\includegraphics[width=9.2cm]{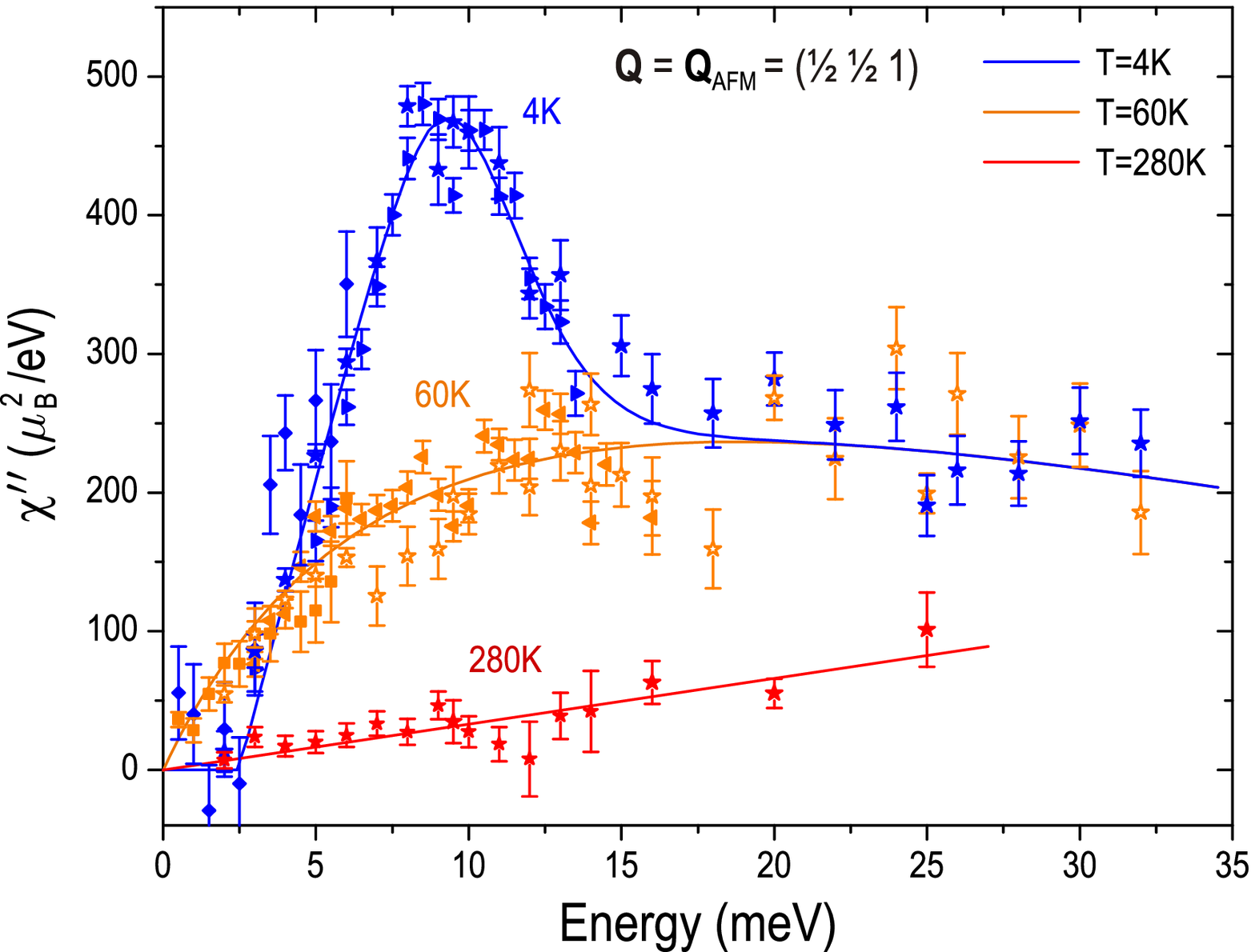}}
\subfigure{
\includegraphics[width=6.8cm]{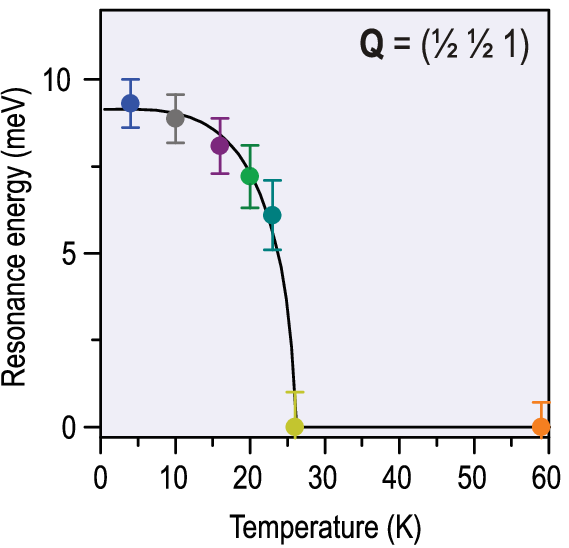}}
\caption{(left) The neutron scattering spin resonance for
BaFe$_{1.85}$Co$_{0.15}$As$_2$(T$_c=26$ K);
(right) The energy of the resonance versus temperature follows a BCS-like curve
(after Inosov et al.\protect\cite{ref:Inosov}).\label{fig:10}}
\end{center}
\end{figure}

Another important experimental observation linking these materials is the appearance
of a neutron scattering spin resonance in the superconducting phase at the
antiferromagnetic or spin-density-wave vector $Q$. This resonance, first observed
in the cuprates \citep{ref:Rossat-Mignod,ref:Mook,ref:Fong1995,ref:Fong1999} and
then discovered in the heavy fermion materials \citep{ref:Stock}, has also recently been observed
in various Fe superconductors \citep{ref:Christiansen,ref:Park,ref:Inosov,ref:Lumsden}. The spin-flip inelastic
scattering rate is proportional to the imaginary part of the spin susceptibility.
Experimental results for $\chi^{\prime\prime}(Q,\omega)$ obtained for CeCoIn$_5$,
Bi$_2$Sr$_2$CaCu$_2$O$_{8+\delta}$ and BaFe$_{1.85}$Co$_{0.15}$As$_2$ are shown in
Figs.~\ref{fig:8a}-\ref{fig:10}.
While the energy of the resonant peak in YBCO is relatively insensitive to
$T/T_c$, the peak in $\rm Ba(Fe_{0.975}Co_{0.125})_2As_2$ was found to follow
the temperature dependence of the superconducting gap obtained from
ARPES \citep{ref:Terashima2,ref:Evtushinsky}.

Although the detailed behavior of the resonance requires a calculation of the
spin susceptibility, the occurrence of the resonance is directly related to the
BCS coherence factor that enters the neutron spin-flip scattering process.
This coherence factor for flipping the spin of a quasi-particle scattered
from $k$ to $k+Q$ is
\begin{equation}
\frac{1}{2}\left(1-\frac{\Delta(k)\Delta(k+Q)}{E(k)E(k+Q)}\right)
\label{eq:2}
\end{equation}
with $E(k)=\sqrt{\varepsilon^2_k+\Delta^2(k)}$ the quasi-particle energy. The
occurrence of a resonance, requires that the gap changes sign between regions
on the Fermi surface or surfaces separated by momentum $Q$ which contribute
significantly to the spin scattering \citep{ref:BulutScaSca,ref:Monthoux}
\begin{equation}
  {\rm Sgn}\left(\Delta(k+Q)\right)=-{\rm Sgn}(\Delta(k))
\label{eq:3}
\end{equation}
In this case the coherence factor Eq.~(\ref{eq:2}) goes to 1 near threshold while
if there were a plus sign in Eq.~(\ref{eq:3}), it would vanish.

Equation~(\ref{eq:3}) defines the class of unconventional superconductors which
are the subject of this review.
Materials in this class have a gap
that changes sign on different parts of the Fermi surface or surfaces separated
by a momentum $Q$ which connects regions which play an important role in the
scattering of the electrons. Thus ``unconventional" as used in this review
is not related to the symmetry of the gap, nor is it determined by
whether the gap has nodes or is nodeless. For example, the gap may have $A_{1g}$
($s$-wave) symmetry but change sign between two different pieces of the Fermi
surface, as the so-called $s^\pm$-gap proposed for the Fe-pnictides \citep{ref:Mazin}.
As discussed in Sec.~\ref{sec:4}, such an $A_{1g}$ gap can also have nodes \citep{ref:Hirschfeld}.
Alternatively, one could have a $B_{1g}$ ($d$-wave) nodeless gap on multiple
Fermi surfaces.

\section{Models}\label{sec:3}

In this section we introduce the basic models that will be discussed. While
these are certainly minimal models, we will argue that they exhibit a number
of the important physical properties which are observed in the actual materials.
On this basis, it is reasonable to examine the structure of the pairing interaction
in these models as will be done in Sec.~\ref{sec:4}.

As illustrated in Fig.~\ref{fig:4}, these materials have crystal structures
consisting of layers containing square planar arrays of $d$- or $f$-electron
cations embedded in an anion lattice. Here we will take a minimal
approach which focuses on the $d$ or $f$ electrons and treats the anion lattice
as providing a crystalline electric field and a hybridization network. This
misses the charge-transfer character \citep{ref:Zaanen} of the CuO$_2$ planes, the dynamic
polarization effects of anions such as As, and the spd conduction bands of the
heavy fermion and actinide anions. However, as we will discuss, we believe that
this approach captures the essential physics that leads to pairing in
these materials.

In outline, this approach begins with the selection of local $d$ or $f$ atomic
states for the (Cu, Fe, Ce, Pu) ions which takes account of the appropriate
crystal-field and spin-orbit couplings. Then these states are hybridized through
the (O, As, In, Ga) anion states, or directly, leading to a tight binding band
or bands. The tight binding hopping parameters are typically adjusted so that
the low energy states fit the results of bandstructure calculations. For the
heavy fermion and actinide systems, one includes a further phenomenological
renormalization. Here one has the Kondo physics to deal with and the approximation
is based on the assumption that just as in the single-ion case, the system
renormalizes to a heavy Fermi liquid. Then an onsite Coulomb interaction and, if there are multiple
orbitals, additional inter-orbital Coulomb and exchange interactions are added.
Even at this level, there are various parameterizations which involve the choice
of basis states for the bandstructure calculation, and the Wannier projection of
the bands in the vicinity of the Fermi energy onto the local orbital
basis \citep{ref:Gunnarsson,ref:Miyake2010,ref:Vildosola}.

Then of course, when a model is selected, one needs to determine its properties.
There have been a number of different theoretical approaches used to determine
the properties of Hubbard models. Analytic or semi-analytic methods have included
random phase approximations (RPA) \citep{ref:Miyake,ref:Scal-Loh-Hirsch,ref:Monthoux1991,ref:Graser2009},
renormalized meanfield theory (RMFT) \citep{ref:Anderson87,ref:Kotliar-prb88,ref:Anderson04},
conserving fluctuation exchange (FLEX) \citep{ref:Bickers,ref:DahmTewordt,ref:Kuroki99},
self-consistent renormalization (SCR) \citep{ref:Moriya03}, two-particle-self-consistent
(TPSC) \citep{ref:Tremblay}, and slave-boson approximations \citep{ref:Coleman,ref:Ruckenstein,ref:Kotliar-prl88}.
Numerical approaches include determinantal quantum Monte Carlo
(DQMC) \citep{ref:Blankenbecker,ref:Hirsch85,ref:Paiva}, variational Monte Carlo
(VMC) \citep{ref:Gros,ref:Paramekanti,ref:Ogata}, a variety of cluster Monte Carlo
(CDMFT \citep{ref:Kotliar01}, DCA \citep{ref:Jarrell},
VCPT \citep{ref:Potthoff}) methods, density matrix renormalization group
(DMRG) \citep{ref:White92} calculations as well as functional renormalization group
(FRG) \citep{ref:HalbothPRL,ref:Honerkamp,ref:Zhai,ref:Platt} studies.
Our goal in this section is to introduce the Hubbard models that have been used to describe the
unconventional superconductors and illustrate some of the results for their
physical properties which have been found from numerical calculations.

\subsection{The cuprates}

To illustrate the type of models that we have in mind, and discuss some of their
properties, we begin with the cuprates. At the Cu site, the crystal field
splitting pushes the Cu $d_{x^2-y^2}$ orbit up in energy so that it contains
the last (3d)$^9$ electron of Cu$^{2+}$.  The undoped system with one hole per
Cu, is a charge-transfer antiferromagnetic insulator with a gap set by the energy
to move the hole from a Cu to a neighboring O. The large onsite Cu Coulomb
interaction leads to well formed $S=1/2$ moments on the Cu which are coupled by
a Cu-O-Cu superexchange interaction \citep{ref:J}. A weak interlayer exchange coupling leads
to a N\'eel transition with a checkerboard antiferromagnetic spin arrangement
in the CuO$_2$ plane. When a material such as La$_{2-x}$Sr$_x$CuO$_4$ is hole
doped by adding Sr, the antiferromagnetism is rapidly suppressed and below a
temperature $T^*$ one enters a pseudogap phase. This phase is believed to reflect
the approach to the Mott state and provides a medium in which a variety
of instabilities can appear as the temperature is lowered. These continue to be
studied and among other correlations are believed to contain fluctuating charge
and $\pi$-phase shifted antiferromagnetic stripes \citep{ref:Emery1999} which at
low temperatures may order leading to a reconstruction of the Fermi
surface \citep{ref:NLM2010,ref:Yao,ref:Moon} or if disordered form a
spin glass \citep{ref:Tranquada99}. While evidence of superlattice order does
appear in some underdoped cuprates (La$_{1.875}$Ba$_{0.125}$CuO$_4$ \citep{ref:QLi}),
there are others, including ordered stoichiometric crystals (YBa$_2$Cu$_4$O$_8$
\citep{ref:Tomeno}) in which a pseudogap appears in the apparent absence of a
translational broken symmetry. This has led to various interesting theoretical
proposals of Fermi surface reconstruction without translational symmetry breaking
\citep{ref:YangRiceZhang,ref:Sachdev2}. In the overdoped regime the system is
metallic with a large Fermi surface and spin-fluctuations.

Early on, \citeauthor*{ref:Anderson87} suggested that a minimal model which contained the
essential cuprate physics was the single band Hubbard model. In this case, one
focuses on the Cu d$_{x^2-y^2}$ orbital and hybridizes it through the O anion
network leading to a single $d_{x^2-y^2}$ band.  Then adding
an onsite Coulomb interaction $U$, one has the well known 2D single band Hubbard
model \citep{ref:Hubbard}.
\begin{equation}
  H=-\sum_{ijs}t_{ij}(d^+_{is}d_{js}+d^+_{js}d_{is})+U\sum_in_{i\uparrow}n_{i\downarrow}
	\label{eq:4}
\end{equation}
Here $t_{ij}$ are tight binding one-electron hopping parameters between sites
$i$ and $j$ which are adjusted to fit the bandstructure and $U$ is an onsite
Coulomb interaction. In Eq.~(\ref{eq:4}), $d^+_{is}$ creates an electron with
spin $s$ in a $d_{x^2-y^2}$ orbital on the $i$th site, $d_{js}$ destroys one on
the $j$th site and $n_{i\uparrow}=d^+_{i\uparrow}d_{i\uparrow}$ is the
occupation number for a spin up electron on the $i$th site.

Although the single-band Hubbard model, Eq.~(\ref{eq:4}), is certainly a minimal
model, it exhibits a number of the basic phenomena which are seen in the cuprate
materials. At half-filling, in the strong coupling limit it maps to the 2D
spin 1/2 Heisenberg model on a square lattice. Numerical studies of the
Heisenberg model \citep{ref:Oitmaa} find evidence of long range antiferromagnetic
order at $T=0$. In addition, analytic calculations \citep{ref:Chakravarty,ref:Arovas}
have provided the basis for understanding a range of experimental results for the
undoped cuprates. Alternatively in weak coupling, it has been shown \citep{ref:Raghu}
that the doped Hubbard model has a transition to a $d_{x^2-y^2}$ superconducting
phase. While this result was obtained in the limit $U/t\to0$, it establishes the
fact that this simple model can exhibit a $d_{x^2-y^2}$ superconducting phase.

\begin{figure}[!htbp]
\includegraphics[height=9cm]{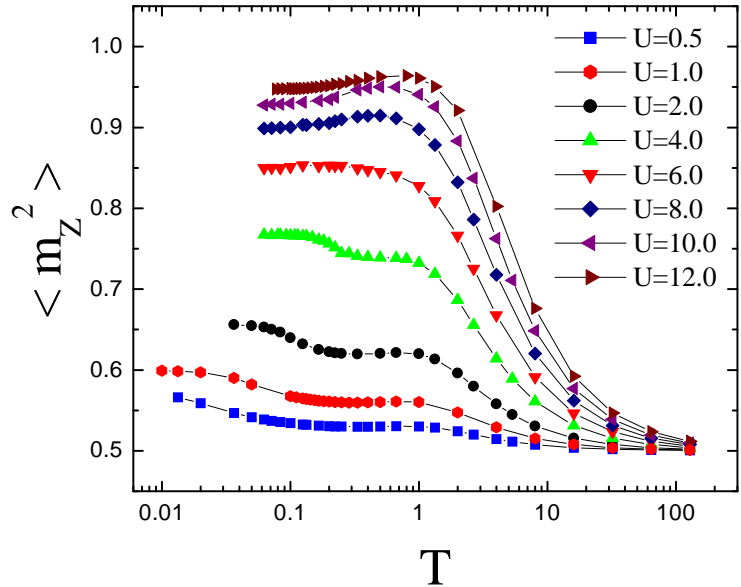}
\caption{The temperature dependence of the square of the local moment for
different values of the on-site Coulomb repulsion $U$ (in units where $t=1$).
As the temperature decreases below $\sim U/2$, local onsite correlations lead to
an increase in $\langle m^2_z\rangle$. Then on a lower temperature scale,
non-local spin correlations develop and for weak coupling $\langle m^2_z\rangle$
increases, while for strong coupling it decreases. This crossover marks a change
from an itinerant to a more local magnetic behavior (after Paiva et al.\protect\cite{ref:Paiva}).
\label{fig:Paiva}}
\end{figure}

As noted, there have been a variety of numerical approaches used to study the Hubbard model.
At half-filling, the particle-hole symmetry eliminates the so-called ``fermion
sign" problem for a Hubbard model with a near-neighbor one-electron hopping.
In this case, determinant quantum Monte Carlo (DQMC) \citep{ref:Blankenbecker}
calculations can be carried out on large lattices down to low temperatures.
These calculations find that the half-filled 2D Hubbard model
with a near neighbor hopping $t$ and an onsite Coulomb interaction $U$ of order
the bandwidth $8t$ is a Mott insulator and has a groundstate with long range
antiferromagnetic order \citep{ref:Hirsch85}. In addition, in this intermediate
coupling regime where $U$ is of order the bandwidth, one sees both the local
and itinerant character of the magnetism. Figure~\ref{fig:Paiva} shows Monte
Carlo results for the square of the $z$-component of the local moment
$m_z(\ell)=n_{\ell\uparrow}-n_{\ell\downarrow}$ versus temperature for a range
of $U/t$ values \citep{ref:Paiva}. As expected, when the temperature decreases
below a scale set by $U$, $\langle m^2_z\rangle$ increases. However, at a lower
temperature scale $\langle m^2_z\rangle$ is found to increase further for weak
coupling, while it decreases for strong coupling. In the weak coupling itinerant
case, this increase is associated with the formation of short range particle-hole
magnetic correlations. In this case, the energy gain at low temperatures is
proportional to $\langle m^2_z\rangle$ so that $\langle m^2_z\rangle$ increases
further as $T$ decreases. Alternately, in the strong coupling case, below an
energy scale $U$ one has well-defined local moments. In this case, as the
temperature decreases further and drops below the exchange energy
$J\sim4t^2/U$, virtual electron transfer associated with $J$ reduces the degree
of localization and $\langle m^2_z\rangle$ decreases. As seen in Fig.~\ref{fig:Paiva}
the crossover between this local moment and itinerant behavior occurs for a
value of $U$ of order the bandwidth. As we will see, it is in this intermediate
coupling parameter regime, where the system has both local and itinerant
characteristics, that the doped system has its highest $T_c$.

For the doped Hubbard model the fermion sign problem limits the temperatures
that are accessible using the DQMC approach and alternative numerical approximations
have been developed. Using a Gutzwiller projected $d$-wave BCS wavefunction
\citep{ref:Anderson87}, variational Monte Carlo (VMC) calculations have been
used to explore the $T=0$ phase diagram of the doped $x=1-\langle n\rangle$ Hubbard
model \citep{ref:Paramekanti}. The groundstate is found to be a $d$-wave
superconductor for $0<x<x_c$ with $x_c\approx0.35$. For $x>x_c$, the groundstate
is a Landau-Fermi liquid. At low doping ($x\ltwid0.1$) Gutzwiller projected
wavefunctions with both $d$-wave and antiferromagnetic variational parameters
have been found to have a lower energy than the $d$-wave alone, providing
evidence for a coexisting antiferromagnetic and $d$-wave superconducting phase
\citep{ref:Ogata}. These VMC calculations find results for the doping dependence
of the coherence length, the penetration depth as well as the momentum distribution
in agreement with experimental observations.

An alternative approach to dealing with the doped case is represented by
various cluster methods. Here, the basic idea is to treat the 
degrees of freedom within a cluster exactly and take into account the
correlations beyond the cluster by introducing a self-consistent dynamic
mean-field. The resulting problem of a cluster embedded in a dynamic meanfield
is then solved by means of exact disgonalization for small clusters or by
various Monte Carlo approaches such as the Hirsch-Fye algorithm \citep{ref:Hirsch86}
for larger clusters. The coupling of the cluster to the self-consistent dynamic
meanfield significantly reduces the fermion sign problem. In the so-called
cellular dynamic mean-field theory (CDMFT) \citep{ref:Kotliar01} and the variational
cluster-perturbation theory (VCPT) \citep{ref:Potthoff} methods, the system is
mapped onto an embedded cluster in real space while in the dynamic cluster
approximation (DCA) \citep{ref:Jarrell} the cluster is embedded in reciprocal
space. This latter scheme keeps the periodic boundary conditions and coarse grains
the Brillouin zone, making it a convenient approach for studying the momentum
dependence of the pairing interaction.

There are also functional renormalization group (FRG) approaches \citep{ref:Salmhofer},
so named because they follow the flow of the four-point vertex function
$\Gamma(k_1,k_2,k_3,k_4)$ for scattering between states on the Fermi surface as
the states outside an energy $\Delta E$ of the Fermi energy are integrated out.
Here, the degrees of freedom are reduced to states in a $\Delta E$-shell around
the Fermi surface. This shell is then discretized into a finite number of Fermi
surface patches which allows one to take into account the tangential momentum
dependence of the effective interaction. In practice, the
renormalization group equations are typically carried out at the one-loop level.
The resulting coupled renormalization group equations are then numerically
integrated to determine the functional renormalization group flow of the scattering
vertex as the energy cut-off $\Delta E$ or temperature is reduced. Although the
one-loop approximation means that it is necessary to start the system off with
appropriate bare interactions and stop the calculations when the renormalized
interaction grows too large, this approach can provide an unbiased treatment of
competing instabilities and can indicate which instability or combination of
instabilities are important. There have also been proposals in which the FRG
is used down to a given cut-off where the most divergent parts of
$\Gamma(k_1,k_2,k_3,k_4)$ are then taken to construct a low energy reduced
Hamiltonian \citep{ref:Lauchli}, which can then be solved using exact
diagonalization.

The density matrix renormalization group (DMRG) \citep{ref:White92} has also
been used to study these models. This approach has primarily been implemented
as a real space renormalization procedure in which degrees of freedom are
iteratively added, for example by increasing the size of the lattice system.
Then the less important degrees of freedom are truncated from the Hilbert space
by keeping only a finite number of the most probable eigenstates of a reduced
density matrix. This iterative, variational method is designed to thin the degrees
of freedom to those which play the dominant role in the ground state. It has
proved particularly effective for one-dimensional ladder models.

\begin{figure}[!htbp]
\includegraphics[width=8cm]{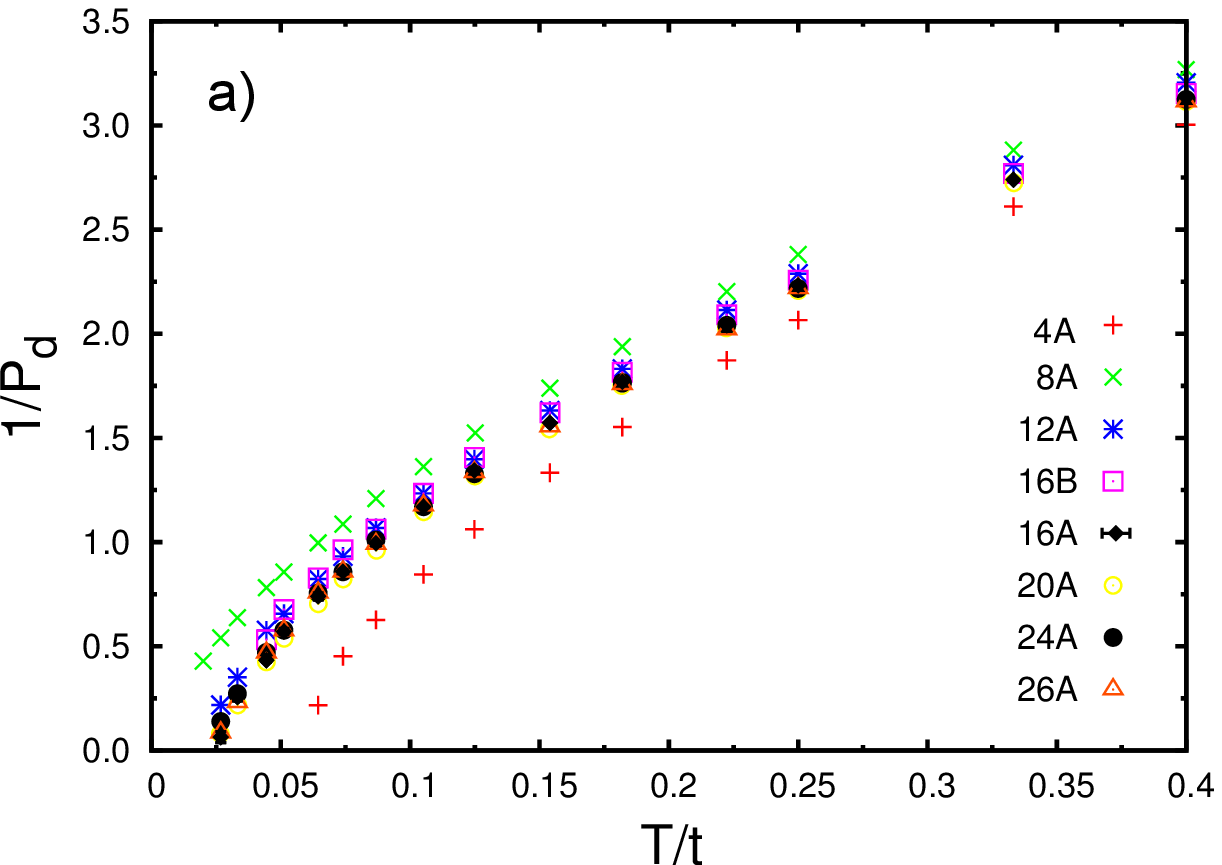}
\newline
\includegraphics[width=8cm]{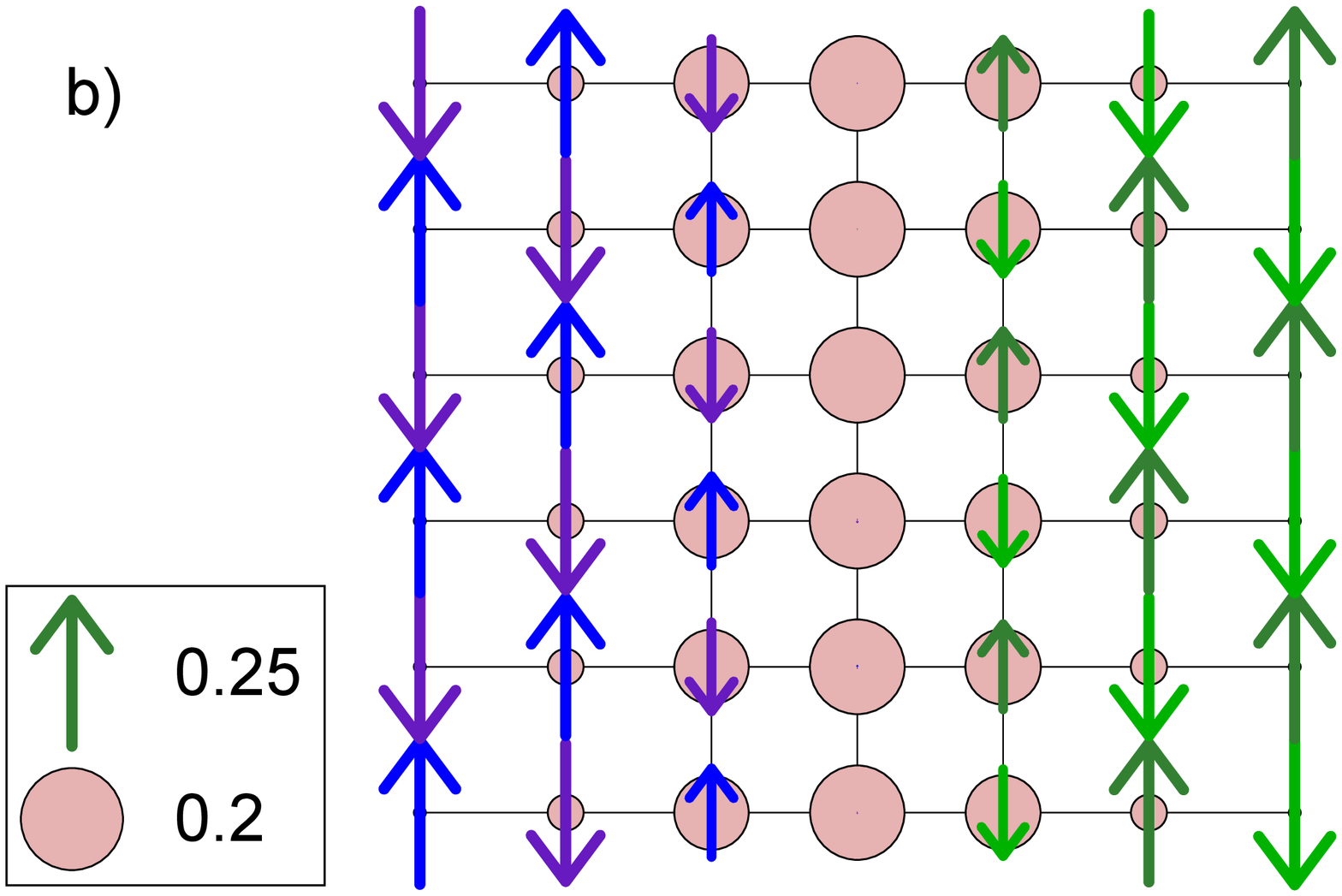}
\vspace{-1cm}
\caption{$d_{x^2-y^2}$ pairing and stripes have been found in various numerical
calculations for the doped Hubbard model. (a) DCA results for the inverse of the
$d_{x^2-y^2}$-wave pairfield susceptibility versus
$T/t$ for various sized Betts\protect\cite{ref:Betts} clusters. Here,
$U/t=4$ and $\langle n\rangle=0.9$ (after Maier et al.\protect\citep{ref:Maier}).
(b) The charge $\langle n_{\rm hole}(\ell)\rangle$ and spin $\langle S^z(\ell)\rangle$
structure seen in a DMRG calculations of a cylindrical 6-leg Hubbard model with
$U/t=12$ (after White et al.\protect\cite{ref:White} and Hager et al.\protect\cite{ref:Schrieffer14}).\label{fig:15a}}
\end{figure}

Using these approaches, further evidence has been found that the Hubbard models
exhibit many of the basic physical properties which characterize the
unconventional superconductors. Specifically, for the doped systems there is
evidence for antiferromagnetic spin-fluctuations, pseudogap behavior, nematic
correlations, $d$-wave or more generally unconventional pairing, as well as
stripes. Real space CDMFT \citep{ref:Senechal} and VCPT \citep{ref:Aichhorn}
cluster calculations find clear signatures of antiferromagnetic, pseudogap and
$d$-wave behavior in the Hubbard model. Including longer range one-electron
hopping, these calculations find ground state phase diagrams and single particle
spectral weights for electron- and hole-doping that are similar to the overall
behavior observed in these materials. A small orthorhombic distortion of the
one-electron hopping is found to lead to a large nematic response \citep{ref:Okamoto}.
Similarly, momentum space DCA calculations find evidence for pseudogap behavior
in the spin susceptibility and the single particle spectral weight \citep{ref:Macridin}
as well as nematic correlations \citep{ref:Su}. Using the DCA and a sequence of
different clusters \citep{ref:Betts}, \citet{ref:Maier} found
evidence shown in Fig.~\ref{fig:15a}a for the divergence of the $d$-wave
pairfield susceptibility
\begin{equation}
  P_d(T)=\int^{1/T}_0\langle\Delta_d(\tau)\Delta^+_d(0)\rangle d\tau
	\label{eq:6a}
\end{equation}
for a doped Hubbard model. Here
$\Delta^+_d=\frac{1}{2\sqrt{N}}\sum_{\ell,\delta}(-1)^\ell d^+_{\ell\uparrow}
d^+_{\ell+\delta\downarrow}$ with $\delta$ summed over the four near neighbor
sites of $\ell$.

FRG studies of the single band Hubbard model with a next near neighbor hopping
$t'$ find dopings for which the interaction vertex flows to antiferromagnetic
or $d$-wave dominated regimes as well as a region of intermediate doping in
which the forward scattering Pomeranchuck Fermi surface instabilities and CDW
as well as nematic fluctuations grew \citep{ref:HalbothPRL,ref:Honerkamp,ref:Zhai}.
In this latter region, umklapp processes are found to play an important role
linking the instabilities in various channels. In the underdoped regime,
\citet{ref:Lauchli} have used the FRG to construct a low-energy
effective Hamiltonian and argue that umklapp processes truncate Fermi surface
segments leading to a psedogap phase. FRG calculations have also been carried
out for the multi-orbital Hubbard models \citep{ref:Zhai,ref:Platt}. Here the
geometry of the electron- and hole-Fermi surfaces (see Fig.~\ref{fig:17a} of
the next section) lead to SDW $(\pi,0)$ and $s^\pm$ pairfield dominated flow
regimes along with other umklapp mediated scattering processes
\citep{ref:Maiti,ref:Fernandes1110}.

Calculations using the density matrix renormalization group (DMRG) to study the
2-leg Hubbard ladder find a spin gapped state at half-filling and power law
$d$-wave-like pairfield correlations for the doped system \citep{ref:Noack}.
As discussed in Sec.~\ref{sec:4}B, a twisted version of this same 2-leg ladder
mimics the SDW stripe structure and $s^\pm$ pairing correlations seen in the
Fe-based superconductors \citep{ref:Berg09}. Calculations for a doped 6-leg
Hubbard ladder \citep{ref:White,ref:Schrieffer14} find striped
charge-density-$\pi$-phase-shifted-antiferromagnetic states like that shown in
Fig.~\ref{fig:15a}b.
While short range $d_{x^2-y^2}$ pairfield
correlations along the stripes were also observed there were no long range
$d$-wave pairing correlations. In this case, periodic boundary conditions were
used for the 6-site direction and open end boundary conditions along the
direction of the legs forming a cylindrical tube. On an 8-leg $t$-$J$ system
\citep{ref:White98} the favored filling was 0.875 and the $\pi$-phase shifted
antiferromagnetic striped structure was similar to that shown Fig.~\ref{fig:15a}b
with each cylindrical stripe containing four holes corresponding to a
half-filled stripe. This is the same pattern
which is observed in La$_{1.875}$Ba$_{0.125}$CuO$_4$ \citep{ref:Fujita}. In
these calculations, the tube-like boundary conditions favor the formation of
cylindrical stripes. The short length of the circumference of the tube
suppresses pair fluctuations between the stripes and leaves only short range
$d$-wave pairing correlations along a stripe. With open boundary conditions
and applied fields to orient the stripes along the long direction of the 6-
and 8-leg ladders that have been studied, pair fluctuations between the stripes
become possible and a stronger $d$-wave pairing response is observed. While
present DMRG calculations find that the antiphase $d$-wave state is slightly
higher in energy than that of the in-phase state, VMC calculations found
parameter ranges in which the antiphase state was stabilized \citep{ref:Himeda}.
There are also calculations for a coupled ladder model that exhibit stripes
with antiphase pairing \citep{ref:Berg07}.

Finally, along with the observations of $d$-wave and stripe
correlations, there is numerical evidence of pseudogap behavior in the
underdoped Hubbard model. A variety of dynamic cluster Monte Carlo calculations
of the single particle spectral weight \citep{ref:Macridin,ref:Kyung2,ref:Aichhorn2}
show the emergence of pseudogap behavior in the underdoped $t$-$t'$-$U$
Hubbard model.  A phenomenological theory of the
pseudogap phase by \citet{ref:YangRiceZhang}
has had success in reproducing many of the observed properties
of the pseudogap regime.

The important point for the present discussion is that while the choice of the
variational wavefunction in the VMC
and finite size effects for the cluster calculations can influence what one
finds, there is overall agreement
among these various approaches that Hubbard models exhibit many of the basic
physical properties which characterize the unconventional superconductors \citep{ref:Scal-handbook,ref:Kancharla}.
There are of course phenomena such as the unusual ordered magnetic phase in the
underdoped cuprates observed in polarized neutron scattering experiments \citep{ref:Fauque}
and dichroic angular resolved photoemission measurements
\citep{ref:Kaminski} which have not yet been found in these basic Hubbard models.
Here we take the view that these phenomena are peripheral to the pairing mechanism.

\subsection{The Fe-pnictides}

The undoped Fe-pnictide materials have partially filled 3d shells and are
antiferromagnetic metals below $T_N$. Their magnetic moments alternate in
alignment row to row creating a stripe-like antiferromagnetic pattern different
from the checkerboard pattern of the cuprates. Just above, or in some cases
coinciding with, $T_N$ there is a tetragonal to orthorhombic lattice
transition. As the system is doped, both the structural and the N\'eel
transitions are suppressed and superconductivity occurs \citep{ref:Johnston}.

For the Fe-pnictide superconductors, photoemission \citep{ref:Malaeb} as well as
band structure calculations \citep{ref:Lebegue,ref:Singh,ref:Cao}
find that the states associated with the pnictide 4p orbitals are located some
2eV or more below the Fermi level. Thus an effective tight binding model based
on the five Fe 3d orbitals can provide a reasonable description of the electronic
states near the Fermi surface. Since the crystal field splitting, as well as the
exchange and spin-orbit splittings of the iron 3d orbitals are small relative to
the bandwidth, all five 3d orbitals need to be taken into account. For the 1111
materials the 3D coupling between the Fe layers is relatively weak and 2D models
have proved useful. Due to the tetrahedral coordination of the pnictide, the
unit cell contains two Fe sites. However, the Fe-pnictide plane is invariant
under a reflection and a translation since each Fe has the same local
arrangement of the surrounding atoms. Thus for the 2D Fe-pnictide layer one can
unfold the Brillouin zone and work with an effective five-orbital model on a
square lattice with one Fe per unit cell \citep{ref:Lee-Wen}. Including one-electron hopping
parameters to describe both the direct Fe-Fe hopping as well as the hybridized
hopping through the pnictide or chalcogen $4p$ orbits, one arrives at a 5-band
model with the one electron part of the Hamiltonian given by \citep{ref:Kuroki2008,ref:Graser2009}
\begin{equation}
  H_0=\sum_{ij}\sum_{\ell n\sigma}t^{\ell n}_{ij}c^+_{i\ell\sigma}c_{jn\sigma}+
	\sum_i\sum_{\ell\sigma}\varepsilon_\ell n_{i\ell\sigma}
	\label{eq:5}
\end{equation}
Here $\ell=(1,2,\dots5)$ denotes the Fe-$d$ orbitals $(d_{xz},d_{yz},d_{xy},
d_{x^2-y^2},d_{3z^2-r^2})$ and $c^+_{i\ell\sigma}$ creates an electron on site
$i$ in the $\ell$th orbit with spin $\sigma$. The tight binding parameters
$t^{\ell n}_{ij}$ describe the one-electron hopping from the $\ell$th orbit on
site $i$ to the $n$th orbit on site $j$ and $\varepsilon_\ell$ is the site
energy of the $\ell$th orbit. The onsite Coulomb and exchange interaction part
of the Hamiltonian is
\begin{eqnarray}
  H_1=\sum_i\biggl(\sum_\ell Un_{i\ell\uparrow}n_{i\ell\downarrow}
	      &+&U'\sum_{\ell'<\ell}n_{i\ell}n_{i\ell'}\nonumber \\
		 -J\sum_{\ell\ne\ell'}{\bf S}_{i\ell}\cdot{\bf S}_{i\ell'}
		    &+&J'\sum_{\ell\ne\ell'}c^+_{i\ell\uparrow}c^+_{i\ell\downarrow}
				c_{i\ell'\downarrow}c_{i\ell'\uparrow}\biggr)
	\label{eq:6}
\end{eqnarray}
with $n_{i\ell}=n_{i\ell\uparrow}+n_{i\ell\downarrow}$ and ${\bf S}_{i\ell}=
\frac{1}{2}c^+_{i\ell\sigma}\boldsymbol{\sigma}_{\sigma\sigma'}c_{i\ell\sigma'}$.
Here $U$ and $U'$ are the intra- and inter-orbital Coulomb interactions, $J$ is
the Hund's rule exchange and $J'$ the so-called pair hopping term. If these
interactions are generated from a two-body term with spin rotational invariance
$U'=U-2J$ and $J'=J$. However, many body interactions can renormalize these
couplings altering these relations. In addition the dressed
interaction terms can in general depend on the orbital indices.

\begin{figure}[!htbp]
\includegraphics[height=8cm,viewport=5pt 5pt 630pt 313pt,clip=true]{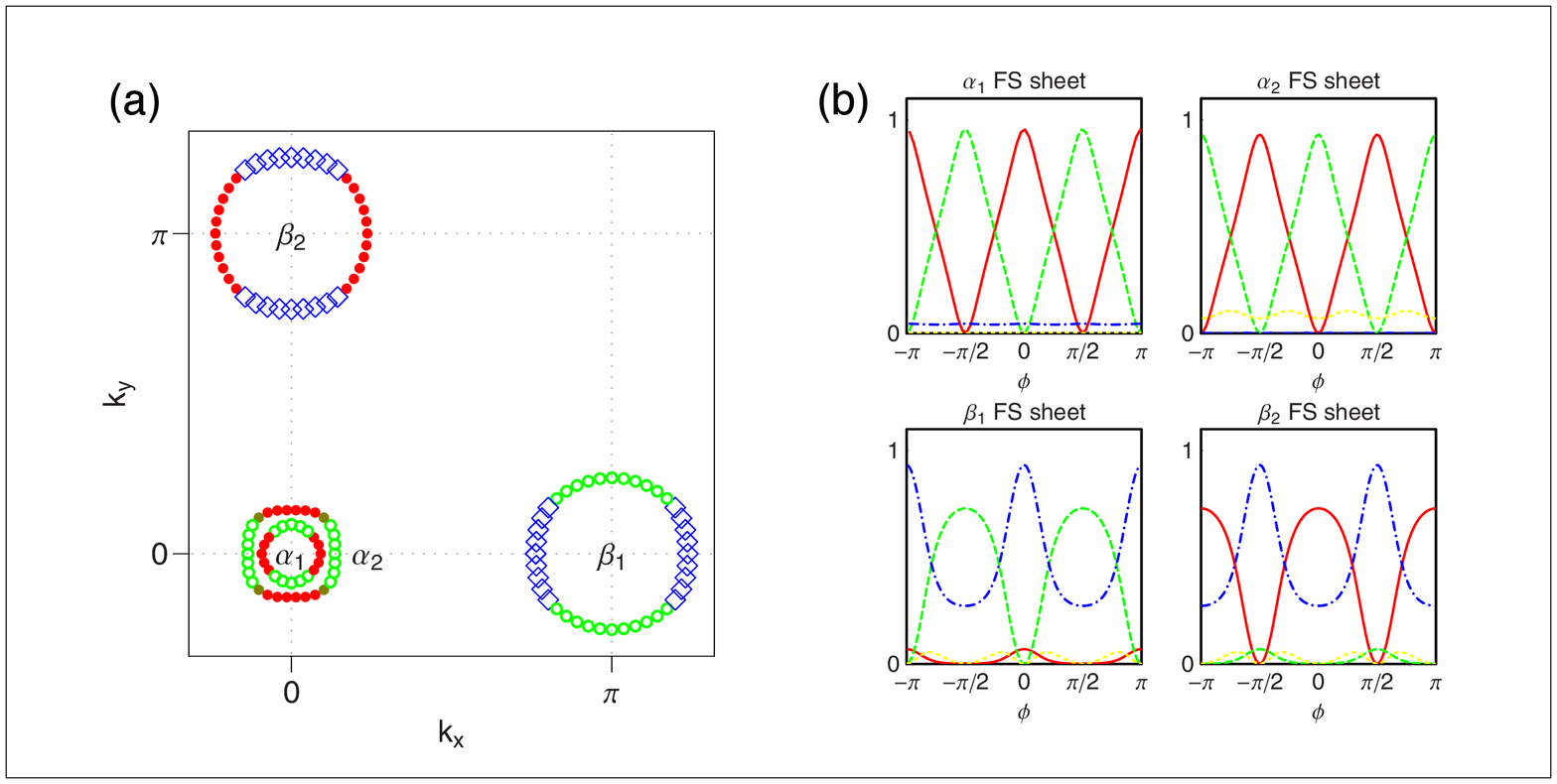}
\caption{This figure illustrates the Fermi surfaces and orbital weight
distributions for a 5-orbital model of the 1111 Fe-based superconductors.
(a) The Fermi surface sheets of a 5-orbital tight binding model of the
1111 Fe-based superconductors. The symbols (color online) denote the orbital
which has the largest orbital weight, with the $d_{xz}$ (red solid circles $\bullet$),
$d_{yz}$ (green open circles $\circ$), and $d_{xy}$ (blue open diamonds $\Diamond$).
(b) The orbital weights as a function of winding angle $\phi$ on the various Fermi
surface sheets with $d_{xz}$ (solid red), $d_{yz}$ (dashed green), $d_{xy}$
(dash-dot blue) and $d_{x^2-y^2}$ (short dashed yellow). The $d_{3z^2-r^2}$
orbital weight is negligible. Here, the $d_{xz}$ and $d_{yz}$ orbitals are
aligned along the Fe-Fe directions (after Graser et al.\protect\cite{ref:Graser2009}).\label{fig:17a}}
\end{figure}

The Fe$^{+2}$ ion separation $\sim2.7^\circ$A is significantly smaller than the
Cu$^{+2}$ separation of $\sim3.8^\circ$A and the direct Fe-Fe hopping along
with the d-p hybridization through the pnictogen or chalcogen anions leads to
a metallic groundstate. Observation of quantum oscillations provide clear
evidence of well defined small Fermi surfaces consistent with a semi-metallic
bandstructure \citep{ref:QoscFe}. The basic structure of the Fermi surfaces of the
Fe-based superconductors consists of two electron cylinders at the zone corner
of the 2Fe per unit cell Brillouin zone compensated by two or three hole
sections around the zone center. The Fermi surface sheets for a two-dimensional
five-orbital tight-binding fit \citep{ref:Graser2009} of the DFT
bandstructure \citep{ref:Cao} of LaOFeAs are shown in Fig.~\ref{fig:17a}a. Here
and in the following an unfolded 1Fe per unit cell Brillouin zone will be used.
Diagonalizing the 5-orbital tight-binding Hamiltonian of Eq.~(\ref{eq:6}), one
has for the Bloch states of the $\nu$th band,
\begin{equation}
  \Psi_{\nu\sigma}(k)=\sum_\ell\langle\nu k|\ell\rangle c_{\ell\sigma}(k)
	\label{eq:9a}
\end{equation}
where, again, $\ell$ sums over the Fe orbitals $(d_{xz},d_{yz},\cdots)$ and
$c_{\ell\sigma}(k)=\sum_{\bf i}c_{{\bf i}\ell\sigma}e^{i({\bf k}\cdot{\bf i})}/\sqrt{N}$.
The main orbital
weight contributions $|\langle\nu k|\ell\rangle|^2$ to the band states that lie
on the various Fermi surfaces are indicated by the colors in Fig.~\ref{fig:17a}a.
A more detailed look at the orbital weights is shown in Fig.~\ref{fig:17a}b,
where they are plotted as a function of the winding angle on the different
Fermi surfaces. Here one sees, for example, that the $d_{yz}$ and $d_{xy}$
orbitals contribute the dominant weights on the $\beta_1$ electron pocket
while it is the $d_{xz}$ and $d_{yz}$ that mainly contribute to the $\alpha$
pockets. These orbital weights play an important role in determining the
strength and structure of the pairing interaction.

While the 1111 materials can be reasonably treated as two-dimensional, the
structure of the 122 systems is such that one needs to take their three
dimensionality into account. The loss of the reflection-translation invariance
of the 2D layer leads to more complex 10-orbital models \citep{ref:Suzuki}.

\subsection{The heavy fermion materials}

The heavy fermion materials have incomplete $f$-shells and there is a
balance between the strong onsite Coulomb interactions which tend to localize
the $f$-electrons and the hybridization with extended bandstates of the ligand
anions which delocalize them. At high temperatures the system exhibits local
moment behavior with magnetic moments of order atomic values while at low
temperatures the system resembles a Fermi liquid with large quasi-particle masses
associated with the hybridized $f$-electrons. In the coexisting state where one has both
SDW antiferromagnetism and superconductivity, the magnitude of the ordered moments
determined from neutron scattering and the effective mass of the paired electrons,
determined from the specific heat jump at $T_c$, are large. Thus the $f$ electrons
play an important role in both the antiferromagnetism and the
superconductivity \citep{ref:Gegenwart,ref:Nair}.

\citeauthor*{ref:Hotta} introduced a minimal model for such an $f$-electron
system based on a $j-j$ coupling scheme since the spin-orbit interaction is large.
In addition they noted that this provided a convenient way to define the
one-electron states that make up the pairs. The resulting Hamiltonian for the
115 Ce heavy fermion superconductors has a form similar to Eqs.~(\ref{eq:5})
and (\ref{eq:6}) but with the one electron
operators describing Kramer's doublets and with $\sigma$ a pseudospin quantum
number. The 14-fold degenerate $f$ electronic states are split by the spin-orbit
coupling into a low lying $j=5/2$ sextet and a higher energy $j=7/2$ octet. For
Ce$^{3+}$ with a $(4f)^1$ configuration, only the $j=5/2$ sextet contributes to
the electronic states near the Fermi energy. The one electron states of the
$j=5/2$ sextet are further split by the crystalline electric field of the In
ligand anions, separating the six $j=5/2$ states into three sets of Kramer's
doublets. For a tetragonal crystal field one has
\begin{equation}
  \begin{array}{lll}
  c^+_{in\sigma}&=pf^+_{i\pm5/2}+qf^+_{i\mp3/2}&n=1\\
	&\ -qf^+_{i\pm5/2}+pf^+_{i\mp3/4}&n=2\\
	&\ f^+_{i\pm1/2}&n=3
  \end{array}
	\label{eq:7}
\end{equation}
Here $f^+_{im}$ creates an electron on the $i$th lattice site in a $j=5/2$
orbital with a $z$-component of total angular momentum $m$. The ``orbital"
index $n=1$, 2 and 3 denotes the $\{\Gamma_7,\Gamma^\prime_7,\Gamma_6\}$
tetragonal field Kramer's doublets, the $q$ and $p$ coefficients in Eq.~(\ref{eq:7})
depend on the tetragonal crystalline field and $\sigma=\pm1$ is the pseudospin
quantum number.

As schematically illustrated in Fig.~\ref{fig:4}, the spacing $\sim4.6^\circ$A
of the Ce$^{3-}$ ions is the largest of the three systems and the $4f$
electrons of Ce$^{3-}$ tend to be localized. Thus as opposed to the itinerant
3d-electrons of the Fe-based materials and the doped cuprates, the $f$-electrons
of the heavy fermion 115 materials are nearly localized. The materials are
metallic because of the $4p$ states of the anions and the dispersion of the $4f$
electrons arises from their hybridization with these $4p$ conduction electrons.
As in both the Fe-pnictide and the doped cuprates, quasi-two-dimensional Fermi surfaces
have been observed in de Haas-van Alphen experiments \citep{ref:Elgazzar} for the Ce compounds.
Similarly to the Fe-based superconductors, the heavy fermion materials have
multiple Fermi surfaces and there are orbital weight factors associated with
the $\Gamma_7$, $\Gamma^\prime_7$ and $\Gamma_6$ orbital states.

As previously discussed, the plutonium intermetallic compounds PuMGa$_5$
have the same tetragonal structure as the cerium-based heavy fermion 115
superconductors.  Electronic structure calculations \citep{ref:Maehira} for
PuCoGa$_5$ show a similarity between the main Fermi surfaces of CeCoIn$_5$ and
PuCoGa$_5$. In particular, there are $f$-electron dominated cylindrical Fermi
surface hole sheets centered at the $\Gamma$ point, and cylindrical electron
sheets centered at the $M$ point of the 1Fe per unit cell Brillouin zone. Using
the $j$-$j$ coupling scheme to construct a low energy model for this actinide
superconductor, \citet{ref:Maehira} noted that the Pu-115
compound is the hole version of Ce-115. That is, the low lying $j=5/2$ sextet
accommodates the one $(4f)$ electron of Ce$^{3+}$ for CeCoIn$_5$, while it has
one hole for the $(5f)^5$ Pu$^{3+}$ ion in PuCoGa$_5$.This picture of the Pu-115
compound being a hole version of the Ce-115 compound is particularly striking for
PuCoIn$_5$ and CeCoIn$_5$.


Finally, while the existence and, to a reasonable degree, the structure of Fermi surfaces
of the heavy fermion \citep{ref:Hall} and the Fe-based \citep{ref:Shishido1,ref:Terashima}
superconductors are well established, the situation for the cuprates is still
debated \citep{ref:Norman}.
In the overdoped single layer cuprate Tl$_2$Ba$_2$CuO$_{6+\delta}$ (Tl2201)
both angle-dependent magnetoresistance \citep{ref:Hussey} and ARPES
measurements \citep{ref:Plate} provide evidence for a large quasi-two-dimensional
Fermi surface in reasonable agreement with bandstructure calculations. More
recently \citep{ref:Vignolle}, the observation of quantum oscillations in the
magnetoresistance and the magnetization of Tl2201 provided direct evidence of
this large hole-like Fermi surface and coherent fermionic excitations. Here,
the observation of quantum oscillations are
important in determining that coherent excitations are present.
Following the development of highly ordered YBa$_2$Cu$_3$O$_{6.5}$ (ortho-II)
crystals \citep{ref:Liang}, quantum oscillations were also observed in the
underdoped regime, both in the Hall resistance \citep{ref:Doiron-Leyraud} and in
the magnetization \citep{ref:Jaudet2008,ref:Sebastian2008}. This showed that the
doped cuprates, just as the heavy fermion and Fe superconducting materials, can
have a Fermi surface with low-lying fermionic excitations, even in the underdoped
regime. The fact that the Hall and Seebeck coefficients are negative indicates that
the observed small Fermi-surface pockets are electron-like
\citep{ref:LeBoeuf2007,ref:Chang}. The large Fermi surface of the
overdoped cuprates must therefore undergo a reconstruction as the doping level
decreases \citep{ref:Taillefer2009}. One mechanism for such a reconstruction is
the occurrence of some new periodicity associated with an ordered phase such
as a spin striped phase \citep{ref:Millis,ref:Moon2009} or a unidirectional
charge density wave \citep{ref:Yao}.
NMR measurements show that high magnetic fields induce charge order without
spin order \citep{ref:Wu2011} which would be consistent with a unidirectional charge
density wave. Some studies
\citep{ref:LeBoeuf,ref:Laliberte,ref:Wu2011} attribute the Fermi-surface
reconstruction in YBa$_2$Cu$_3$O$_y$ to a form of stripe order similar to
that observed in La$_2$CuO$_4$-based cuprates \citep{ref:Tranquada} and there is
evidence for a phase transition at $T^*$ associated with some form
of density wave \citep{ref:Chang} or nematic \citep{ref:Daou2010} order
leading to a pseudogap phase \citep{ref:He}. A recent compilation
\citep{ref:Sebastian1112} of ARPES measurements, high magnetic field quantum
oscillation studies and transport experiments suggests that a small $Q$ wave-vector
bidirectional charge density wave provides an explanation for the nodal Fermi
surfaces which is consistent with a wide variety of complementary measurements.


\section{The Pairing Interaction\label{sec:4}}

In this section, we examine the structure of the pairing interaction for the
models discussed in Sec.~\ref{sec:3}. The pairing interaction is given by the
irreducible particle-particle four-point vertex. As discussed in the appendix,
for the conventional superconductors this interaction is well described by a
phonon exchange and screened Coulomb interaction. In general, for spin rotationally invariant
models, the irreducible particle-particle vertex can be separated into a fully
irreducible vertex and $S=1$ spin and $S=0$ charge (particle-hole) exchange
channels. For the 2D Hubbard model near half-filling, DCA calculations find
that the $S=1$ spin channel gives the dominant contribution to the pairing.
Similarly, for the two-layer Hubbard model introduced in this section, it is
the $S=1$ spin fluctuation channel that leads to pairing. However, as
discussed, it can lead to $B_{1g}$ ($d$-wave) or $A_{1g}$ ($s$-wave) pairing
depending upon the structure of the Fermi surface. This bilayer Hubbard model,
as well as a ``twisted ladder" model discussed in this section, illustrate the
link between the cuprate and Fe-based superconductors. For the multi-band Fe-based superconductors
one has only weak coupling results, but here the resulting phenomenology
provides evidence that the pairing is driven by the spin fluctuations and
similarly for the heavy fermion models where it is the pseudo-spin fluctuations.
The conclusion is that the pairing in the models of Sec.~\ref{sec:3}
is mediated by spin-fluctuations.

As discussed in Appendix A, the momentum and frequency dependence of the
superconducting gap provide information on the space-time structure of the pairing
interaction \citep{ref:ScalRandom}. For conventional superconductors such as Pb
or Hg, the gap is weakly dependent upon momentum but strongly frequency
dependent, implying that the pairing interaction is short range and has a
retarded part. As is well known, electron tunneling \citep{ref:McMillan}
and optical absorption \citep{ref:Farnworth} measurements of the frequency
dependence of the gap for the low $T_c$ materials identify the pairing
interaction as arising from a retarded phonon-mediated contribution and an
``instantaneous" repulsive screened Coulomb term. For the unconventional
superconductors, a determination of both the momentum and frequency dependence
of the gap are important. Here a wide variety of experiments have been used to
probe the momentum dependence of the gap. These include ARPES
\citep{ref:Damascelli,ref:Argonne,ref:Valla,ref:Kordyuk,ref:Yun}, phase
sensitive tunneling experiments \citep{ref:VanHarlingen,ref:Tsuei,ref:Hanaguri},
Raman scattering \citep{ref:Muschler,ref:Caprara}, low temperature thermal
conductivity \citep{ref:Sutherland} and directional magnetic field specific heat
measurements \citep{ref:CeRh}. There have also been various
tunneling \citep{ref:LeeNature,ref:Pasupathy,ref:Jenkins,ref:Ahmadi} and
optical studies of the frequency dependence of the gap
\citep{ref:Basov,ref:vanHeumen,ref:Carbotte}.
Thus, at present, there are a range
of experimental results and interpretations. From many of these it appears that
for the unconventional superconductors one is dealing with a pairing interaction
that peaks at a large momentum transfer characteristic of the near-neighbor
antiferromagnetic or SDW correlations and which has a frequency response
characteristic of the spectrum of the antiferromagnetic spin fluctuations.
However, there are questions and controversies regarding this
\citep{ref:Zhou,ref:Bok,ref:LiNature,ref:Giannetti}
and it remains a challenge to obtain the close interplay
between experiment and theory that was the hallmark for the traditional
superconductors. Furthermore, a complete range of measurements for the heavy
fermion and Fe-based materials, comparable to the results for the cuprates,
are not yet available.

With this in mind, this review has the more limited goal of understanding the
momentum, frequency and orbital structure of the interaction that is
responsible for pairing in the models discussed in Sec.~\ref{sec:3}. To the
extent that these models exhibit the basic low energy properties which are found
in these materials, one can argue that the interaction responsible for pairing
in the models will reflect the pairing interaction in the real materials.

In this section, we will show dynamic cluster approximation (DCA) results for
the pairing interaction. The basic assumption of the DCA is that the self-energy
and irreducible vertex functions are short-ranged and can be well represented by
a finite size cluster. Under this assumption, one sets up an effective cluster
problem as an approximation for the bulk thermodynamic limit in order to calculate
these quantities. This is done by representing the bulk lattice by an effective
cluster embedded in a mean-field bath, which is designed to represent the remaining
degrees of freedom and is determined self-consistently. In contrast to other finite size
methods, in which one carries out calculations on finite size lattices and then
tries to scale up in size, the DCA, for a given cluster size, gives approximate
results for the bulk thermodynamic limit.

The DCA treats spatial correlations on length scales within the cluster accurately
and non-perturbatively and describes longer-ranged correlations on a mean-field
level. It becomes exact in both the weak-coupling ($U/t=0$) and strong-coupling
($t/U=0$) limits. For finite $U/t$, one can in principle obtain exact results by
carrying out calculations for different size clusters and then extrapolating to
infinite cluster size. Convergence with cluster size depends on the specific
problem, but is usually faster than with finite size methods, because of the
inclusion of the remaining degrees of freedom in terms of a mean-field. This was
discussed for the 3D half-filled Hubbard model in \citet{ref:Kent} and \citet{ref:Fuchs},
where the accuracy of the DCA was benchmarked against finite size methods for several
different quantities. In particular, it was shown that well converged results for
the antiferromagnetic $T_N$ versus $U$ phase diagram can be obtained from
relatively small clusters. As noted, in this approach the cluster is embedded
in reciprocal space and one obtains momentum space results on a coarse grained
Brillouin zone. It is convenient to work in momentum space and since the
pairing interaction is expected to be short-ranged it is actually more amenable
to cluster calculations than the long-range pairfield correlations. Like the
FRG calculations, the DCA provides an unbiased treatment of the competing
instabilities. In addition, it takes account of self-energy and interaction
effects within the cluster while treating the remaining degrees of freedom
within a dynamic meanfield.

\subsection{The single band Hubbard model}

\begin{figure}[!htbp]
\includegraphics[width=13cm]{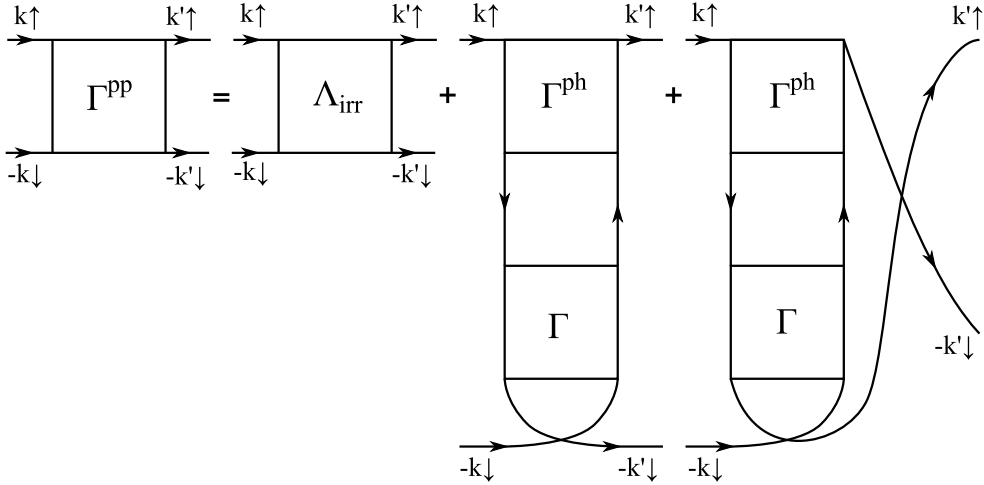}
\caption{The pairing interaction is given by the irreducible particle-particle
vertex $\Gamma^{pp}$. Here $\Gamma^{pp}$ is decomposed into a fully irreducible
two-fermion vertex $\Lambda_{\rm irr}$ plus contributions from the $S=1$ and
$S=0$  particle-hole channels. $\Gamma^{ph}$ are irreducible particle-hole
vertices, $\Gamma$ is the full vertex and the solid lines are fully dressed
single particle propagators.\label{fig:12}}
\end{figure}

\begin{figure}[!htbp]
\includegraphics[height=18cm]{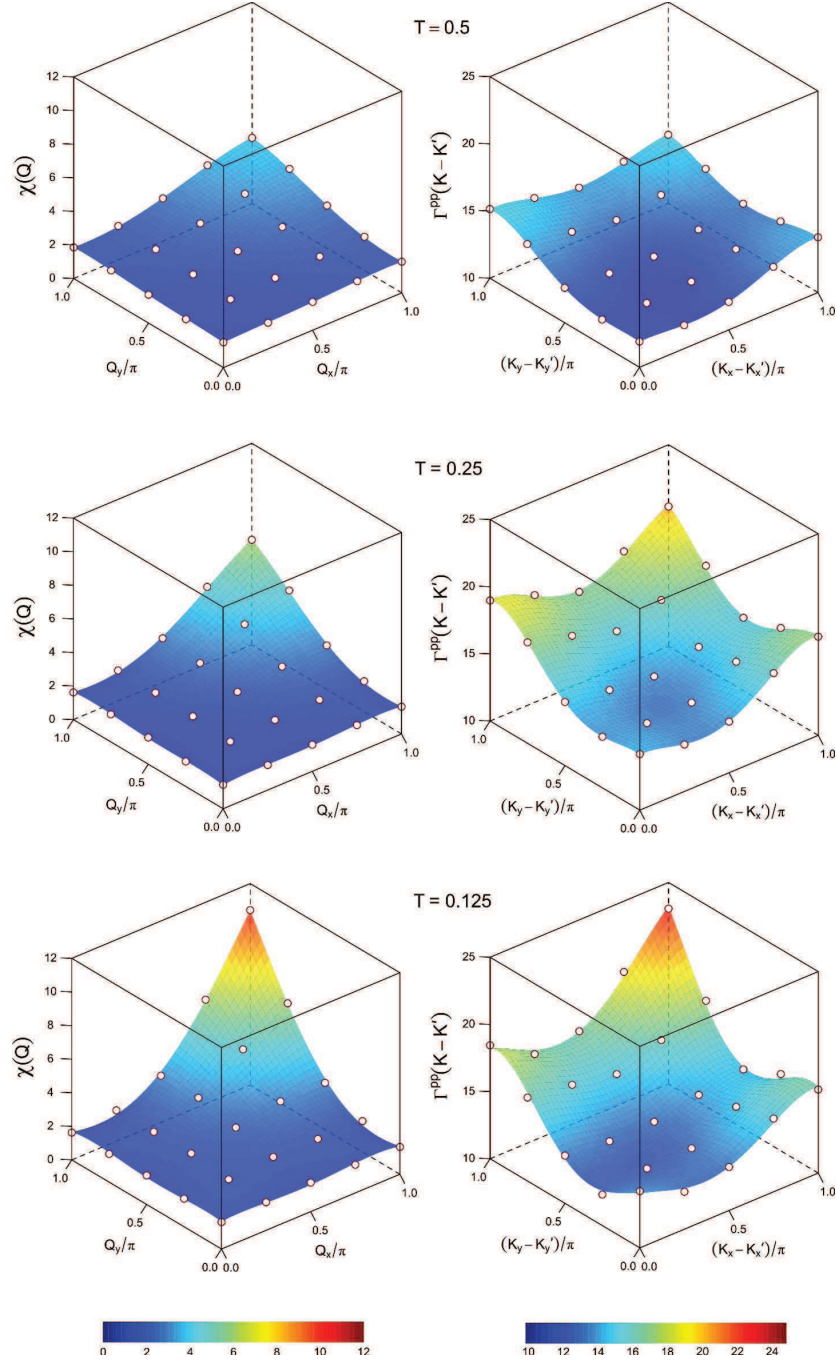}
\caption{The spin susceptibility $\chi(q)$ and the pairing interaction
$\Gamma^{pp}(K,K')$ for $U=4t$ and $\langle n\rangle=0.85$ are compared at
various temperatures. As the temperature is reduced a peak develops in $\Gamma^{pp}$
reflecting the peak in $\chi$. This repulsive peak is the origin of the
unconventional superconductivity discussed in this review.\label{fig:chigamma}}
\end{figure}

\begin{figure}[!htbp]
\includegraphics[width=11cm]{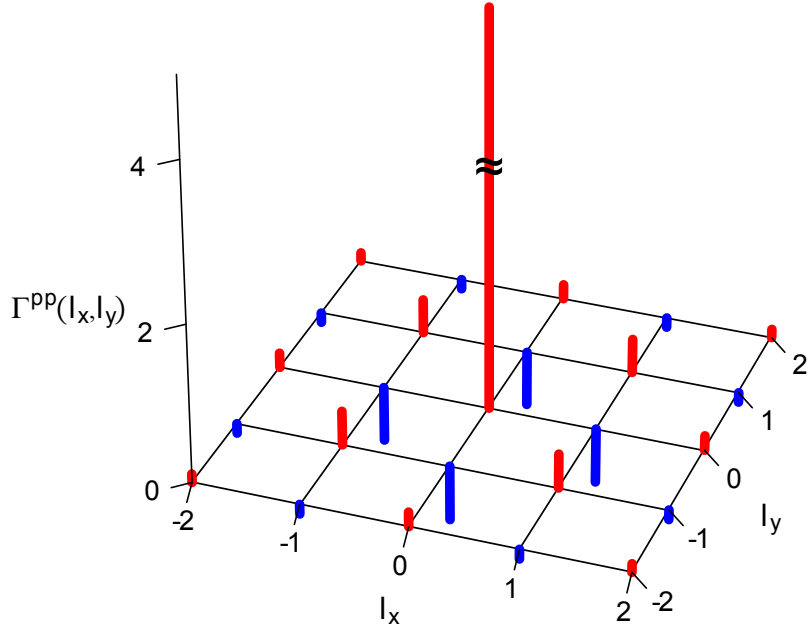}
\caption{The real space structure of the pairing interaction obtained from the
Fourier transform Eq.~(\protect\ref{eq:8}) of $\Gamma^{pp}(k,k')$ at a temperature $T=0.125t$ for $U=4t$
and $\langle n\rangle=0.85$. Here red indicates a repulsive and blue an
attractive pairing interaction for a singlet formed between an electron at the
origin and an electron at site $(\ell_x,\ell_y)$. The peak in $\Gamma^{pp}$
shown in Fig.~\protect\ref{fig:chigamma} leads to a pairing interaction which
oscillates in space.\label{fig:15}}
\end{figure}

For the single band Hubbard model DCA numerical simulations have been used to
determine the momentum and frequency dependence of the pairing interaction \citep{ref:Maier1}.
Formally, this interaction is given by the irreducible particle-particle scattering vertex
$\Gamma^{pp}(k,k')$ shown on the left-hand side of Fig.~\ref{fig:12}.
It consists of all Feynman diagrams that
can not be separated into two parts by cutting just two particle lines. Here,
$k=({\bf k},i\omega_n)$ with $\omega_n=(2n+1)\pi T$ a fermion Matsubara frequency
and one is interested in the scattering of a pair in a singlet, zero center-of-mass
momentum and energy state with relative momentum and Matsubara frequency $k=({\bf k},i\omega_n)$
to a final state with $k'=({\bf k'},i\omega_{n'})$. Results obtained from a
64-site $8\times8$ numerical dynamic cluster approximation (DCA) for
$\Gamma^{pp}(k,k')$ with $\omega_n=\omega_{n'}=\pi T$ at a filling
$\langle n\rangle=0.85$ and $U=4t$ are shown on the right hand side of
Fig.~\ref{fig:chigamma}.\footnote{Just as the electron-phonon interaction strength
is characterized by $\int\frac{d\omega}{\pi}|g_q|^2\frac{{\rm Im}{\cal D}(q,\omega)}{\omega}
=|g_q|^2{\rm Re}{\cal D}(q,0)=\frac{-2|g_q|^2}{\omega_q}$ and a cut-off frequency
of order $\omega_{\cal D}$, the pairing interaction strength for the Hubbard model
is given by $\Gamma(k,k')$ with $\omega_n=\omega_{n'}=\pi T$. The cut-off in the
Matsubara frequency is set by the spin-fluctuation spectrum as shown in Fig.~\protect\ref{fig:18}.}
Here one sees that as the temperature is lowered, the singlet pairing
interaction increases for large momentum transfers. This is a reflection of the
growth of the short range antiferromagnetic spin-fluctuations as seen in a
similar plot of the spin susceptibility $\chi(q)$ shown on the left hand side of Fig.~\ref{fig:chigamma}.
Taking the Fourier transform of $\Gamma^{pp}(k,k')$
\begin{equation}
  \Gamma^{pp}(\ell_x,\ell_y)=\frac{1}{N}\sum_{kk'}e^{ik\cdot\ell}\Gamma^{pp}(k,k')
	e^{ik'\cdot\ell}
  \label{eq:8}
\end{equation}
leads to the real space picture of the pairing interaction illustrated in
Fig.~\ref{fig:15}.
Here $\Gamma^{pp}(\ell_x,\ell_y)$ is the strength of the $\omega_n=\omega_{n'}=\pi T$
pairing interaction between a singlet formed with one electron at the origin and
the other at site $(\ell_x,\ell_y)$. It is large and repulsive if the electrons
occupy the same site but attractive if they are on near neighbor sites
reflecting the peaking of $\Gamma^{pp}(k,k')$ for $k-k'\sim(\pi,\pi)$.

\begin{figure}[!htbp]
\includegraphics[width=16cm]{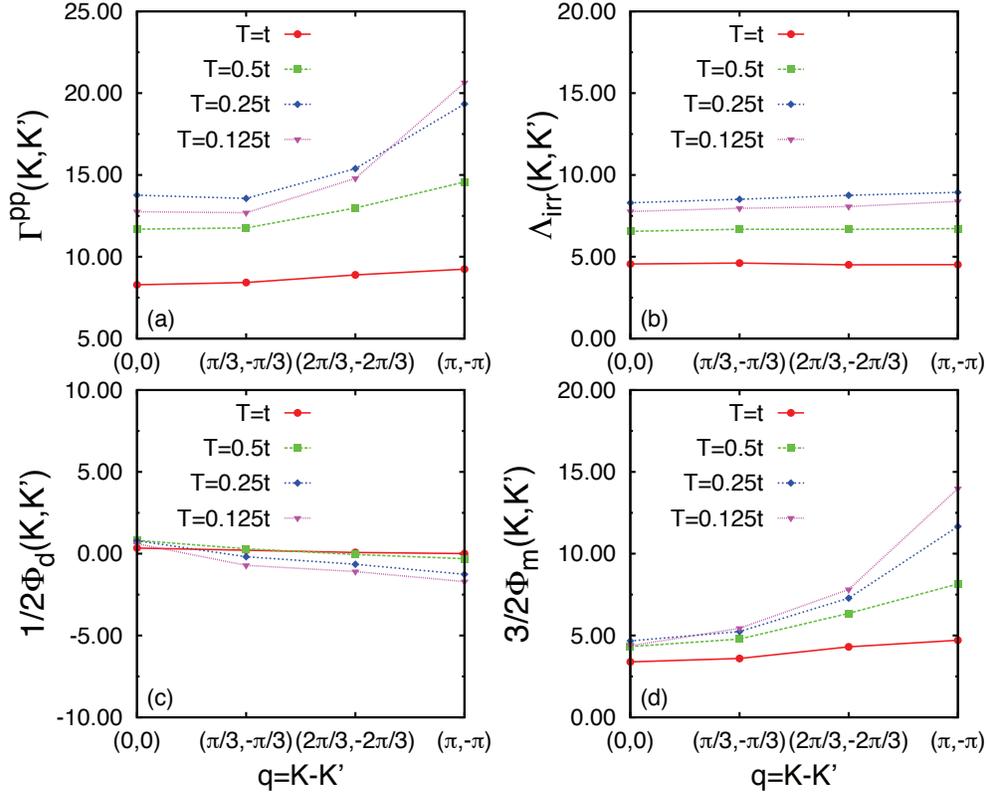}
\caption{This figure illustrates the momentum dependence of the various
contributions that make up the irreducible particle-particle pairing vertex
$\Gamma^{pp}$. (a) The irreducible particle-particle vertex $\Gamma^{pp}$ versus
$q=K-K'$ for various temperatures with $\omega_n=\omega_{n'}=\pi T$. Here,
$K=(\pi,0)$ and $K'$ moves along the momentum values of the 24-site cluster
which lay on the dashed line shown in the inset of Fig.~\protect\ref{fig:17}.
Note that the interaction increases with the momentum transfer as expected for
a $d$-wave pairing interaction. (b) The $q$-dependence of the fully irreducible
two-fermion vertex $\Lambda_{\rm irr}$. (c) The $q$-dependence of the charge
density $(S=0)$ channel $\frac{1}{2}\Phi_d$ for the same set of temperatures.
(d) The $q$-dependence of the magnetic $(S=1)$ channel $\frac{3}{2}\Phi_m$.
Here, one sees that the increase in $\Gamma^{pp}$ with momentum transfer arises
from the $S=1$ particle-hole channel (after Maier et al.\protect\cite{ref:Maier1}).
\label{fig:16}}
\end{figure}

As shown in Fig.~\ref{fig:12}, the pairing interaction $\Gamma^{pp}{(k,k')}$ can be separated into a
fully irreducible two-fermion vertex $\Lambda_{\rm irr}$ and partially reducible
particle-hole exchange contributions. Here the fully irreducible part
$\Lambda_{\rm irr}$ is defined as the sum of all diagrams that can not be
separated into two pieces by cutting any combination of two lines (particle or
hole). For a spin rotationally invariant system, the particle-hole exchange
contributions appearing on the right hand side of Fig.~\ref{fig:12} can be
combined into an $S=1$ magnetic spin fluctuation piece
$\frac{3}{2}\Phi_m$ and a spin $S=0$ charge density fluctuation contribution
$\frac{1}{2}\Phi_d$.
\begin{equation}
  \Gamma^{pp}(k,k')=\Lambda_{\rm irr}(k,k')+\frac{3}{2}\Phi_m(k,k')+\frac{1}{2}\Phi_d
	(k,k')
  \label{eq:9}
\end{equation}
Carrying out a DCA calculation, one can evaluate the individual terms that enter
Eq.~(\ref{eq:9}). The upper left panel of Fig.~\ref{fig:16}
shows the pairing interaction $\Gamma(k,k')$ versus momentum transfers along the
diagonal $(k_x-k'_x,k_y-k'_y)$ of Fig.~\ref{fig:chigamma} for
$\langle n\rangle=0.85$ and $U/t=4$ as the temperature is reduced. The remaining
panels of Fig.~\ref{fig:16} show the contributions of the fully irreducible vertex
$\Lambda_{\rm irr}$, the $S=0$ charge-fluctuations $\frac{1}{2}\Phi_d$ and the $S=1$ spin-fluctuations
$\frac{3}{2}\Phi_m$. As noted, it is the increase of $\Gamma$ with momentum transfer
that gives rise to the attractive near-neighbor pairing and it is clear from
Fig.~\ref{fig:16}, that this comes from the $S=1$ part of the interaction.
The fully irreducible vertex is essentially independent of momentum transfer and so
it only contributes to the on-site repulsion, while the $S=0$ charge part decreases
at large momentum giving rise to a small repulsive near neighbor interaction.

\begin{figure}[!htbp]
\includegraphics[height=7.5cm]{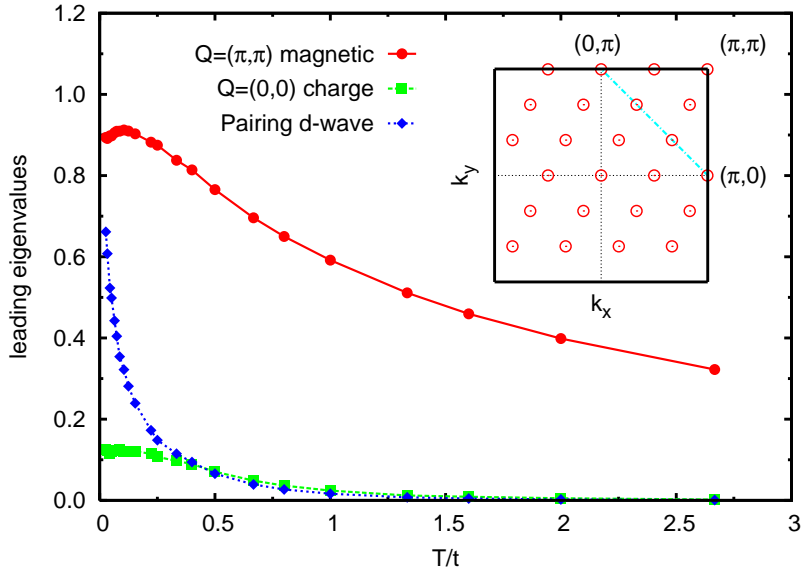}
\caption{Leading eigenvalues of the Bethe-Salpeter equation in various
channels for $U/t=4$ and a site occupation $\langle n\rangle=0.85$. The
$Q=(\pi,\pi)$, $\omega_m=0$, $S=1$ magnetic eigenvalue is seen to saturate
at low temperatures. The leading eigenvalue in the singlet $Q=(0,0)$,
$\omega_m=0$ particle-particle channel has $d_{x^2-y^2}$ symmetry and
increases toward 1 at low temperatures. The largest charge density eigenvalue
occurs in the $Q=(0,0)$, $\omega_m=0$ channel and saturates at a small value.
The inset shows the distribution of $k$-points for the 24-site cluster
(after Maier et al.\protect\cite{ref:Maier1}).\label{fig:17}}
\end{figure}

\begin{figure}[!htbp]
\includegraphics[height=7.5cm]{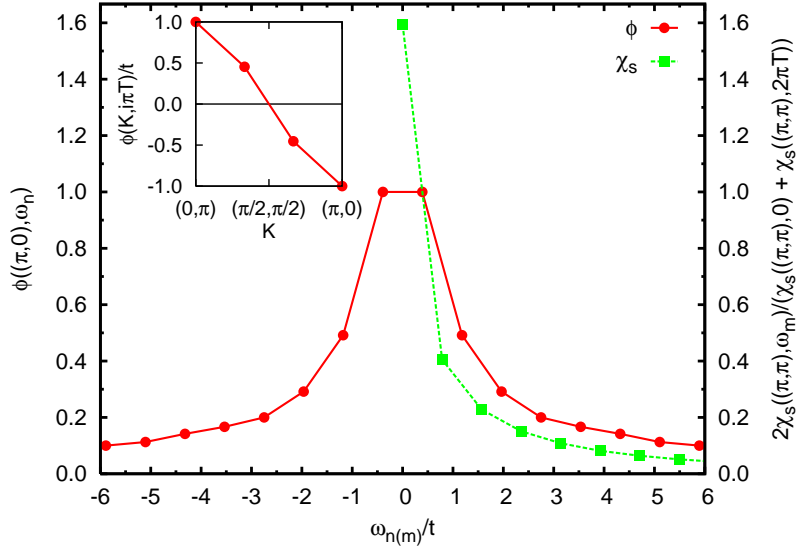}
\caption{The Matsubara frequency dependence of the eigenfunction
$\phi_{d_{x^2-y^2}}(K,\omega_n)$ of the leading particle-particle eigenvalue
of Fig.~\protect\ref{fig:17} for $K=(\pi,0)$ normalized to $\phi(K,\pi T)$
(red, solid). Here, $\omega_n=(2n+1)\pi T$ with $T=0.125t$. The Matsubara
frequency dependence of the normalized magnetic spin susceptibility
$2\chi(Q,\omega_m)/[\chi(Q,0)+\chi(Q,2\pi T)]$ for $Q=(\pi,\pi)$ versus
$\omega_m=2m\pi T$ (green, dashed). The Matsubara frequency dependence of
$\phi_{d_{x^2-y^2}}$ and the normalized spin $Q$ susceptibility are similar.
Inset: The momentum dependence of the eigenfunction $\phi_{d_{x^2-y^2}}(K,\pi T)$ normalized to
$\phi_{d_{x^2-y^2}}((0,\pi),\pi T)$ shows its $d_{x^2-y^2}$ symmetry. Here,
$\omega_n=\pi T$ and the momentum values correspond to values of $K$ which
lay along the dashed line shown in the inset of Fig.~\protect\ref{fig:17}
(after Maier et al.\protect\cite{ref:Maier1}).\label{fig:18}}
\end{figure}

In these numerical calculations, one also obtains the dressed single particle
Green's function $G(k,i\omega_n)$. Given $G$ and $\Gamma^{pp}$, one can determine the
Bethe-Salpeter eigenvalues and eigenfunction in the particle-particle channel
by solving
\begin{equation}
  -\frac{T}{N}\sum_{k'}\Gamma^{PP}(k,k')G_\uparrow(k')G_\downarrow(-k')
	\phi_\alpha(k')=\lambda_\alpha\phi_\alpha(k).	
	\label{eq:10}
\end{equation}
This is basically the fully dressed BCS gap equation and when the leading
eigenvalue goes to 1 the system becomes superconducting. One can also
construct similar Bethe-Salpeter equations for the charge and magnetic
particle-hole channels. Figure~\ref{fig:17} shows a plot of the leading
eigenvalues associated with the particle-particle pairing channel and the
particle-hole charge $S=0$ and spin $S=1$ channels for $U/t=4$ and a filling
$\langle n\rangle=0.85$. As the temperature is lowered, the particle-hole
$S=1$ antiferromagnetic channel with center of mass momentum $Q=(\pi,\pi)$ is
initially dominant. However, at low temperatures the $Q=0$ pairing channel rises
rapidly and the divergence of the antiferromagnetic channel saturates. The
charge channel eigenvalue remains small. Thus one concludes that the pairing
interaction arises from the exchange of $S=1$ particle-hole fluctuations.

\begin{figure}[!htbp]
\includegraphics[height=12cm]{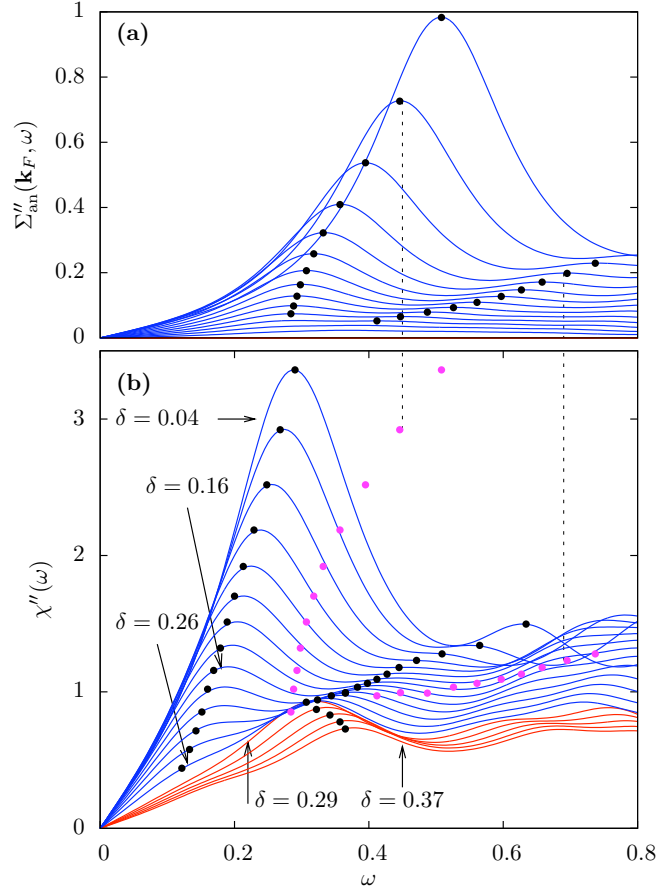}
\caption{This figure provides evidence linking the frequency dependence of the
imaginary part of the gap function $\phi_d(\omega,k_F)$, which is called
$\Sigma''_{\rm an}(\omega,k_F)$ in this figure, to the frequency
dependence of the spin fluctuation spectral weight $\chi''(\omega)$.
(a) The imaginary part of the gap function $\Sigma''_{\rm an}(\omega,k_F)$
at a wave vector $k_F$ near the antinode is plotted versus $\omega$ for various
dopings $\langle n\rangle=1-\delta$. (b) The imaginary part $\chi''(\omega)$ of
the local spin susceptibility versus $\omega$ for the same set of dopings. The
black dots in (a) and (b) identify peaks. The position of the peaks of
$\Sigma''_{\rm an}$ in (a) are shown as the magenta dots in (b) at the same
height as the corresponding $\chi''$ to illustrate their correspondence. One
can see that the upward frequency shift of the $\Sigma''_{\rm an}$ peaks relative
to the $\chi''$ peaks decreases with the doping like the single particle gap.
The red curves are for the normal state. Here, $U=8t$, $t'=-0.3t'$, $t''=-0.08t$
and a Lorentzian broadening of $0.125t$ was used for an embedded $2\times2$
plaquette (after Kyung et al.\protect\cite{ref:Kyung}).\label{fig:19}}
\end{figure}

The momentum dependence of the leading pairing eigenfunction $\varphi_\alpha(k)$
is shown in the inset of Fig.~\ref{fig:18} and corresponds to a $d_{x^2-y^2}$-wave.
The Matsubara frequency dependence of this eigenfunction, shown in Fig.~\ref{fig:18},
has a similar decay to that of the spin susceptibility. However, as one knows,
it is difficult to determine the real frequency response from limited numerical
Matsubara data. Recent cellular dynamic meanfield studies by
\citet{ref:Kyung} for real frequencies find a correspondence
between the frequency dependence of the gap function and the local spin
susceptibility as shown in Fig.~\ref{fig:19}.
The frequency dependence of the interaction has also been discussed by
\citet{ref:Maier2008} and \citet{ref:Hanke} who find that the
dominant part of the interaction comes from the spectral region associated with
spin fluctuations with an additional small contribution coming from high
frequency excitations. All of these dynamic calculations are for small clusters
so that it will be useful to have further work on the dynamics for larger
clusters since it provides an important fingerprint of the pairing interaction.

At low temperatures where the leading eigenvalue $\lambda_\alpha$ of Eq.~(\ref{eq:10})
approaches 1, the pairing interaction $\Gamma^{pp}(k,k')$ can be approximated as
\begin{equation}
 \Gamma^{pp}(k,k')\cong\varphi_\alpha(k)V_\alpha\varphi_\alpha(k')
 \label{eq:a}
\end{equation}
with a pairing strength $V_\alpha$
\begin{equation}
 V_\alpha=\frac{\sum_{k,k'}\varphi_\alpha(k)\Gamma^{pp}(k,k')\varphi_\alpha(k')}
                          {\left(\sum_{k}\varphi^2_\alpha(k)\right)^2}.
 \label{eq:b}
\end{equation}
Using Eq.~(\ref{eq:a}), the inverse of the pairfield susceptibility is approximately given by
\begin{equation}
 P^{-1}_\alpha\cong P^{-1}_{0\alpha}+V_\alpha
 \label{eq:c}
\end{equation}
with
\begin{equation}
 P_{0\alpha}=\frac{T}{N}\sum_kG(k)G(-k)\varphi^2_\alpha(k).
 \label{eq:d}
\end{equation}
Here $G(k)$ is the dressed single particle Green's function. For $d_{x^2-y^2}$-wave
pairing one has $\phi_\alpha(k)\sim(\cos k_x-\cos k_y)$ with a Matsubara frequency
cut-off as seen in Fig.~\ref{fig:18}. As seen in Fig.~\ref{fig:chigamma},
$\Gamma^{pp}(k,k')$ peaks for {\bf k}-{\bf k}$'\sim(\pi,\pi)$ so that $V_d$ given
by Eq.~(\ref{eq:b}) is negative. One can think of $P_{0\alpha}$
as the ``intrinsic" $\alpha$-pairfield susceptibility of the interacting system.

\begin{figure}[!htbp]
\includegraphics[height=12cm]{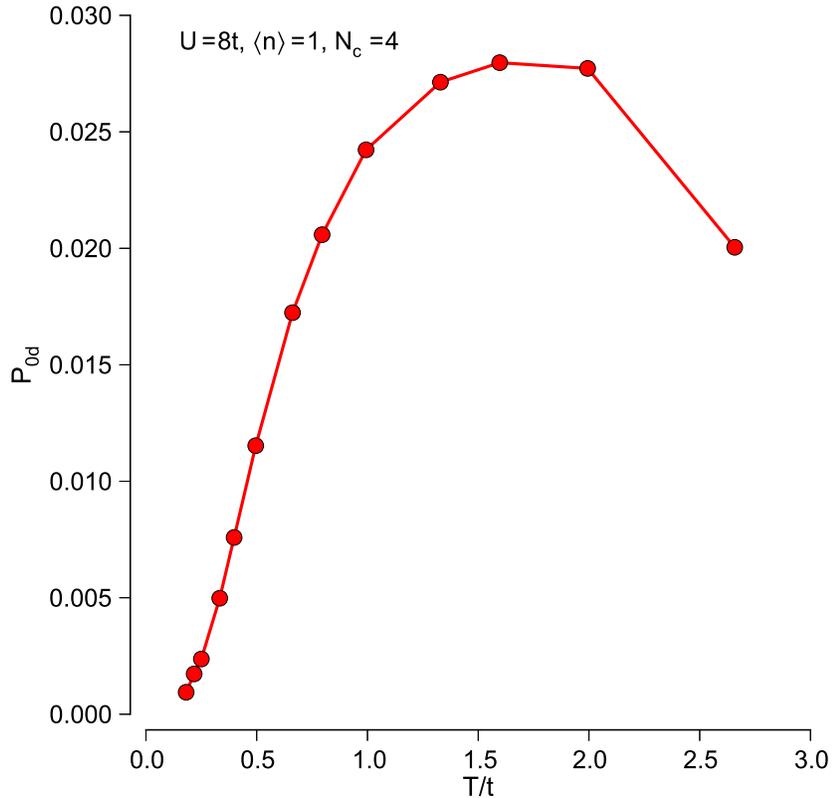}
\caption{The intrinsic pairfield susceptibility $P_{0d}(T)$ for $U=8t$ and
$\langle n\rangle=1$ is suppressed as $T$ goes to zero
(after Maier et al.\protect\cite{ref:Maier2006}).\label{fig:24}}
\end{figure}

In the traditional phonon mediated case, the pairing strength $V_\alpha$ is
essentially independent of temperature once the ionic lattice is formed.
Then the $N(0)$ log $(\omega_D/T)$ divergence of $P_{0\alpha}$ gives a transition
temperature $T_c\sim\omega_De^{-1/N(0)|V_\alpha|}$ where $P^{-1}_\alpha(T_c)=0$.
For a strongly interacting system, both $P_{0\alpha}$ and $V_\alpha$ are functions
of temperature. As seen from the temperature dependence of $\Gamma^{pp}(k,k')$ in
Fig.~\ref{fig:chigamma}, the strength $|V_d(T)|$ of the interaction will increase as the temperature
is lowered and $\chi(Q,T)$ increases. For the doped system, away from the
antiferromagnetic instability, $|V_d(T)|$ will saturate to a constant value at low
temperatures. However, as the doping $x$ goes to zero, it will continue to
increase as the temperature decreases. In this case for $\langle n\rangle=1$, $P_{0d}(T)$,
Fig.~\ref{fig:24}, will be suppressed at low temperatures due to the vanishing of
the quasi-particle weight as well as phase fluctuations \citep{ref:Emery1999}
and $T_c$ will go to zero \citep{ref:Maier2006}.
The interplay of the pairing strength $V_\alpha$, as $\langle n\rangle$ goes
to 1, and the intrinsic pairfield susceptibility $P_{0d}$, which is
suppressed as $\langle n\rangle$ goes to 1, leads to a dome-shaped $T_c$ versus doping
behavior. Thus while the strength of the pairing interaction can increase,
the increased scattering leads to a reduction of the quasiparticle weight.
In addition, it is important to remember that the pairing interaction is
short range, of order the near-neighbor spacing. This is reflected in the
$(\cos k_x-\cos k_y)$ structure of the gap. Thus it is not the correlation
length of the antiferromagnetic correlations but rather having the spectral
weight of the interaction in the right momentum and energy regime that
determines the pairing strength.

The interplay of $P_{0d}$ and the pairing interaction strength is of
particular interest near a quantum critical point
\citep{ref:Si,ref:Sachdev2,ref:Metlitski2}. \citet{ref:Abanov2}
have argued that the pseudogap phase reflects aspects of the pairing in the
quantum-critical regime near the antiferromagnetic QCP. Recently,
\citet{ref:Metlitski2} have discussed the special role played by
the competition between the spin density wave, Fermi surface structure and
superconducting order in the two-dimensional system. In this case, while the
quasiparticle spectral weight is suppressed at ``hot spots" on the Fermi surface
where $\varepsilon_{k+Q}=\varepsilon_k$, they find that the pairing interaction
slightly away from the hot spots is strong and combined with a finite
quasiparticle spectral weight can lead to high $T_c$ superconductivity. 

Based on the similarity of the momentum and frequency dependence of $\Gamma^{pp}$ to
that of the spin susceptibility $\chi$, approximate pairing interactions have been used
in which
\begin{equation}
  \Gamma^{pp}(k,k')\simeq\frac{3}{2}\bar U^2\chi(k-k').
	\label{eq:11}
\end{equation}
Here $\bar U$ is treated as an adjustable parameter and $\chi$ is numerically
calculated \citep{ref:MJS}, approximated by a phenomenological RPA-like
function \citep{ref:Monthoux1991} or determined experimentally from neutron
scattering \citep{ref:Dahm} or inelastic x-ray scattering (RIXS) data \citep{ref:LeTacon}.
These calculations find that with reasonable coupling strengths the spin-fluctuation
interaction given by Eq.~(\ref{eq:11}) can account for the scale of the observed
transition temperatures. Note that when one speaks of pairing mediated by spin-fluctuations
one is not thinking of an exchange of some boson with a sharp well defined
$\omega(q)$ dispersion. Rather what is meant is that the dominant pairing
interaction arises from the $S=1$ part of the particle-hole exchange contributions
to $\Gamma^{pp}$. While this particle-hole exchange has some of the characteristics of
a spin 1 boson, its spectral weight is spread out in momentum and frequency.
This is clearly seen in the numerical calculations of $\Gamma^{pp}$ and to the
extent that the spin susceptibility provides an approximation for the $\Gamma^{pp}$,
it is seen directly in experimental measurements of $\chi''(q,\omega)$. Finally,
it is important to keep in mind that low frequency spin fluctuations are pair
breaking \citep{ref:Varma} and the optimal spin-fluctuation spectral weight for
pairing occurs in a frequency range larger than twice the maximum value of the
gap \citep{ref:MonthouxScal1994}.

This aspect of the dynamics of the pairing interaction is reflected in the
rapid increase in $\Delta_{\rm Max}(T)$ as $T$ decreases below $T_c$ as well as
large $2\Delta_{\rm Max}(0)/kT_c$ ratios \citep{ref:Monthoux,ref:MonthouxScal1994,ref:Pao}.
As the gap opens the low frequency pair breaking spin fluctuation spectral weight
is shifted to higher energies where it contributes to the pairing, increasing
the gap. The increase in the gap in turn leads to a further suppression of the
low-frequency interaction spectral weight producing a positive feedback and a
rapid increase of $\Delta_{\rm Max}(T)$ as $T$ drops below $T_c$. Finally, at
low temperatures one finds a large $2\Delta_{\rm Max}(0)/kT_c$ ratio. This is
due to the altered spin-fluctuation spectral weight in the superconducting state
which gives rise to a stronger pairing interaction than the normal state. In
principle, if one could create a spin-fluctuation spectral weight in the normal
state which had the same structure that it has deep in the superconducting state,
one would find a significant increase in $T_c$.

\subsection{The bilayer Hubbard model}

\begin{figure}[!htbp]
\begin{center}
\includegraphics[width=15cm,viewport=5pt 5pt 1035pt 501pt,clip=true]{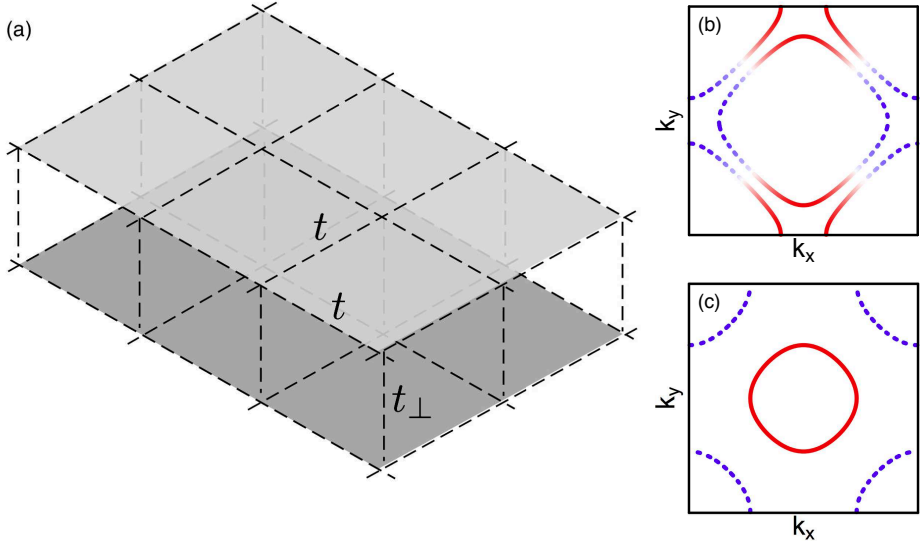}
\caption{The bilayer Hubbard model illustrates the close connection between the
$d$- and $s^\pm$-wave states. (a) The bilayer Hubbard lattice with a near neighbor intra-layer hopping
$t$ and an inter-layer hopping $t_\perp$. (b) The bonding ($k_z=0$) and antibonding
($k_z=\pi$) Fermi surfaces for $t_\perp=0.5$ (upper) and 2.0 (lower) for a filling
$\langle n\rangle=0.95$. A $d_{x^2-y^2}$ gap structure is illustrated for the
$t_\perp=0.5$ Fermi surface and an $s^\pm$ gap is shown for $t_\perp/t=2.0$.
Here, a solid (red) line denotes a positive gap and a dashed (blue) line a
negative gap. The intensity of the line denotes the $d$-wave ($\cos k_x-\cos k_y$)
like variation of the gap (color online).
\label{fig:bilayer}}
\end{center}
\end{figure}

Another variation of the Hubbard model, the bilayer Hubbard model, provides an
interesting link between the single- and multi-orbital models. It shows how
the structure of the Fermi surface or surfaces can alter the spin fluctuations
and change the gap symmetry from $B_{1g}$ ($d$-wave) to $A_{1g}$ ($s^\pm$-wave).
It is an example which illustrates how the spin fluctuation interaction can
give rise to the different gap structures seen in the cuprate and iron-based
superconductors. As shown in Fig.~\ref{fig:bilayer}(a) in the bilayer Hubbard
model, two 2D Hubbard layers are coupled by a one-electron
inter-layer hopping $t_\perp$.
For a doping near half-filling, the topological character of the non-interacting
Fermi surface changes as $t_\perp/t$ is turned on. For example, for
$\langle n\rangle=0.95$ and $t_\perp/t\ltwid0.07$, the system has two electron
Fermi surfaces around the origin. Then when $t_\perp/t\gtwid0.07$, the Fermi
surface topology changes to one in which there is one electron and one
hole-like Fermi surface as shown for $t_\perp/t=0.5$ and 2.0 in Fig.~\ref{fig:bilayer}
(b) and (c), respectively.
This Fermi surface structure is a simplified version of the multi-Fermi surfaces found from
bandstructure calculations for the Fe-based superconductors shown in Fig.~\ref{fig:17a}a.

This model, originally studied using determinant quantum Monte Carlo
\citep{ref:BulutScaSca,ref:Scalettar1994,ref:Hetzel,ref:Bouadim}
has also been studied using fluctuation exchange (FLEX) \citep{ref:Kuroki2002},
phenomenological spin fluctuation approximations \citep{ref:Liechtenstein}, 
FRG \citep{ref:Zhai} and DCA \citep{ref:MaierScal1107.0401} methods. One finds
that for $t_\perp/t$ less
than of order one, the most divergent pairfield correlations occur in the
$d_{x^2-y^2}$ channel while for $t_\perp/t$ larger they occur in an $A_{1g}$
channel in which the gap has one sign on the antibonding Fermi surface and the
opposite sign on the bonding Fermi surface, as schematically illustrated in
Fig.~\ref{fig:bilayer}. This gap, which changes sign
between the two Fermi surfaces, is an $s^\pm$-like gap.

At half-filling, determinental Quantum Monte Carlo (DQMC) calculations showed that the
ground state for $U=6$ had AF long-range order for $t_\perp/t\ltwid2$. For
larger values of $t_\perp/t$, the system enters a disordered valence bond phase
with singlet correlations between electrons on opposite sites of the two layers.
In the doped system, there is a cross-over in which the intra-layer AF fluctuations
decrease and the inter-layer spin fluctuations increase as $t_\perp/t$ is initially
increased. Then at still larger values of $t_\perp/t$ the low energy interlayer
spin fluctuations become gapped and the superconducting pairing is suppressed.

\begin{figure}[!htbp]
\includegraphics[height=10cm]{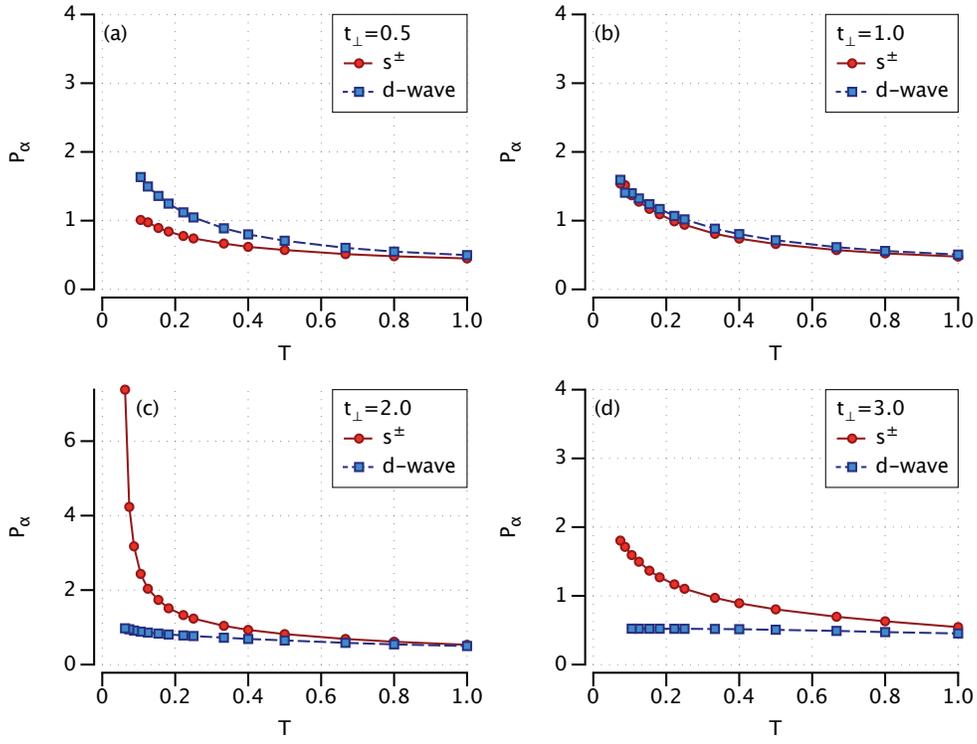}
\caption{The $d_{x^2-y^2}$ and $s^\pm$ pairfield susceptibilities $P_\alpha$
versus temperature $T$ for various values of the inter-layer hopping $t_\perp$.
These DCA results were for a $(4\times4)\times2$ cluster and we have set the
intra-layer hopping $t=1$. One sees that as $t_\perp/t$ increases there is a
crossover from $d_{x^2-y^2}$ pairing to $s^\pm$ pairing
(after Maier and Scalapino\protect\cite{ref:MaierScal1107.0401}).
\label{fig:palpha-vs-t}}
\end{figure}

For the two-layer system, the two pairfield susceptibilities that are of interest
are given by
\begin{equation}
  P_\alpha(T)=\int^\beta_0d\tau\langle\Delta_\alpha(T)\Delta^+_\alpha(0)\rangle
	\label{eq:new14}
\end{equation}
with
\begin{equation}
  \Delta_{x^2-y^2}=\frac{1}{\sqrt N}\sum_k(\cos k_x-\cos k_y)c^+_{k\uparrow}c^+_{-k\downarrow}
	\label{eq:new15}
\end{equation}
and
\begin{equation}
  \Delta_{s^\pm}=\frac{1}{\sqrt N}\sum_k\cos k_zc^+_{k\uparrow}c^+_{-k\downarrow}.
	\label{eq:new16}
\end{equation}
Here for the two-layer model, $k_z=0$ (bonding) and $k_z=\pi$ (antibonding).
For $U=6$ and $\langle n\rangle=0.95$, Fig.~\ref{fig:palpha-vs-t} shows DCA
results for $P_\alpha(T)$ for both the $d_{x^2-y^2}$ case and the $s^\pm$ case.
For $t_\perp/t=0.5$ where there are strong AF planar spin-fluctuations, the
dominant pairing occurs in the $d_{x^2-y^2}$ channel. However, as $t_\perp/t$
increases, the $s^\pm$ response increases and for $t_\perp/t\gtwid1$, it becomes
dominant with the response peaking for $t_\perp/t\approxeq2$. At half-filling with
$U/t=6$, DQMC calculations \citep{ref:Bouadim} find a QCP for $t_\perp/t\approx2$
which separates an antiferromagnetic phase from a valence bond phase \citep{ref:Sachdev3}.
Finally, for $t_\perp/t=3$ one finds that the pairing becomes weaker
as the inter-layer valence bonds become stronger.

\begin{figure}[!htbp]
\includegraphics[height=8cm]{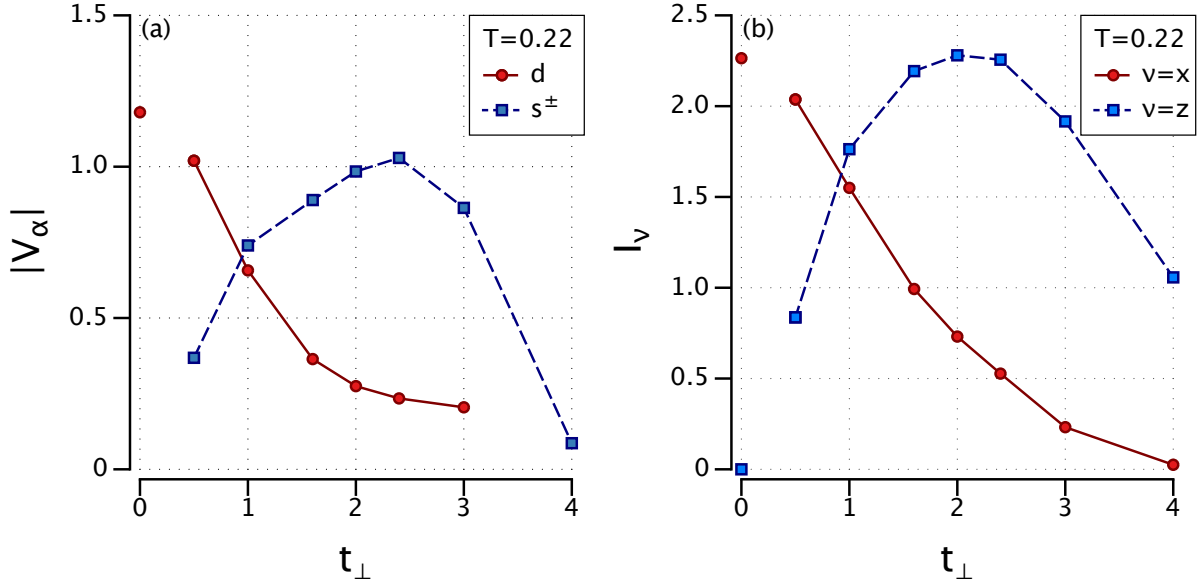}
\caption{The pairing interaction strength in the $d$ and $s^\pm$ channels
reflects the spatial structure of the local spin fluctuations.
(a) The strength of the pairing interactions $V_\alpha$ and (b) the
integrated spectral weights $I_\nu$ versus $t_\perp$ for $k_\nu=k_x$ and
$k_z$. The strength $|V_\alpha|$ of the pairing interaction for $d_{x^2-y^2}$
pairing is correlated with the intra-layer near neighbor spin fluctuation spectral weight,
while the $s^\pm$ pairing strength reflects that of the inter-layer spin fluctuations
(after Maier and Scalapino\protect\cite{ref:MaierScal1107.0401}).\label{fig:valph-ialph-vs-tperp}}
\end{figure}

Just as the pairing interaction $\Gamma^{PP}(k,k')$ was analyzed for the
single layer Hubbard model, one can examine how the bilayer pairing interaction
is related to the underlying spin correlations of the system. A
useful measure of the strength of the pairing interaction for a given channel is
$|V_\alpha|$ given by Eq.~(\ref{eq:b}).
Results for $|V_\alpha|$ versus $t_\perp$ for $\alpha=d_{x^2-y^2}$ and $s^\pm$ are
shown in Fig.~\ref{fig:valph-ialph-vs-tperp}.
Also plotted in this figure are the integrated spectral weights for the intra- and
inter-layer near-neighbor spin fluctuations
\begin{equation}
  I_\nu=\frac{1}{N}\sum_k\int\frac{d\omega}{\pi}\frac{{\rm Im}\chi(k,\omega)}{\omega}
  \cos k_\nu=\frac{1}{N}\sum_k{\rm Re}\chi(k,0)\cos k_\nu\label{eq:new18}
\end{equation}
with $k_\nu=k_x$ and $k_z$ for the intra- and inter-layer spin-fluctuation
weights, respectively. In Fig.~\ref{fig:valph-ialph-vs-tperp}, one sees that
the $d_{x^2-y^2}$ pairing strength
is correlated with the near-neighbor planar spin fluctuations while the
$s^\pm$ pairing strength reflects the inter-layer spin fluctuation strength.

The bilayer Hubbard model is clearly simpler than the five-orbital Fe models.
However, it has the advantage that one can carry out numerical calculations
and examine the relationship between the pairfield structure, the pairing
interaction strengths and the spin correlations. The fact that one
can change a one-electron hopping parameter $t_\perp$ and observe that the system evolves
from a $d_{x^2-y^2}$ to an $s^\pm$ pairing phase provides further
evidence supporting the notion of a commonality between the cuprate and
Fe-based superconductors.

A similar relationship between $d$-wave and $s^\pm$ pairing is seen in
density matrix renormalization (DMRG) studies of a two-leg ladder \citep{ref:Berg09}.
In this case, the DMRG method has been used to study a caricature of the Fe-pnictide
problem which focuses on the $d_{xz}$ orbital pair scattering process associated
with the $k_y=0$ and $k_y=\pi$ states near the $\alpha_1$ and $\beta_2$ Fermi
surfaces shown in Fig.~\ref{fig:17a}a. These scattering processes can be
described by the Hamiltonian for a 2-leg ladder
\begin{figure}[!htbp]
\includegraphics[width=14cm]{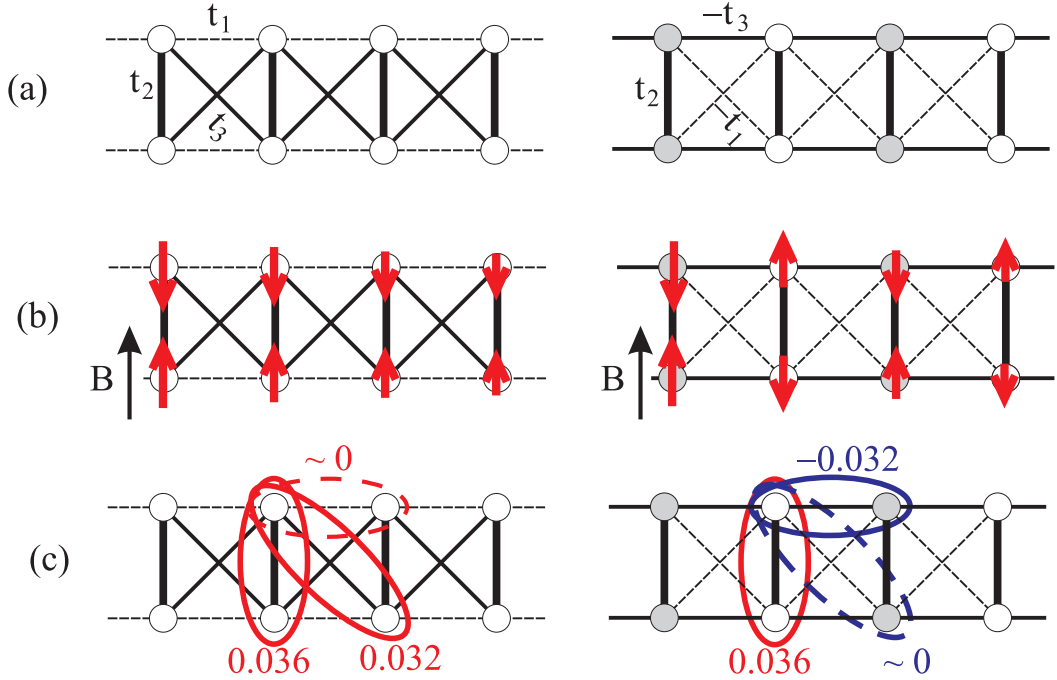}
\caption{An ``Fe-ladder" is simply a unitary transformation of a ``Cu-ladder."
The left hand side shows (a) a 2-leg Fe ladder, (b) schematic illustration of the
spin structure $\langle S^z(\ell_x,\ell_y)\rangle$ induced by applying an external
magnetic field to the lower left hand site, and (c) the singlet pairfield
$\langle\Delta_{ij}\rangle$ induced at a distance 10 sites removed from the end
of a $32\times2$ ladder with a unit external pairfield applied to the end
rung. On the right hand side, every other rung has been twisted by $180^\circ$
and the phase of the orbitals denoted by the open circles have been changed
by $\pi$. As discussed in the text, this ``twisted Fe-ladder" corresponds to
the well-studied cuprate ladder (after Berg et al.\protect\cite{ref:Berg09}).\label{fig:28}}
\end{figure}
\begin{eqnarray}
  H=-t_1\sum_{i\ell\sigma}c^+_{i\ell\sigma}c_{i+1\ell\sigma}&-&2t_2\displaystyle{\sum_{i\sigma}}c^+_{i1\sigma}c_{i2\sigma}\nonumber \\
	&-&2t_3\displaystyle{\sum_{i\sigma}}(c^+_{i1\sigma}c_{i+12\sigma}+c^+_{i+12\sigma}c_{i1\sigma})+U\displaystyle{\sum_{i\ell\sigma}}
	n_{i\ell\uparrow}n_{i\ell\downarrow}
  \label{eq:2leg-ham}
\end{eqnarray}
with the tight binding parameters illustrated on the left-hand side of Fig.~\ref{fig:28}a.
Here, $\ell=1,2$ is the leg index, there are leg $t_1$, rung $t_2$ and diagonal
$t_3$ one-electron hopping matrix elements and an on-site Coulomb interaction
$U$. The factors of 2 in front of $t_2$ and $t_3$ takes into account the
periodic boundary conditions which have been used in the transverse direction.
As discussed in \citet{ref:Berg09}, the hopping parameters $t_1=-0.32$ and
$t_3=-0.57$ measured in units of $t_2=1$, were taken to fit the Fe-pnictide (1111) DFT band
structure near the $\alpha_1$ and $\beta_2$ Fermi surfaces for $k_x$ cuts
through $k_y=0$ and $k_y=\pi$, respectively. As seen in Fig.~\ref{fig:17a}a,
at these points the Bloch wave functions have $d_{xz}$ character.

With $U=3$, DMRG calculations for the half-filled case with an external magnetic
field applied to the first site of the lower leg gave the spin pattern shown on
the left-hand side of Fig.~\ref{fig:28}b. This spin pattern has a striped-like
SDW structure similar to the magnetic structure seen in the Fe-pnictides. The 2-leg
system was found to have a spin gap $\Delta_s=0.14$ corresponding to a spin
correlation length of approximately four sites. For the doped system with
$\langle n\rangle=0.94$, a pairfield boundary term
\[H_1=\Delta_1(P^+_1+{\rm h.c.})\]
with $\Delta_1=1$ and
\[P^+_1=(d^+_{11\uparrow}d^+_{12\downarrow}-d^+_{11\downarrow}d^+_{12\uparrow})\]
was added. This term acts as a proximity coupling to the rung at the left-hand
end of the ladder. Then the expectation values of the resulting induced singlet
pairfield was measured on the rung as well as the diagonal and the leg near
neighbor sites at positions further down the ladder. The values of this
induced pairfield 10 sites away from site $\ell=1$ are shown on
the left-hand side of Fig.~\ref{fig:28}c.

This result is directly related to the 2-leg ladder cuprate model shown on the
right-hand side of Fig.~\ref{fig:28}. Here, every other rung of the left-hand
ladder has been twisted by 180$^\circ$ and the phase of the $d_{xz}$-orbit has
been changed by $\pi$ on each of the open sites of the twisted rungs. In this
way, the rung hopping matrix element remains $t_2$, but the leg and diagonal
hoppings are changed to $-t_3$ and $-t_1$, respectively. Then with the parameters
that have been used, the dominant hoppings on this ``twisted Fe-ladder" are along
the legs and rungs with only a weak diagonal hopping. These are typical parameters
for a cuprate ladder. Furthermore, as shown on the right-hand side of
Fig.~\ref{fig:28}b and c, the resulting spin and pairfield correlations
of the original Fe ladder have turned into the spin gapped $(\pi,\pi)$
antiferromagnetic and the familiar $d$-wave like pairing correlations \citep{ref:Noack}.
Thus, similar to the 2-layer Hubbard model, the 2-leg ladder illustrates
the close connection that exists between the cuprates and the Fe-based materials.

\subsection{Multi-orbital models}

\begin{figure}[!htbp]
\includegraphics[height=7cm]{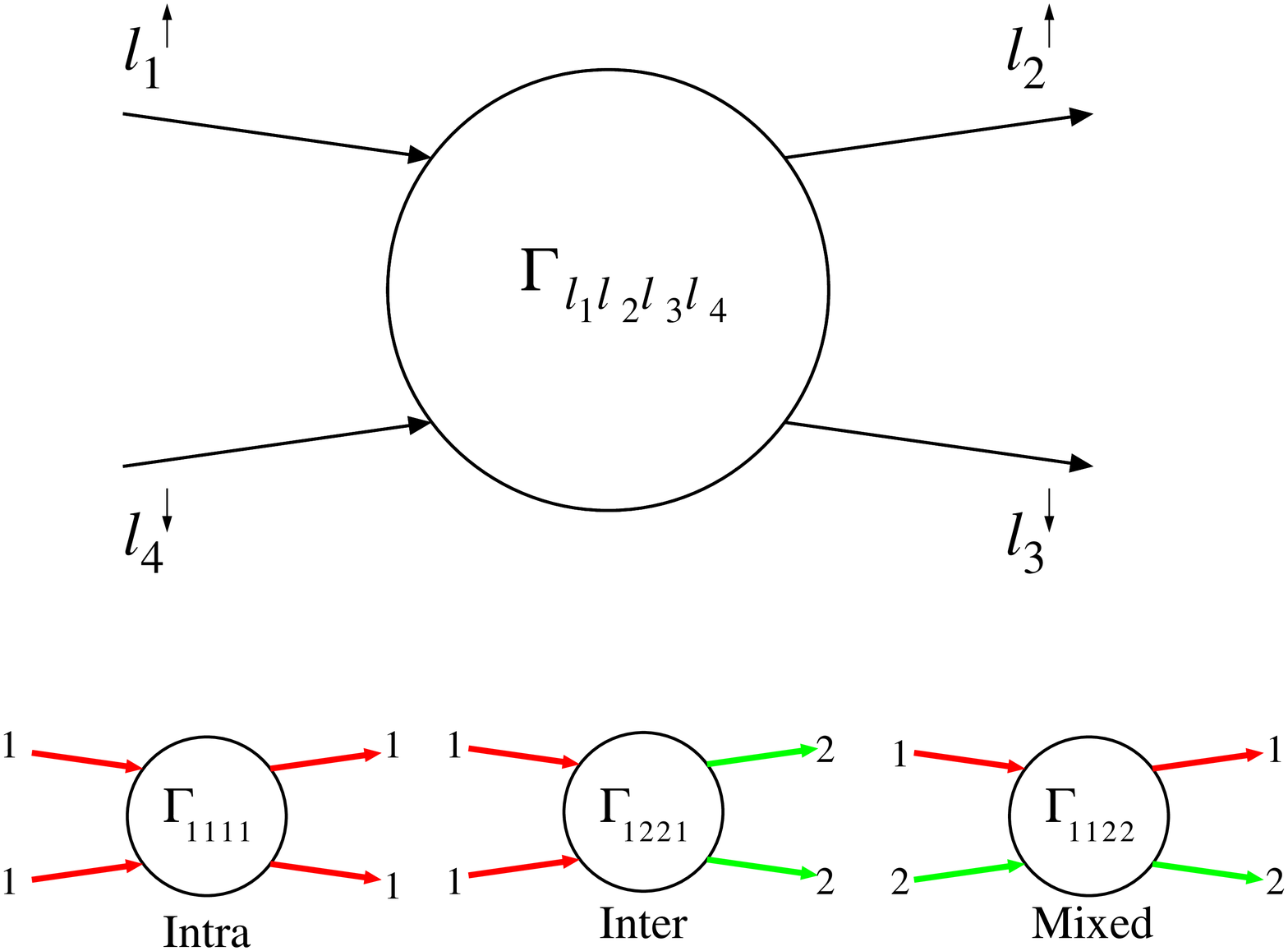}
\caption{The orbital dependent pairing interaction
$\Gamma^{(k,k')}_{\ell_1\ell_2\ell_3\ell_4}$ defined in
terms of orbital states $\ell_i$ of incoming and outgoing electrons. The lower
diagrams illustrate intra-orbital, inter-orbital and mixed orbital scattering
processes (after Kemper et al.\protect\cite{ref:Kemper}).
\label{fig:20}}
\end{figure}

In general, for the multi-orbital models, the orbital structure of the pairing
interaction is important and one introduces an orbital dependent pairing interaction
$\Gamma_{\ell_1\ell_2\ell_3\ell_4}$ illustrated in Fig.~\ref{fig:20}, which describes
the irreducible particle-particle scattering of electrons in orbitals $\ell_1,\ell_4$
with momentum $k$ and $-k$ into orbitals $\ell_2,\ell_3$ with momentum $k'$ and
$-k'$. In terms of this vertex, the effective pairing
interaction for scattering a $(k'\uparrow,-k'\downarrow)$ pair on the $\nu_j$
Fermi surface to a $(k\uparrow,-k\downarrow)$ pair on the $\nu_i$ Fermi surface is
\begin{equation}
  \Gamma_{ij}(k,k')=\sum_{\ell_1\ell_2\ell_3\ell_4}a^{{\ell_2}^*}_{\nu_i}(k)
	a^{{\ell_3}^*}_{\nu_i}(-k)\Gamma_{\ell_1\ell_2\ell_3\ell_4}(k,k')
	a^{\ell_1}_{\nu_j}(k')a^{\ell_4}_{\nu_j}(-k')
	\label{eq:13}
\end{equation}
with $a^{\ell_1}_{\nu_j}(k)$ the orbital matrix element $\langle\nu_jk|\ell_1\rangle$
given in Eq.~(\ref{eq:9a}).

Besides the numerical calculations for the two-layer (effective two-obital)
Hubbard model discussed above, there have been some quantum Monte Carlo
\citep{ref:White1989,ref:Dopf} and cluster studies \citep{ref:Hanke} for the
three-orbital CuO$_2$ model. These calculations show that the undoped state is
a charge-transfer anti-ferromagnetic insulator rather than a Mott-Hubbard
anti-ferromagnetic insulator. However, the anti-ferromagnetic and
$d_{x^2-y^2}$-pairing correlations in the doped state of these models are
remarkably similar to those found for the doped single band Hubbard model.

The main studies of the multiple-orbital models which have been carried out for
the heavy fermion and Fe-based materials have been based upon weak coupling
random phase (RPA) \citep{ref:Kuroki2008,ref:Graser2009,ref:Kuroki,ref:Chubukov},
fluctuation-exchange (FLEX) \citep{ref:IAK} or functional renormalization
group (FRG) methods \citep{ref:Wang,ref:Platt,ref:Zhai,ref:Uebelacker}.
Just as the Monte Carlo calculations \citep{ref:Maier1} of the four-point vertex
allow one to study the interplay of the various spin, charge and pairing
correlations on an equal footing as the temperature is reduced (see for example
Fig.~\ref{fig:17}), the FRG provides an unbiased approach for monitoring the
strength of the various scattering processes as an energy cutoff is reduced.
Of course the FRG calculations are typically one-loop approximations, suitable
for weaker coupled systems. Nevertheless, the FRG calculations for the multi-band
Hubbard models find that spin-density-wave (SDW) scattering processes grow in
strength as the renormalization energy cutoff is reduced, driving an increase
in the pair scattering strength. In addition, just as for the single-band Hubbard
model, strong SDW fluctuations also drive other pairing, Pomeranchuk and CDW
channels.
The same electrons are involved in both
the spin-fluctuation and these channels.

In the RPA and FLEX approaches, the orbital dependent vertex is approximated by
\begin{eqnarray}
  \Gamma_{\ell_1\ell_2\ell_3\ell_4}({\bf k},{\bf k}',\omega)&=& \biggl[\frac{3}{2}
	U^S\chi^{\rm RPA}_1({\bf k}-{\bf k}',\omega)\ U^S-\nonumber \\
	& &\phantom{\biggl[}\frac{1}{2}\ U^C\chi^{\rm RPA}_0({\bf k}-{\bf k}',\omega)\ U^C+\frac{1}{2}
	\ (U^S+U^C)\biggr]_{\ell_3\ell_4\ell_1\ell_2},
	\label{eq:14}
\end{eqnarray}
with
\begin{equation}
  \chi^{\rm RPA}_1(q)=\chi^0(q)[1-U^S\chi^0(q)]^{-1}
  \label{eq:15}
\end{equation}
and
\begin{equation}
  \chi^{\rm RPA}_0(q)=\chi^0(q)[1+U^C\chi^0(q)]^{-1}
  \label{eq:16}
\end{equation}
Here the quantities $U^S$, $U^C$, and the one-loop susceptibility
$\chi^0$ are represented by matrices in the orbital space. Details of this can be found
in the literature \citep{ref:Takimoto1}. Here we note that the basic structure of the pairing
interaction is similar to Eq.~(\ref{eq:9}) with
\begin{equation}
  \Lambda_{\rm irr}\sim\frac{1}{2}(U^S+U^C)\hskip 5mm
	\phi_m\sim U^S\chi^{\rm RPA}_1U^S\hskip 5mm
	\Phi_d=-U^C\chi^{\rm RPA}_0U^C.
  \label{eq:17}
\end{equation}
While this represents a weak coupling approximation, we know from numerical
studies \citep{ref:MJS} of the
single band Hubbard model that by treating the interaction parameters
phenomenologically, RPA and FLEX approximations can provide reasonable
descriptions of the pairing interaction for intermediate coupling.

From Eq.~(\ref{eq:13}) one sees that the effective pairing interaction
$\Gamma_{ij}(k,k')$ for a multi-orbital system depends upon the number of
Fermi surfaces and their shapes as well as the orbital matrix elements.
In general, these matrix elements act to suppress the mixed pair orbital vertex
contributions in which $\ell_1\ne\ell_4$ and $\ell_2\ne\ell_3$ (lower right hand
diagram shown in Fig.~\ref{fig:20}). For spin rotational interaction parameters
the dominant contributions to the pairing interaction $\Gamma_{ij}(k,k')$ comes
from intra-orbital ($\ell_1=\ell_2=\ell_3=\ell_4$) scattering processes with
weaker contributions from the inter-orbital processes ($\ell_1=\ell_4\ne\ell_2=\ell_3$).
The number, the shape and the location of the various Fermi surfaces also play
a key role in determining the strength of the pairing interaction and the
structure of the gap $\Delta(k)$.

\begin{figure}[!htbp]
\includegraphics[width=16cm]{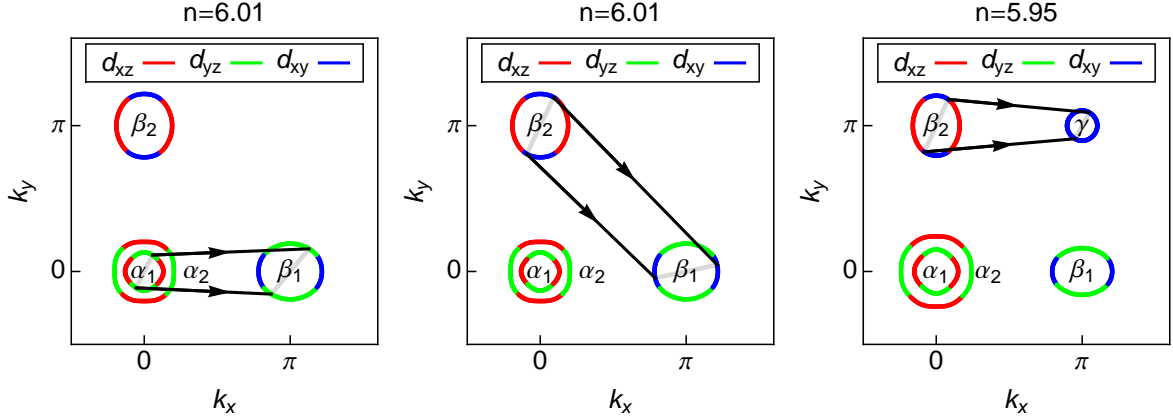}
\caption{For a filling $n=6.01$, the scattering of a pair from the $\alpha_1$
hole Fermi surface to the $\beta_1$ electron Fermi surface shown in the left
hand figure favors pairing in which there is a sign change of the gap between
$\alpha_1$ and $\beta_1$. A similar pair scattering process between $\alpha_1$
and $\beta_2$ leads to a gap which has the same sign on $\beta_1$ and $\beta_2$.
However, the $\beta_2$--$\beta_1$ pair scattering shown in the middle figure
tends to frustrate this, since they favor a gap which has opposite signs on the
$\beta_2$ and $\beta_1$ Fermi surfaces. As shown in the right hand figure, for a
filling $\langle n\rangle=5.95$, an additional hole pocket $\gamma$ appears and
$\beta_2$--$\gamma$, as well as $\beta_1$--$\gamma$, pair scattering processes
stabilize the $s^\pm$ gap.\label{fig:21}}
\end{figure}

\begin{figure}[!htbp]
\includegraphics[width=15cm]{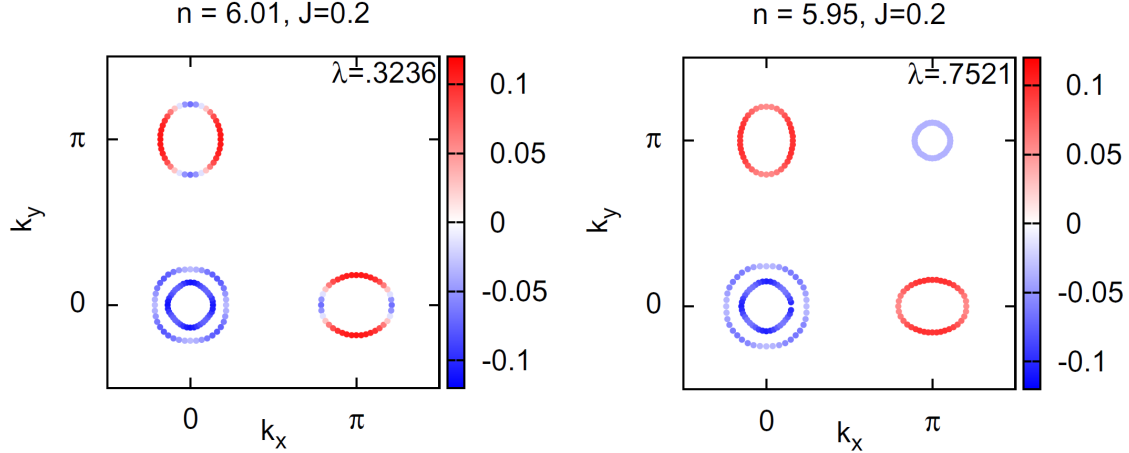}
\caption{The gap eigenfunctions $g(k)$ for a spin rotationally invariant
parameter set $\bar U=1.3$, $\bar U'=0.9$, $\bar J=\bar J'=0.2$, for dopings
$n=6.01$ (left) and $n=5.95$ (right). Here, one sees how the $s^\pm$ gap is
stabilized by the $\beta_1$--$\gamma$ and $\beta_2$--$\gamma$ pair scattering
processes shown in the right hand portion of Fig.~\protect\ref{fig:21}
(after Kemper et al.\protect\cite{ref:Kemper}).\label{fig:22}}
\end{figure}

\begin{figure}[!htbp]
\vskip 5mm
\includegraphics[width=16cm]{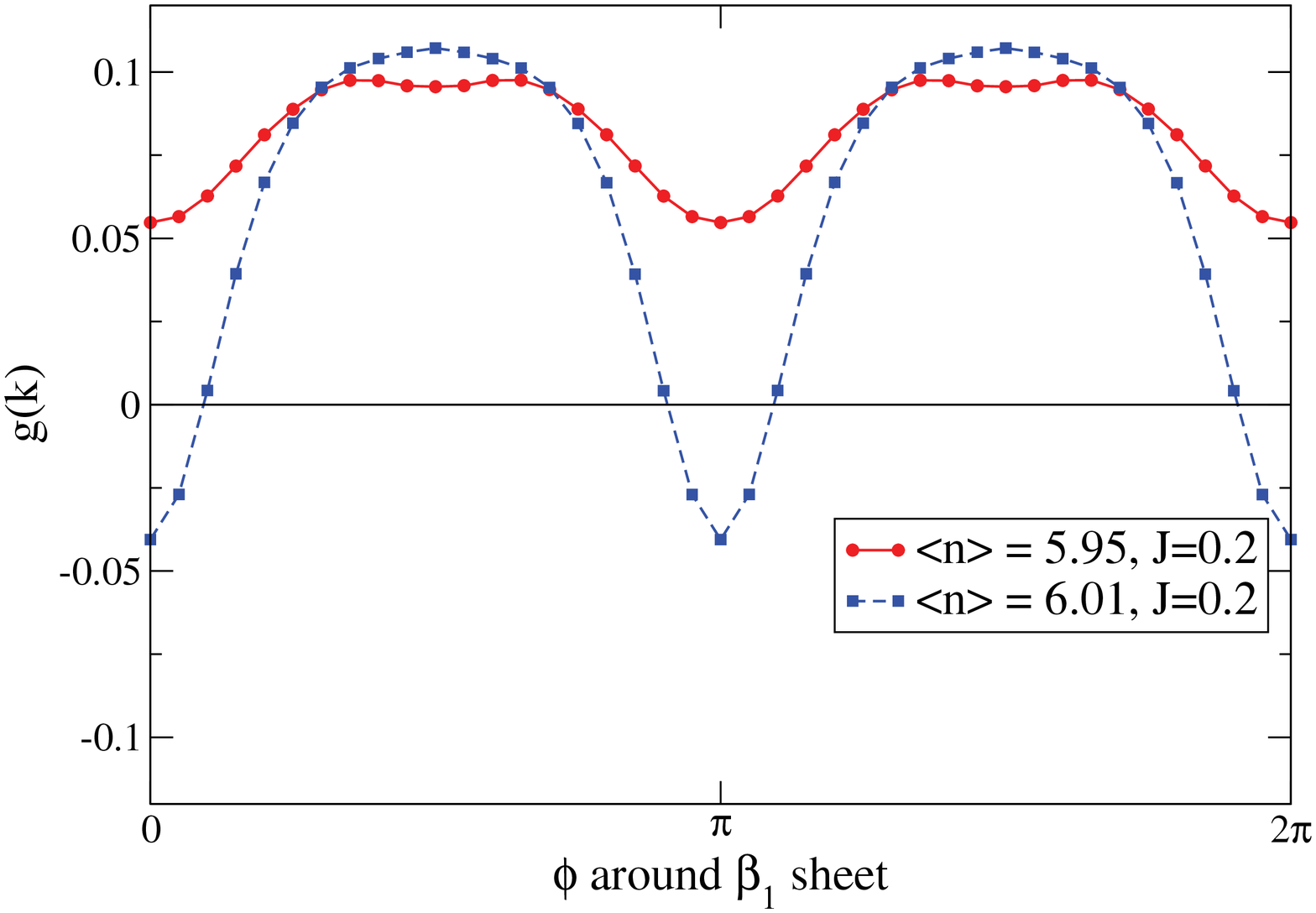}
\caption{The gap function $g(k)$ on the $\beta_1$ pocket for $n=5.95$ solid
(red) and $n=6.01$ dashed (blue) from Fig.~\protect\ref{fig:22}. Here the angle
$\phi$ is measured from the $k_x$-axis (after Kemper et al.\protect\cite{ref:Kemper}).\label{fig:23}}
\end{figure}

As noted by \citet{ref:Kuroki} for the 1111 Fe material, depending upon the height of
the pnictide and the doping, an additional hole Fermi surface with $d_{xy}$
orbital character may be present around the $(\pi,\pi)$ point of the unfolded
Brillouin zone. Figure~\ref{fig:21} shows the Fermi surfaces at two different
fillings for a tight binding parameterization of the 1111 Fe material.
In this
case, for a filling $\langle n\rangle=6.01$, there are two hole Fermi surfaces
around the $\Gamma$ point and two electron Fermi surfaces around $(\pi,0)$ and
$(0,\pi)$ in the unfolded 1 Fe/cell Brillouin zone. However, for the hole doped
system with $\langle n\rangle=5.95$, an additional hole Fermi surface appears
around the $(\pi,\pi)$ point. The dominant orbital weight along the Fermi
surfaces are also indicated along with various intra-orbital pair scattering
processes. The lefthand figure shows a pair scattering from the $\alpha_1$ hole
Fermi surface around the $\Gamma$ point to a pair on the electron Fermi
surface $\beta_1$ centered at $(\pi,0)$. Here, electrons in states $k$ and $-k$
on the $\alpha_1$ Fermi surface are scattered to states $k'$ and $-k'$ on the
$\beta_1$ Fermi surface. This process is illustrated in Fig.~\ref{fig:21}
using an extended Brillouin zone in which $-k'$ is replaced by $-k'+(2\pi,0)$.
The orbital weight on both Fermi
surfaces is dominantly $d_{yz}(\ell=2)$ over the regions in which there is a
reasonable nesting giving rise to a peak in $\Gamma_{2222}$ for a momentum
transfer $q\sim(\pi,0)$. There are similar intra-orbital $d_{xz}$ scattering
processes between $\alpha_1$ and the electron $\beta_2$ Fermi surface which
give rise to a peak in $\Gamma_{1111}$ for $q\sim(0,\pi)$. These processes
lead to a $\Gamma_{ij}(k,k')$ interaction which favors an $A_{1g}$ $s^\pm$ gap
which switches sign between the $\alpha_1$ and the $(\beta_1,\beta_2)$ Fermi
surface. However, as shown in the middle diagram of Fig.~\ref{fig:21}, there
are inter-orbital $d_{xz}$--$d_{xy}$ pair scattering processes between $\beta_2$ and
$\beta_1$. These act to frustrate a uniform $s^\pm$ state. This same behavior
is seen in the FRNG calculations \citep{ref:Zhai,ref:Thomale}. In addition, unless
the Fermi surface areas weighted by $v^{-1}_F(k)$ are such that the electron
and hole regions exactly balance, the short range Coulomb interaction can be
reduced by an anisotropic $A_{1g}$ gap. As a consequence, for a filling
$\langle n\rangle=6.01$ and a typical set of interaction parameters, one finds
the $A_{1g}$ gap structure shown on the left of Fig.~\ref{fig:22} and as the
blue curve in Fig.~\ref{fig:23}. Here the gap has nodes on the $\beta$ electron
Fermi surfaces.
The possibility of such accidental nodes in the $A_{1g}$ state is consistent
with the linear low temperature $T$ dependence seen in the penetration depth of
LaFePO \citep{ref:Hicks}.

The gap $\Delta(k)$ for $\langle n\rangle=6.01$ and $\langle n\rangle=5.95$ is
shown in Fig.~\ref{fig:22}. For $\langle n\rangle=6.01$ the $(\pi,\pi)$ Fermi
surface is absent while for a doping $\langle n\rangle=5.95$,
there is an additional hole Fermi surface around the $(\pi,\pi)$ point of
the 1 Fe/cell Brillouin zone. In this latter case, intra-orbital $d_{xy}$ pair
scattering processes like the one shown in the right panel of Fig.~\ref{fig:21}
favor a more uniform $s^\pm A_{1g}$ state and as shown on the right hand side of
Fig.~\ref{fig:22} and the red curve in Fig.~\ref{fig:23}, the nodes on the
$\beta$ Fermi surfaces are lifted. In addition the overall pairing strength is
larger when the extra hole $(\pi,\pi)$ Fermi surface is present. In similar
calculations in which the bandstructure parameters were changed so that the
$(\pi,\pi)$ Fermi surface had dominant $d_{3z^2-r^2}$ weight, the nodes of the
gap were not lifted. Thus the orbital weights as well as the Fermi surface
topology play an important role in determining the gap structure as well as
$T_c$ \citep{ref:Kuroki,ref:Kemper,ref:Platt2011,ref:Thomale,ref:Uebelacker}.

\section{Summary and Outlook}
\label{sec:5}
Here it has been proposed that the interaction which is responsible for pairing
in some families of heavy fermion materials, the 115 Pu actinides, the high $T_c$
cuprates and the Fe-based superconductors arises from
the exchange of spin-fluctuations. Just as different materials ranging from Hg
and Pb to Nb$_3$Sn and MgB$_2$ have a phonon mediated pairing interaction, the
suggestion is that this class of unconventional superconducting materials,
though clearly different from each other, share a common pairing mechanism.
As noted earlier, one should also include the organic Bechgaard salts
\citep{ref:Bechgaard,ref:Taill28,ref:Taillefer,ref:Doiron} in this group.\footnote{Spin-fluctuations
are also believed to give rise to pairing in Sr$_2$RuO$_4$ \protect\citep{ref:Maeno}.
Here, \protect\citet{ref:Rice} have proposed that the pairing
is associated with small momentum transfer ferromagnetic fluctuations while
\protect\citet{ref:RaghuKK} have suggested that the pairing is
driven by large momentum spin-fluctuations associated with the quasi-1D band
structure of Sr$_2$RuO$_4$. If the latter mechanism is correct, one would group
Sr$_2$RuO$_4$ with the class of superconductors discussed in this review.} Looking back
with this perspective, one would say that this class of antiferromagnetic
spin-fluctuation mediated
superconductors began with the seminal discoveries of superconductivity in the
heavy fermion material CeCu$_2$Si$_2$ by \citet{ref:Steglich} and
in the organic material (TMTSF)$_2$PF$_6$ by \citet{ref:Jerome}.

Theoretical proposals that spin-fluctuations near a spin-density-wave instability
could give rise to unconventional pairing in some organic Bechgaard salts and
some heavy fermion materials were made in 1986 (\citet{ref:Emery}, \citet{ref:Cyrot}, \citet{ref:Miyake}, \citet{ref:Scal-Loh-Hirsch}).
Then, following the discovery of the cuprate superconductors various suggestions
were made to also include the cuprates in this
group \citep{ref:ScalRep,ref:Moriya03,ref:MonthouxNature}. However, while the
antiferromagnetism and $d$-wave superconductivity appeared in close proximity
in the phase diagrams of the electron doped cuprates, in the hole doped cuprates,
a pseudogap phase appears adjacent to the superconducting phase. Furthermore,
the undoped cuprates are antiferromagnetic charge-transfer Mott insulators.
Thus there were arguments made that superconductivity in the high $T_c$
cuprates arose from a different underlying mechanism, and that it was
inappropriate to speak of a spin-fluctuation pairing glue \citep{ref:DJS}. Now
the question of whether there is a pairing glue is basically a question
regarding the dynamics of the pairing interaction \citep{ref:ScalSciMag}.
As discussed in Sec.~\ref{sec:4}, numerical calculations of the pairing
interaction for the Hubbard model provide evidence which supports the view that
its dynamics dominantly reflects that of the dynamic spin susceptibility
\citep{ref:Maier2008,ref:Kyung,ref:Hanke}. Thus there is pairing
glue in the Hubbard models and the question becomes ``Should one speak of a
spin-fluctuation pairing glue for this class of real materials?"

The discovery of the Fe-based superconductors \citep{ref:LaOFeP,ref:LaFeAsO}
provided renewed support for the idea that indeed there exists a class of
materials in which superconductivity does not arise from the traditional phonon
exchange mechanism \citep{ref:NormanScience}. In addition, as noted in
Sec.~\ref{sec:2}, a variety of measurements show that antiferromagnetic
spin-density-wave-like fluctuations are ubiquitous in these materials and are
the primary excitations which scatter the electrons. Now in principle, one
would like to determine the $k$ and $\omega$ dependence of the normal and
anomalous (gap) self-energies and from these infer the structure and origin of
the pairing interaction. In particular, the $k$-dependence of the gap on the
multi-Fermi surfaces of the Fe-based superconductors can provide a more detailed
probe of the $k$-dependence of the pairing interaction providing a test of
different pairing mechanisms. For example, the spin-fluctuation theory finds
that there can be a near-degeneracy between an anisotropic sign-changing $s$-wave
($A_{1g}$) state and a $d_{x^2-y^2}$ ($B_{1g}$) state due to the near nesting of
Fermi surface sheets \citep{ref:Graser2009,ref:Thomale2}. This is also clearly
seen in the DCA results for the bilayer model discussed in Sec.~\ref{sec:4}.
Thus the $k$-dependence of the gap on the multi-Fermi surfaces of the
Fe-based superconductors can provide a test of the theory.
In addition, as dicussed in Sec.~\ref{sec:4},
there are a number of experiments which are exploring the $\omega$ dependence of
the gap. The recent progress in material quality,
the increase in the frequency and momentum resolution of ARPES, neutron
scattering and RIXS, along with tunneling and STMS hold the promise of providing
the kind of detailed information that will be needed.
There will also be support for these ideas if they
can provide guidance in the search for new and possibly higher $T_c$
superconductors. This review concludes by summarizing some of the ideas which have
been discussed that may help in this search.

The numerical calculations 
for the doped single band Hubbard model with a near neighbor hopping $t$ and an
onsite Coulomb interaction $U$, show that $T_c$ is maximized for $U$ of order 
the bandwidth $8t$. As $U$ increases beyond the bandwidth, the characteristic
energy of the spin fluctuations is suppressed and $T_c$ decreases. In addition,
$T_c$ is found to decrease in the underdoped regime. Here, the superfluid
stiffness tends to zero as the Mott state is approached \citep{ref:EmKiv}. In
addition, there is the reduction of the quasi-particle weight due to the Mott
correlations which suppress the intrinsic pairfield
susceptibility $P_{0d}$. Thus optimal superconductivity is obtained by doping
the single layer Hubbard model away from half-filling. In the doped bilayer
case, $T_c$ is enhanced when $t_\perp/t$ is increased and in this way the
system is again moved away from the Mott regime to a semi-metallic state. Thus
optimal superconductivity in these models is expected to be found at
intermediate coupling away from the Mott regime. In this regime, the
fluctuation-exchange (FLEX) approximation \citep{ref:Bickers} gives results in
reasonable agreement with the numerical calculations and it has been used to
address further issues.

\begin{figure}[!htbp]
\includegraphics[height=8.5cm]{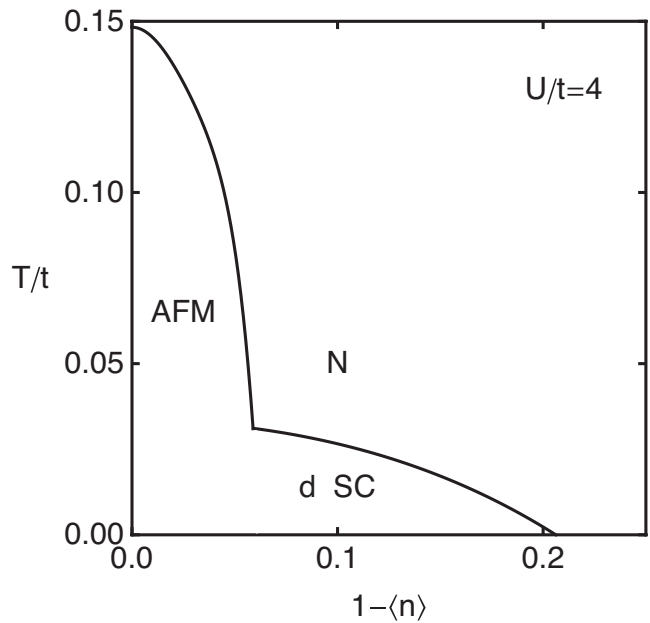}
\caption{Phase diagram for a two-dimensional
Hubbard model with $U/t=4$ calculated within the fluctuation-exchange
approximation. As the system is doped away from half-filling, the N\'eel
temperature is suppressed and a $d_{x^2-y^2}$ superconducting phase appears
(after Bickers et al.\protect\cite{ref:Bickers}).\label{fig:16a}}
\end{figure}

The phase diagram obtained for a two-dimensional Hubbard model with $U/t=4$
using FLEX is shown in Fig.~\ref{fig:16a}. Here one sees that as the system is
doped, the SDW antiferromagnetic phase is suppressed and $d_{x^2-y^2}$-wave
superconductivity appears. As discussed by \citet{ref:Vorontsov}
and \citet{ref:Fernandes} there can be a coexistence region near
the intersection of the antiferromagnetic and superconducting transitions.
As the doping increases, in the absence of the superconducting transition,
the antiferromagnetic transition is suppressed towards $T=0$ giving rise to a
quantum critical point (QCP) \citep{ref:Si,ref:Sachdev3}. The shape of the phase
boundaries as well as the temperature dependence of the transport properties
reflect the antiferromagnetic spin fluctuations associated with the QCP
\citep{ref:Daou2009,ref:Abanov,ref:Metlitski}. The precise role of the QCP
remains under study. Within the framework of FLEX calculations, the characteristic
antiferromagnetic energy at zero doping $T_N(x=0)$ is large compared with $T_c$.
In this case, to optimize $T_c$ one changes the doping $x$ so as to reduce the
frequency of the antiferromagnetic fluctuations to some multiple of $T_c$ in
order to optimize the pairing. Since $T_N\gg T_c$, this means that one will
indeed have to tune the doping $x$ close to the critical concentration $x_c$
where $T_N(x_c)$ would vanish in the absence of superconductivity.

\begin{figure}[!htbp]
\includegraphics[height=12cm]{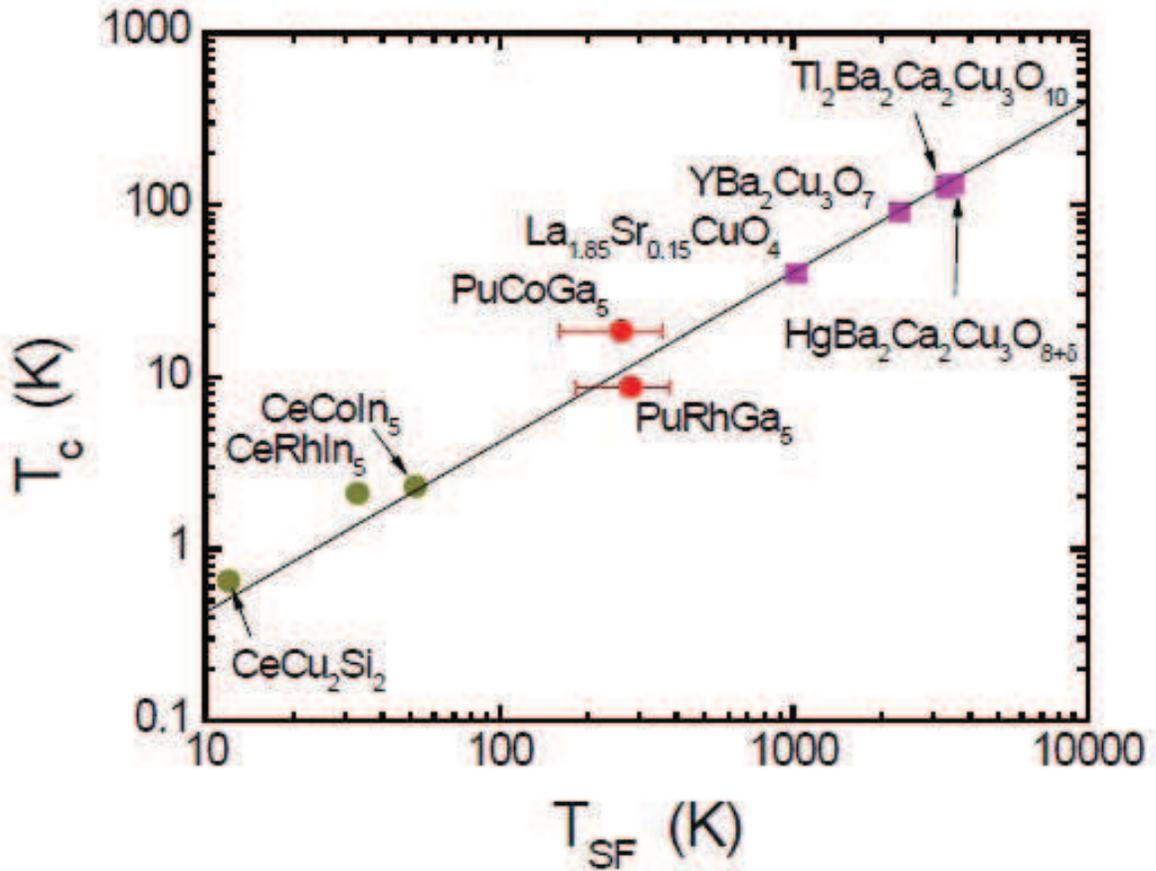}
\caption{A Moriya-Ueda like plot of the temperatures of various unconventional
superconductors plotted against $T_{\rm SF}$, a characteristic temperature indicating
the energy spread of the wave vector-dependent part of the spin-fluctuations
(after Curro et al.\protect\cite{ref:Curro}).\label{fig:Moriya-Ueda}}
\end{figure}

With a near neighbor hopping $t$, a nominal filling $\langle n\rangle\sim0.85$
and $U/t$ fixed, the size of the transition temperature $T_c$ scales with the
energy scale $t$. In this framework then, the range of $T_c$ values found
between the heavy fermion materials and the cuprates is seen as a reflection of
their electronic energy scales. This notion, that the variation of $T_c$
depended on a basic electronic energy scale of the material, was considered within a
fluctuation-exchange treatment of the single-band Hubbard model by
\citet{ref:Moriya03} who related this scale to a spin fluctuation energy
$T_{\rm SF}$. In their approach $T_{\rm SF}\simeq1.25\times10^4/\gamma$ with the
specific heat $\gamma$ measured in $mJ/{\rm mol}$ K$^2$ and the spin-fluctuation
cut-off wave vector taken to be of order the zone boundary wave vector. Based
on these results, they proposed a unified picture in which
$T_c$ varied as $T_{\rm SF}$. Alternatively, \citet{ref:Uemura} has used
an effective Fermi energy obtained from the penetration depth in place of $T_{\rm SF}$.
The basic idea is similar to what one finds in the Hubbard model where with $U$
and $\langle n\rangle$ optimized, $T_c$ is set by the energy scale $t$.
Figure~\ref{fig:Moriya-Ueda} shows this type of Moriya-Ueda plot with the
addition of the 115 Pu actinides.
\citet{ref:Curro} noted that the 115 Pu actinides could be added
to this group of materials providing a natural bridge between the heavy fermions
and the high $T_c$ cuprates. In this case, the larger $T_c$ values of the 115
Pu compounds relative to the 115 Ce systems is a reflection of the larger 
hybridization among the $5f$ electrons of the 115 Pu compounds and hence to a
larger value of the basic energy scale \citep{ref:Takimoto,ref:Hotta}. In a
similar way, the unit cell volume of PuCoIn$_5$ is nearly 28\% larger than that
of PuCoGa$_5$ \citep{ref:Zhu}, leading to a weaker hybridization and a reduced $T_c$.

In addition to the intermediate coupling requirement and the size of a basic
energy scale, the topology of the Fermi surface as well as the orbital weights
on the Fermi surface play an important role in determining $T_c$.
As noted in Sec.~\ref{sec:4}, the lattice structure and/or doping can alter the
number of Fermi surfaces of the Fe-pnictide materials. \citet{ref:Kuroki}
have suggested that the pnictogen height $h_{\rm pn}$ above the Fe layer controls the
appearance of a $d_{xy}$ hole pocket around the $(\pi,\pi)$ point of the unfolded
1 Fe/cell Brillouin zone. They noted that when $h_{\rm pn}$ is such that the
pnictogen ions form a nearly regular tetrahedron as in NdFeAsO$(T_c\sim50$K),
the nearest-neighbor hopping for the $d_{xy}$ orbital (here $x,y,z$ refer to
the single Fe/cell lattice) decreases and an additional $d_{xy}(\pi,\pi)$
hole pocket appears. Spin fluctuation mediated scattering of pairs between this
pocket and the $d_{xy}$ regions of the $\beta_1$ and $\beta_2$ electron pockets
at $(\pi,0)$ and $(0,\pi)$ lead to a nodeless $A_{1g}$ gap. However for LaFePO,
the pnictide P is closer to the Fe plane and the Fe-pnictogen-Fe angle is
considerably larger than that of a regular tetrahedron. In this case, the
$(\pi,\pi)$ hole Fermi pocket is absent and as discussed in Sec.~\ref{sec:4},
the spin-fluctuation and the Coulomb interaction favor a nodal $A_{1g}$ gap
which has a lower $T_c$. Similarly, as discussed by \citeauthor*{ref:Usui},
for the 1111 Fe-pnictide structure, if the Fe-pnictogen-Fe angle becomes small
relative to the regular tetrahedron, the $\alpha_1$ hole Fermi surface disappears
and $T_c$ decreases.

Multi-orbital effects also appear to play a role in the relative $T_c$ values of the
cuprates. Based on electronic structure calculations,
\citet{ref:Pavarini} observed that the $T_c$ of the hole cuprate
materials was related to the energy of a hybrid orbital formed between the
apical-oxygen and the planar coopers. They noted that the axial orbital
controlled the range $r$ of the intralayer hoppings
and $T_c$ was found to increase with $r$. This range parameter $r$ was
found to increase as the apical O moved away from the CuO$_2$ plane. It was
also suggested by \citet{ref:Ohta} that $T_c$ of the hole doped
cuprates was correlated with the energy difference between the apical O $p_z$
and planar O $p_\sigma$ orbitals. Recently, \citet{ref:Sakakibara}
argued that these correlations could be understood in terms of a two orbital
Hubbard model that included in addition to the $d_{x^2-y^2}$ Cu orbit of the
standard one-band Hubbard model an additional $d_{3z^2-r^2}$ orbit. They
focused on the question of why the superconducting transition temperature of
the single layer HgBa$_2$CuO$_{4+\delta}$ $(T_c\sim90K)$ is significantly
higher than the single layer La$_{2-x}$(Sr/Ba)$_x$CuO$_4$ $(T_c\sim40K)$.
Within the fluctuation-exchange approximation, they found that the
eigenvalue of the Bethe-Salpeter equation (\ref{eq:10}) decreased when the
$d_{x^2-y^2}$ orbital weight on the Fermi surface was reduced by an admixture
of $d_{3z^2-r^2}$ orbital weight. They noted that the $d_{3z^2-r^2}$
orbital weight was controlled by the height of the apex oxygen and the Madelung
potential difference between the planar and apical oxygens, in agreement
with the earlier proposals. The reduction of the pairing strength arising from
the admixture of other orbitals was also found in FRG calculations \citep{ref:Uebelacker}.
Similarly, the level splitting of a two orbital model of the 115 CeCoIn$_5$ and CeRhIn$_5$
heavy fermion materials has also been used to discuss their $T_c$ differences
\citep{ref:Takimoto1}. Here the $\Gamma_\alpha$ levels are split by the
tetragonal crystal field and $T_c$ was found to increase with this splitting.

With respect to guidance in the search for new and possibly higher temperature
superconductors, these results suggest that one is looking for materials
containing quasi 2D layers of 3d ions. One wants magnetic ions to boost the
amplitude of the spin fluctuations and 3d ions rather than 4d or 5d ions which
have a smaller effective Coulomb interaction or $4f$ or $5f$ ions
which have a narrower bandwidth and hence a smaller basic energy scale. One wants 2D
layers so that the antiferromagnetic order is suppressed and the spectral weight of the
spin fluctuations is in a frequency range several times the maximum gap where it
is most effective in pairing. In addition, in 2D it is possible that a larger
fraction of a cylindrical Fermi surface or surfaces can simultaneously be ``optimized" with
respect to the pairing \citep{ref:Monthoux2001}. The Fe-pnictides suggest a further
optimization scheme in which adding an additional Fermi surface
\citep{ref:Kuroki,ref:Usui} with a particular orbital character
allows for additional scattering processes leading to a higher $T_c$. Here, as discussed not
only the presence of the additional Fermi surface is important but it must
have the right orbital character. It is generally better
with respect to both the pairing strength and $T_c$ to have a nodeless gap
instead of a nodal gap, and therefore a multi-Fermi surface system is favored.

Finally, it may be possible to find structures which have spatial or dynamic
properties which enhance $T_c$. Here one has the idea of optimal inhomogeneity
in which a composite material consisting of a ``pairing region" with a large
gap scale is coupled to a ``metallic region" which provides phase stiffness
\citep{ref:KivFrad}. Examples of this range from weakly coupled two-leg
ladder systems \citep{ref:Arrigoni} which could have a period 4 bond-centered
stripe structure to layered materials \citep{ref:Berg08}. As noted in Sec.~\ref{sec:4},
one might also wonder whether it might be possible to alter the dynamic
structure of the spin-fluctuation spectrum in a manner that would increase $T_c$.
Here the idea would be to move the low frequency spin fluctuations to higher
frequency in the normal system so as to obtain the increase in the pairing
strength that is ultimately available in the usual superconducting state in
which the pairing gap has opened. Here of course one would need to do this
without suppressing the intrinsic pairfield susceptibility.

So we will end this review as it began by noting that while, in principle, the
momentum and frequency dependence of the superconducting gap can provide a
fingerprint to identify the pairing interaction, it will be the material record \citep{ref:Fisk}
that will tell us whether these ideas proved useful in providing guidance in
the search for new superconductors.

\appendix
\section{The Structure of Two Pairing Interactions}

As discussed in Sec.~\ref{sec:4}, the Coulomb interaction $U$ gives rise to
short-range antiferromagnetic spin fluctuations which produce a pairing
interaction that is non-local in space and retarded. In particular, as illustrated
in Fig.~\ref{fig:15}, this pairing
interaction is repulsive for two electrons on the same site but attractive if
the electrons are on near neighbor sites. Thus if the paired electrons are
spatially correlated so as to avoid occupying the same site, they can take
advantage of the non-local near-neighbor attractive part of the interaction.
This spatial non-local nature of the Hubbard model pairing interaction has an
analogy with the temporal, retarded nature of the familiar electron-phonon screened
Coulomb pairing interaction. In this appendix, the structure of the traditional
electron-phonon screened Coulomb interaction will be compared with the structure
of the spin-fluctuation interaction. Here to ease the notation, we will drop the
superscript index $pp$ and $\Gamma$ will denote the irreducible particle-particle
vertex which we will call the pairing vertex.

\subsection{The Electron-phonon Screened Coulomb Pairing Interaction}

\begin{figure}[!htbp]
  \vskip 5mm
  \includegraphics[width=15cm]{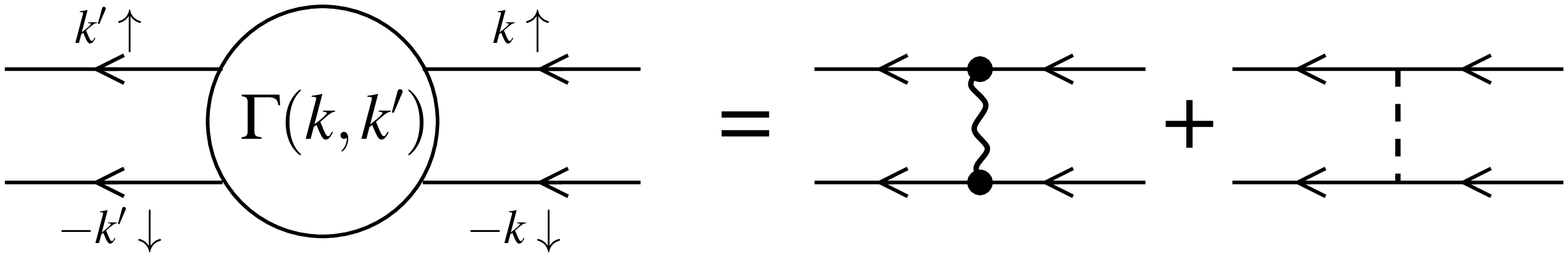}
  \caption{The traditional approximation to the pairing vertex $\Gamma(k,k')$
	for the electron-phonon screened Coulomb model. Here the wavy line represents
	the dressed phonon propagator, the dots the dressed electron-phonon couplings and
	the dashed line a screened Coulomb interaction.\label{appfig:1}}
\end{figure}

To begin, consider the well-known approximation of the pairing vertex for the
traditional electron-phonon screened Coulomb model \citep{appref:3} illustrated
in Fig.~\ref{appfig:1}
\begin{equation}
  \Gamma(q,\omega_m)=-\frac{|g_q|^22\omega_q}{\omega^2_m+\omega^2_q}+
	\frac{4\pi e^2}{q^2+\kappa^2}.
  \label{appeq:1}
\end{equation}
Here $q=k'-k$ and $\omega_m=\omega_{n'}-\omega_n$ are the momentum and Matsubara
energy transferred in the scattering, and we have omitted a sum over the phonon
polarizations. The first term in Eq.~(\ref{appeq:1}) is the phonon-exchange term
with $g_q$ the electron-phonon coupling constant and $\omega_q$ the phonon energy.
The second term is the screened Coulomb interaction with $\kappa$ the Thomas-Fermi
screening wavevector. This form of the vertex, with the phonon frequencies and
the electron-phonon coupling determined from bandstructure and linear response
calculations, has provided a useful approximation for the conventional
superconductors \citep{appref:4}. In this case, as discussed by \citeauthor*{appref:5}
and \citeauthor*{appref:6}, vertex corrections to the electron-phonon term
are of order the ratio of the Debye energy to the Fermi energy and can be
neglected. Furthermore, for materials with negligible magnetic correlations,
the screened Coulomb term (which ultimately is replaced by a Coulomb
pseudopotential $\mu^*$ \citep{appref:7,appref:8}) has proved an adequate
representation of the Coulomb interaction.

Continuing with the traditional approach, we note that the important pair
scattering processes take place on the Fermi surface and the dominant part of
the phase space is associated with large momentum transfers of order $2p_F$.
For these large momentum transfers, $g_q$ and $\omega_q$ are slowly varying
functions of $q$, as is the screened Coulomb interaction. This means that the
interaction is local in space but retarded in time. Averaging the momentum
transfer over the Fermi surface, and taking an Einstein spectrum $\omega_q=\Omega$
for the phonons, the pairing interaction becomes
\begin{equation}
  \Gamma(\omega_m)\approxeq-\frac{2|g|^2\Omega}{\omega^2_m+\Omega^2}+V_c
  \label{appeq:2}
\end{equation}
Here,
\begin{equation}
  V_c=\frac{\left\langle\frac{4\pi e^2}{q^2+\kappa^2}\right\rangle_{\rm FS}}{N(0)}
  \label{appeq:3}
\end{equation}
with $N(0)$ the single spin density of states at the Fermi surface.

\begin{figure}[!htbp]
  \vskip 5mm
  \includegraphics[width=12cm]{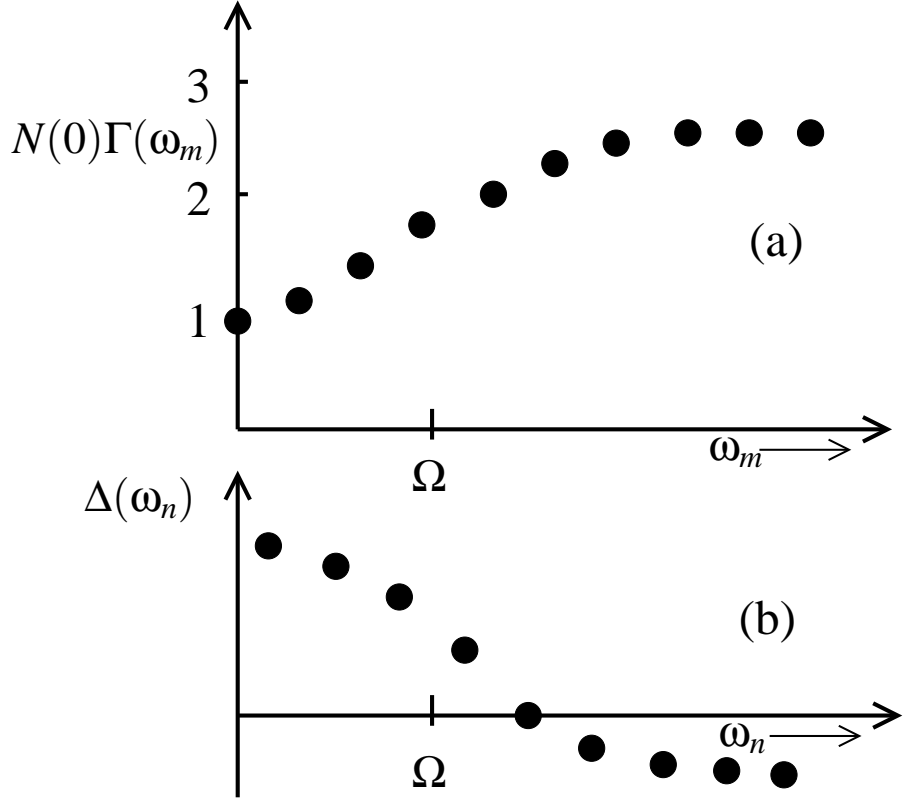}
  \caption{(a) The vertex $\Gamma(\omega_m)$ multiplied by the single particle
	density of states $N(0)$ versus $\omega_m=2m\pi T$. (b) The resulting gap
	$\Delta(\omega_n)$ associated with $\Gamma(\omega_m)$ versus
	$\omega_n=(2n+1)\pi T$. The change in sign of $\Delta(\omega_n)$ is such that
	the gap Eq.~(\protect\ref{appeq:4}) can be satisfied even though $N(0)\Gamma(\omega_m)$
	is positive for all $\omega_m$.\label{appfig:2}}
\end{figure}

A plot of $\Gamma(\omega_m)$ is shown in Fig.~\ref{appfig:2}a for a typical set
of parameters for which $-\frac{2|g|^2}{\Omega}+V_c>0$. In this case, the effective
pairing interaction is positive for all Matsubara frequencies $\omega_m$ and
might na\"\i vely appear to be repulsive.\footnote{In numerical solutions of the
Eliashberg equations it is convenient to cut off the frequency integrals at a
frequency $\omega_c$ of order five times the Debye frequency and replace
$\mu=N(0)V_s$ by a renormalized pseudo-potential
$\mu^*=\mu\left(1+\mu\ln\left(\frac{\mu_F}{\omega_c}\right)\right)^{-1}$
\protect\citep{appref:7,appref:8}. This renormalization takes into account the
fact that by decreasing the energy cut-off from $\mu_F$ to $\omega_c$ one has
eliminated Coulomb scattering processes which keep the electrons apart. The
phonon mediated part of the interaction is unchanged since $\omega_c$ is well
above the dynamic range of the phonons. From a renormalization point of view,
as the cut-off frequency is reduced $-\frac{2|g|^2N(0)}{\Omega}+\mu^*$ becomes
negative and one has an effective low energy theory with an attractive pairing
interaction. In this appendix, we are looking at the dynamics that underlies this
renormalization.} Nevertheless, at a critical temperature $T_c$ one finds that
there is a solution $\Delta(\omega_n)$ of the linearized BCS gap equation
\begin{equation}
  -T_c\sum_{n'}\frac{\pi N(0)\Gamma(\omega_n-\omega_{n'})}{|\omega_{n'}|}
	\Delta(\omega_{n'})=\Delta(\omega_n)
  \label{appeq:4}
\end{equation}
This is because, while $\Gamma(\omega_m)$ is a positive function of $\omega_m$,
it increases over an energy scale set by the characteristic phonon frequency
$\Omega$. In this case, the pair scattering strength is large and positive for
processes in which a pair is scattered from a smaller Matsubara frequency
$\omega_{n'}$ to a larger one $\omega_n$ such that $|\omega_n-\omega_{n'}|>\Omega$.
Then if $\Delta(\omega_{n'})$ is positive, the gap equation (\ref{appeq:4}) can
be satisfied provided $\Delta(\omega_n)$ is negative as shown in Fig.~\ref{appfig:2}b.
This ``sign-changing" frequency structure of the gap reflects the internal structure
of a pair in which the electrons are dynamically correlated to avoid the
``instantaneous" screened Coulomb interaction while taking advantage of the
retarded phonon mediated attraction.

\begin{figure}[!htbp]
  \vskip 5mm
  \includegraphics[width=10cm]{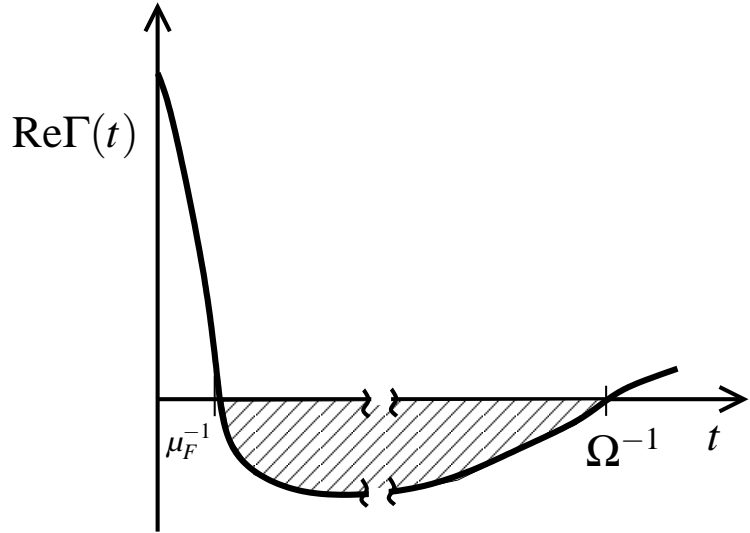}
  \caption{Schematic plot of ${\rm Re}\Gamma(t)$ versus $t$. The interaction is
	repulsive for times less than of order $\mu^{-1}_F$ and then attractive for
	times between $\mu^{-1}_F$ and the inverse of a typical phonon frequency
	$\Omega^{-1}$.\label{appfig:3}}
\end{figure}

Another way to see that $\Gamma(\omega_m)$ describes an attractive pairing
interaction is to replace $i\omega_m$ by $\omega+i\delta$ and take the
Fourier transform to determine the time dependence of the pairing
interaction \citep{ref:ScalRandom}
\begin{equation}
  \Gamma(t)=\int\frac{d\omega}{2\pi}e^{-i\omega t}
	\left(\frac{2|g|^2\Omega}{(\omega+i\delta)^2-\Omega^2}+V_c\right)
  \label{appeq:5}
\end{equation}
then
\begin{equation}
  {\rm Re}\Gamma(t)=-|g|^2\sin\Omega te^{-\delta t}+V_c\not\delta(t)
  \label{appeq:6}
\end{equation}
with $\not\delta(t)$ a broadened $\delta$-function of width $\mu^{-1}_F$.
For a more general phonon spectrum peaked at $\Omega$ with a width $\Delta\Omega$,
the first term decays for times larger than $\sim\Delta\Omega^{-1}$.
Taking these features into account, Fig.~\ref{appfig:3} shows a schematic plot
of Re$\Gamma(t)$ in which one sees that the repulsive Coulomb interaction lasts
for only a brief time of order the inverse of the Fermi energy while the
attractive part of the interaction lasts for a much longer time set by the
phonon spectral weight.

\subsection{The Spin-Fluctuation Exchange Pairing Interaction}

\begin{figure}[!htbp]
  \vskip 5mm
  \includegraphics[width=16cm]{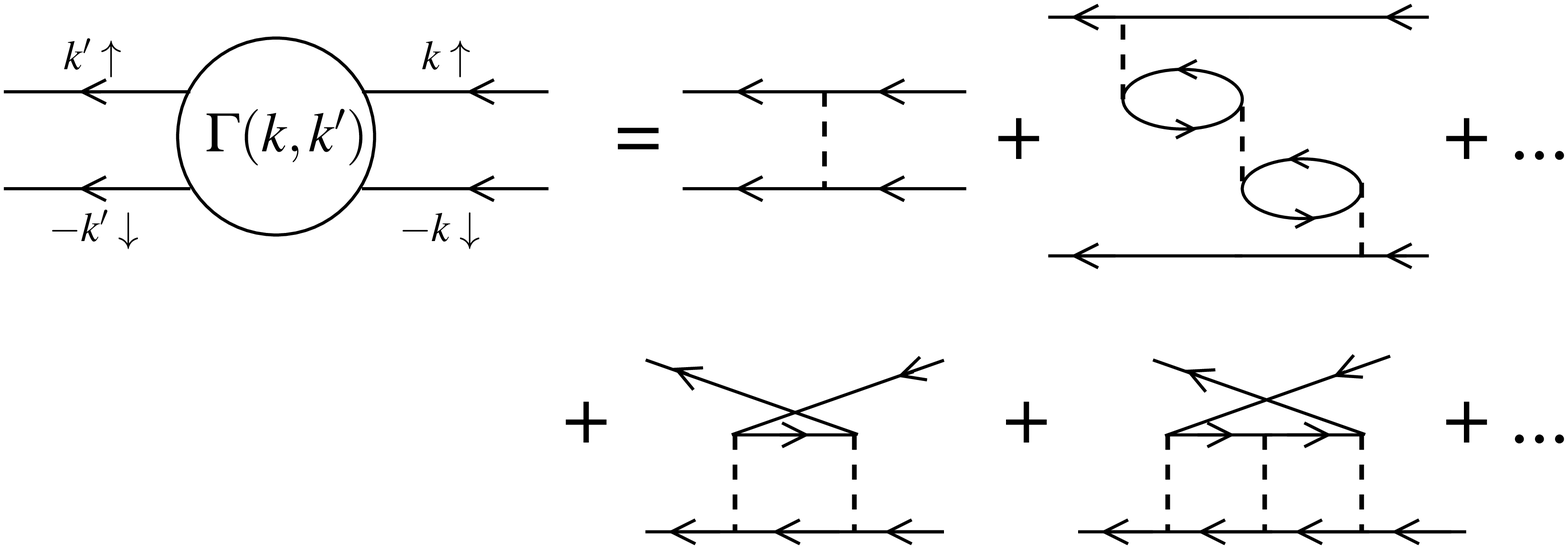}
  \caption{The RPA diagrams for the Hubbard model pairing interaction. The solid
	lines are bare single-particle Green's functions and the dashed lines the
	interaction $U$. Here one
	clearly sees that the electrons which make up the spin-fluctuation pairing
	interaction are the same electrons that pair.\label{appfig:4}}
\end{figure}

In weak coupling, the leading RPA diagrams for the irreducible singlet
particle-particle scattering vertex $\Gamma$ are shown in Fig.~\ref{appfig:4}.
These give
\begin{equation}
  \Gamma(k,k')=\frac{U}{1-U^2\chi^2_0(k'+k)}+
	\frac{U^2\chi_0(k'-k)}{1-U\chi_0(k'-k)}
  \label{appeq:7}
\end{equation}
Here $k=({\bf k},i\omega_n)$ and $k'=({\bf k'},i\omega_{n'})$ and
\begin{equation}
  \chi_0(q,i\omega_m)=\frac{1}{N}\sum_k
	\frac{f(\varepsilon_{k+q})-f(\varepsilon_k)}{i\omega_m-\varepsilon_{k+q}+\varepsilon_k}.
  \label{appeq:8}
\end{equation}
For a single, even frequency pair, the gap function is even under ${\bf k}$ goes
to $-{\bf k}$, so that one can replace $k'+k$ by $k'-k$ in the first term of
Eq.~(\ref{appeq:7}). Then, rearranging the terms in Eq.~(\ref{appeq:7}) gives
\begin{equation}
  \Gamma(k,k')=\frac{3}{2}U^2\frac{\chi_0(k'-k)}{1-U\chi_0(k'-k)}+
	\frac{U^2}{2}\frac{\chi_0(k'-k)}{1+U\chi_0(k'-k)}+U.
  \label{appeq:9}
\end{equation}
The first term is the contribution of the spin fluctuations with $\chi_0(1-U\chi_0)^{-1}$
the RPA spin susceptibility. The second term represents the charge fluctuations
and $U$ is the onsite Coulomb interaction. This interaction was first used by
\citeauthor*{appref:9} to describe the depression of $T_c$ due to
spin-fluctuations for $s$-wave superconductivity in Pd.

\begin{figure}[!htbp]
  \vskip 5mm
  \includegraphics[width=10cm]{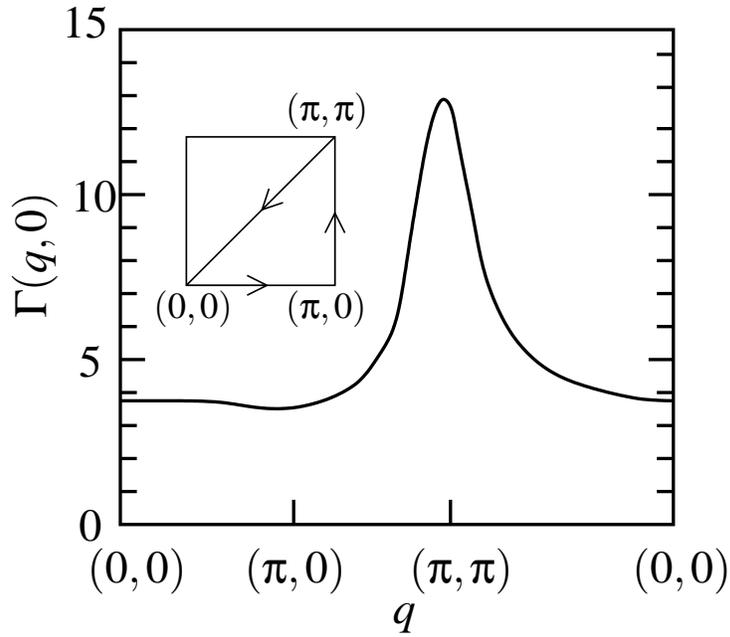}
  \caption{$\Gamma(q,0)$ versus $q$ for $q$ along a path in the Brillouin zone
	which is shown in the inset. An effective interaction that is peaked
	at a large momentum transfer is the origin of the unconventional superconductivity
	discussed in this review. Here $U=2t$, $t'=0$, $\langle n\rangle=0.87$ and
	$T=0.25t$.\label{appfig:5}}
\end{figure}

\begin{figure}[!htbp]
  \vskip 5mm
  \includegraphics[width=10cm]{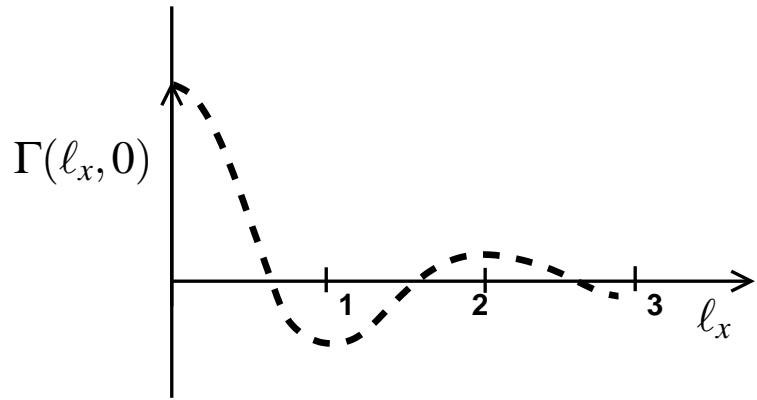}
  \caption{The spatial Fourier transform $\Gamma(\ell_x,\ell_y=0)$ versus
	$\ell_x$. Here one member of a pair is at the origin and $\Gamma(\ell_x,0)$
	is the interaction energy when a second electron is added in a single state at
	site $\ell_x$.\label{appfig:6}}
\end{figure}

For the 2D Hubbard model doped near half-filling, the dominant contribution to
$\Gamma$ comes from the first term which peaks near $(\pi,\pi)$ reflecting the
short range antiferromagnetic correlations. A plot of $\Gamma(q,0)$
versus momentum transfer $q$ is given in Fig.~\ref{appfig:5} for $q$ along a path
in the Brillouin zone shown in the inset. This interaction is positive for all
momentum transfers. Therefore, for there to be a transition to a
superconducting state, the gap function $\Delta(k)$ must have a change of sign
on the Fermi surface in order to satisfy the BCS equation.
\begin{equation}
  \Delta(k)=-\frac{1}{N}\sum_{k'}\frac{\Gamma(k-k')\Delta(k')}{2\varepsilon_k}
	\tanh(\beta_c\varepsilon_k/2).
  \label{appeq:10}
\end{equation}
For the nearly half-filled 2D Hubbard model, Eq.~(\ref{appeq:10}) leads to the
well-known $\Delta(k)=\Delta_0(\cos k_x-\cos k_y)$ $d_{x^2-y^2}$ gap. In this
case, $(k\uparrow,-k\downarrow)$ pairs with $k$ near $(\pi,0)$ which have a
negative gap are strongly scattered by the antiferromagnetic spin fluctuations
to $(k'\uparrow,-k'\downarrow)$ pairs with $k'$ near $(0,\pi)$ which have a
positive gap, satisfying Eq.~(\ref{appeq:10}). This sign change in the momentum
dependence of $\Delta(k)$ reflects an internal structure of a pair in which the
electrons are spatially correlated such that they avoid occupying the same site
while taking advantage of the non-local attractive regions of the interaction.
It is a $d_{x^2-y^2}$ pair rather than an extended $s$-wave $(\cos k_x+\cos k_y)$
pair because it is made up from states near the nearly half-filled Fermi surface.
This structure of the interaction is
illustrated in Fig.~\ref{appfig:6}, which shows the spatial Fourier transform
of $\Gamma(q)$. Here one member of the pair is located at the origin and
another at site $(\ell_x,0)$.

Thus both the conventional and the unconventional superconductors have ``sign
changing gaps." For the conventional case this sign change occurs in the
frequency dependence of the gap and reflects the dynamic correlations of
the electrons which form the Cooper pairs. In the case of the unconventional
superconductors, the sign change occurs in the momentum dependence of the gap
and reflects the spatial correlations of the paired electrons. Naturally,
there are also dynamic correlations since the spin-fluctuations are retarded
and similarly in the phonon case there are some spatial correlations due to the
momentum dependence of the electron-phonon interaction. However, the
characteristic feature of the spin-fluctuation interaction is its momentum
dependence which leads to a spatially non-local pairing interaction, while the
characteristic feature of the phonon mediated pairing interaction is its
frequency dependence which leads to a retarded pairing interaction.

\newpage
\section*{Acknowledgments}

This review article represents a more detailed account of talks given at the
M2S09 conference and a subsequent CIFAR meeting. The author would especially
like to thank CIFAR and the members of the Quantum Materials Program for
providing an environment where ideas can be discussed, debated and criticized.
I would like to acknowledge the close collaboration that I have enjoyed working
with T.A.~Maier, and the many useful discussions with A.D.~Christianson,
E.~Dagotto, M.~Lumsden, D.~Mandrus, A.~Moreo, H.A.~Mook and D.~Singh at ORNL.
I would like to thank E.~Berg, S.~Graser, P.~Hirschfeld, A.F.~Kemper,
S.A.~Kivelson, S.~Raghu and S.R.~White for joint work and many discussions.
I am grateful to Z.~Fisk, R.~Greene, W.~Hanke, D.~Hone, B.~Keimer, K.~Kuroki,
T.M.~Rice, S.~Sachdev, Z.-X.~Shen, L.~Taillefer and J.~Thompson for sharing
their insights. Finally, I want to thank N.~Ghimire, S.~Graser, and R.~Melko
for help with some of the figures and express my gratitude to D.L.~Ceder for her
patience and dedication in putting this manuscript together. The author
acknowledges the support of the Center for Nanophase Materials Science at ORNL,
which is sponsored by the Division of Scientific User Facilities, U.S.~DOE and
thanks the Stanford Institute of Theoretical Physics for their hospitality.

\bibliography{commonthread-varXiv}


\end{document}